\documentclass[final,3p,times,twocolumn,authoryear]{elsarticle}

\usepackage{amssymb}
\usepackage{amsmath}
\usepackage{amsthm}
\newtheorem{hypothesis}{Hypothesis}
\usepackage{booktabs}
\usepackage{tabularx}
\usepackage{dblfloatfix}
\usepackage{multirow}
\usepackage{threeparttable}
\usepackage{doi}
\usepackage{xcolor}
\usepackage{amssymb}
\usepackage{pifont}
\newcommand{\cmark}{\ding{51}}%
\newcommand{\xmark}{\ding{55}}%

\begin{document}

\begin{frontmatter}



\title{Neither Replacement nor Panacea: Comparing LLM-Based Conversational and Graphical Decision Support in Industrial Tasks} 


\author[pisa]{Roberto Figliè\corref{cor1}}
\ead{roberto.figlie@phd.unipi.it}

\author[pisa]{Simone Caputo}

\author[brunel]{Alan Serrano}

\author[brunel]{Daria Mikhaylova}

\author[pisa]{Tommaso Turchi}

\author[pisa]{Daniele Mazzei}

\cortext[cor1]{Corresponding author}

\affiliation[pisa]{
    organization={Department of Computer Science, University of Pisa},
    city={Pisa},
    country={Italy}
}

\affiliation[brunel]{
    organization={Department of Computer Science, Brunel University of London},
    city={Uxbridge},
    country={United Kingdom}
}

\begin{abstract}
Managers in manufacturing settings rely on digital interfaces to interpret operational data for decision-making, but growing data volume and complexity can make relevant insights difficult to identify efficiently. While dashboards remain dominant in industrial contexts, Large Language Model (LLM)-based conversational agents (CAs), accessed through conversational user interfaces (CUIs), may provide more direct access to such data. However, their effectiveness may depend on the information-processing demands of the task. This study compares an LLM-based CA delivered through a CUI with a dashboard in a manufacturing decision-support scenario. In a mixed factorial experiment with a 2 $\times$ 3 design, 134 industrial decision-makers were assigned to one interface condition and completed three tasks of increasing complexity. We examined perceived Mental Workload (MWL), decision accuracy, completion time, and intended reliance, and tested self-reported data literacy as a moderator. Results showed that the CUI reduced perceived MWL overall and supported faster completion in less demanding tasks, but both advantages diminished as task complexity increased. Neither interface produced a consistent overall advantage in decision accuracy, and the CUI was not preferred as a sole basis for subsequent decisions. Furthermore, data literacy did not reliably moderate interface effects. These findings indicate that conversational interaction offers conditional rather than universal benefits for industrial decision support. LLM-based CAs may reduce information-access effort, whereas complex decisions continue to benefit from persistent, inspectable visual representations. Future systems should therefore combine conversational access with graphical overview and verifiable evidence rather than treating either interaction style as a replacement for the other, while future research should examine whether these findings generalize to other decision-making contexts and domains.
\end{abstract}







\end{frontmatter}


\section{Introduction}
\label{intro}

Industrial decision-making is increasingly shaped by data. Under the broad umbrella of smart manufacturing and Industry 4.0, organizations collect and integrate growing volumes of heterogeneous information from production systems, sensors, enterprise platforms, quality records, and supply-chain processes \citep{allen_data_2021}. Yet, despite this expansion in data availability, the dominant way of accessing and acting on such information remains largely unchanged: in manufacturing and operations, decision support is still commonly mediated through graphical, dashboard-based interfaces \citep{lindner_behavioral_2025}.

This predominance of dashboards is understandable. Visual interfaces can condense large amounts of information, support monitoring, and provide decision-makers with an overview of operational conditions through charts, tables, and other visual summaries \citep{yigitbasioglu_review_2012,hjelle_organizational_2024}. In industrial settings, they have become a practical interface layer through which data is turned into action. However, the same visual paradigm can also become demanding as data environments grow in scale and heterogeneity. Decision-makers may need to navigate across multiple views, compare distributed cues, and piece together relevant evidence before reaching a conclusion. In such contexts, the challenge is accessing and using data in a form that supports decision work efficiently and sustainably. As data assume a more central role in Industry 4.0 environments, this challenge also makes users' proficiency in working with data and with data-driven systems increasingly consequential \citep{cezar_cognitive_2023}. Put differently, the effectiveness of industrial decision support may depend not only on the interface itself, but also on the extent to which users are able to interpret, navigate, and act on the information it provides.

Recent advances in generative AI have made this issue more salient. Large Language Models (LLMs), in the form of Conversational Agents (CAs), are increasingly being proposed as a new access layer for data and analytical systems, including in manufacturing and related industrial domains \citep{handler_large_2024,elbasheer_natural_2025,cimino_integrating_2025}. Their appeal lies in the possibility of interacting with complex systems through Natural Language (NL): rather than navigating a dashboard, users can ask questions directly, request clarifications, and potentially obtain synthesized answers that reduce some of the effort involved in locating and combining information. More broadly, this has renewed interest in conversational user interfaces as a possible alternative, or complement, to traditional graphical interfaces for decision support \citep{schobel_charting_2024}.

At the same time, conversational access should not be assumed to be uniformly superior. In fact, dashboards and conversational interfaces embody different interaction logics. On the one hand, dashboards offer persistent overview, direct manipulation and simultaneous visibility of multiple information elements, and direct inspectability of the data representation. By contrast, LLM-based conversational interfaces  mediate access to information through language providing stepwise dialogical exchange. The practical value of one or the other therefore depends on how well each interaction paradigm supports the kind of decision work users must perform. Or, in other words, on the fit between the interaction format and the cognitive demands of the task. This expectation is consistent with Cognitive Fit Theory, which argues that problem-solving performance is shaped by the match between the problem representation and the strategies required by the task \citep{vessey_cognitive_1991}.

This issue is especially relevant in industrial contexts, where decision tasks can vary substantially in the amount and type of information they require users to retrieve, compare, and integrate. 

Despite growing interest in LLM-based systems, direct empirical evidence comparing conversational and dashboard-based interaction for industrial decision support remains limited. 
To address this gap, this paper presents a quantitative confirmatory study comparing an LLM-based conversational interface and a dashboard in industrial decision tasks based on the exploratory mixed-approach results obtained in a previous work. The study tests whether interface type is associated with differences in users' task performance and perceived demands, and whether these effects vary as task demands increase. By doing so, it aims to provide more rigorous evidence on when conversational access to industrial data may constitute a meaningful alternative to conventional dashboard-based interaction.

\section{Theoretical Background}


\subsection{Information Overload}

Information Overload (IO) has been defined as the exceeding of processing capabilities given an amount of information, specifically occurring when information-processing requirements surpass an individual’s cognitive capacity \citep{arnold_dealing_2023, eppler_concept_2004, roetzel_information_2019}. Historically, this hiatus between cognitive limits and information volume was routinely lamented, but as technology progressed, these gaps were managed, reduced, or brought to higher levels of abstraction  \citep{blair_information_2011}. Such a concept becomes even more relevant in the context of the hyper-connected world of Industry 4.0, where huge data flows must be processed and visualized; here, IO acts as a primary stressor in the "information society". Managers are now called to pursue decision-making through a data-driven approach granted by the Internet of Things (IoT) and Cloud Computing, yet IO remains a pervasive threat at the interface level \citep{shahrzadi_causes_2024}. 

This phenomenon is commonly represented with the inverted U-curve inspired from the Yerkes-Dodson law, which shows that decision accuracy (and therefore its quality) declines after reaching a specific threshold of information load \citep{eppler_concept_2004}. Beyond this critical point, the burden of a heavy information load confuses the individual, impairs the ability to set priorities, and makes prior information more difficult to recall. In managerial decision-making, \cite{phillips-wren_decision_2020} categorized IO as a "Decision Stressor" (together with time pressure, complexity, and uncertainty) that is specific to the problem at hand and significantly impacts the quality of outcomes. 

Modern research emphasizes that IO is not merely a product of information quantity but is exacerbated by information complexity, low data quality, and the "technostress" induced by constant digital availability \citep{arnold_dealing_2023, shahrzadi_causes_2024}. Consequently, achieving objective and effective decision-making in modern industrial settings requires mitigating these cognitive pressures through structural preventions, such as visualization dashboards in Decision Support Systems (DSS), intelligent filtering agents, and the enhancement of organizational information management \citep{arnold_dealing_2023}. 

As IO cannot be measured directly with an objective instrument and is strongly context-dependent, studies typically assess it via either subjective measures, performance-based measures, or a combination of them \citep{kock_information_2011}. Previous research also shows that IO could be related to MWL \citep{eppler_concept_2004, speier_influence_1999},  as it is reasonable to think that as information load increases, the need for cognitive processing capabilities grows accordingly. In this work, we induce higher vs. lower information load by varying the number of information cues required to reach a decision (i.e., manipulating task complexity), and we assess its impact through performance (decision accuracy and completion time) and subjective demand (MWL) as downstream indicators.

\subsection{Mental Workload}
Mental workload (MWL) is a core Human Factors and Ergonomics construct concerned with the demands imposed by a task and system, the effort required from the operator to meet those demands, and the potential consequences for performance \citep{babaei2025nasatlx}. It is often used to understand when task demands and system characteristics risk pushing operators toward degraded performance, either because demands exceed available resources (overload) \citep{young_state_2015}. Operationally, there is no single technique that can unambiguously indicate whether workload is “too high” or “too low,” and definitional debates have persisted for decades \citep{casner_measuring_2010, young_state_2015}.

Originally conceived for aviation contexts, MWL has been increasingly used in HCI to quantify cognitive demands imposed by interactive systems, but this import has been uneven. \cite{kosch_survey_2023} note a “vague understanding” of cognitive workload in HCI and the absence of a standard definition, motivating more explicit construct specification in studies. Accordingly, in this study we adopt the integrative definition of MWL by \citep{longo_human_2022}, as the degree of activation of a finite pool of cognitive resources over time while cognitively processing the primary task, expressed through devoted effort and attention to cope with task demands, and mediated by stochastic environmental/situational factors as well as internal operator characteristics. 

This definitional plurality has direct consequences for measurement choices. MWL (similarly to IO) has been assessed through subjective self-report questionnaires, such as NASA-TLX, the Bedford scale, and RSME; psychophysiological measures, such as eye tracking, pupil dilation, skin conductance, and electroencephalogram; performance measures, such as time, errors, and accuracy; and analytical approaches, each embedding different assumptions about what workload is \citep{longo_human_2022}. 

Recent critiques argue that NASA-TLX’s dominance in HCI is not automatically justified \citep{babaei2025nasatlx, mosaferchi_simple_2025}, as it was not designed for modern digital interaction contexts and its ease of use can encourage uncritical deployment without clarity on what it actually measures. Moreover, variations in instrument versions/modifications and analysis decisions (e.g., weighting, aggregate vs subscale interpretation) can further undermine validity if not explicitly reported and theoretically justified. 

From a Cognitive Fit Theory (CFT) perspective---i.e., the idea that problem-solving performance results from the interaction between the problem representation and the task---MWL can also be treated as a cognitive cost arising when the system's external representation does not align with the representation required by the task. When either representation is poorly suited to the task, users may need to invest additional effort in translating, integrating, or re-encoding information before they can reason with it \citep{vessey_theory_2007, speier_influence_2003, padilla_decision_2018}. Such misfit is expected to increase MWL and, in turn, slow decisions and/or reduce accuracy. Prior work links differences in information representation and interface design to both decision performance and subjective workload \citep{speier_influence_2003, nuamah_evaluating_2020}. Although CFT was originally developed around graphical and tabular representations, the same rationale can be extended to LLM-based CUIs: dashboards externalize information in a persistent visual space, whereas CUIs mediate access to information through sequential, language-based exchange. This distinction is consistent with the contrast between diagrammatic and sentential representations described by \citet{larkin_why_1987}.  This motivates our comparison of a visualization-heavy dashboard and an LLM-based conversational interface, whose distinct representational forms may alter MWL even under comparable information demands.

Against this background, in this study we treat MWL as a construct capturing the cognitive costs associated with experimentally manipulated information load and interface type. For subjective MWL assessment in an unmoderated online scenario like ours, we use the NASA-TLX because it remains the de facto standard in HCI \citep{kosch_survey_2023}, enabling comparison with prior work while making known methodological pitfalls well documented. We use the unweighted “Raw TLX" aggregate score, as the weighting procedure adds participant burden and has limited evidence of added diagnostic value  \citep{hart_nasa-task_2006, hertzum_reference_2021}.

\subsection{Task Complexity }
\label{sec:task_complexity}
Task complexity has no single, universally accepted definition, consequently it is typically specified in terms of the properties that make a task complex. Reviews across information-systems and ergonomics traditions converge in treating task complexity as multidimensional rather than a single scalar attribute. For example, \citet{liu_task_2012} synthesize prior work into structuralist, resource-requirement, and interaction perspectives, highlighting that complexity may stem from task structure (e.g., number and arrangement of elements), the cognitive resources required, and the dynamic interplay between task, environment, and operator.

\cite{gill_task_2006} also identify discretion (i.e., multiple viable strategies) as one of the main elements supporting task complexity (alongside task and problem space). This is particularly important because different levels of discretion for a task can lead to very different complexities. In the case of high discretion a person can find different ways to cope with the task requirements. For example, when choosing between two similarly rated headphones on an e-commerce site, an individual might flip a coin instead of reading a detailed comparison. This illustrates why experiments that manipulate complexity must constrain discretion when feasible to ensure participants engage with the intended complexity. 

Relatedly, \citet{liu_task_2012} distinguish \emph{objective} from \emph{subjective} task complexity. While in objective computations task complexity is determined by a number of a priori quantifiable elements, subjective complexity takes a constructivist stance stating that the complexity of a task is not an inherent property of the task itself, rather it is contingent on the individual performing the task. This matters because task complexity is typically treated as a task-side antecedent of MWL \citep{longo_human_2022}. 

However, for experimental work, an explicit \emph{objective} formalization can be preferred as it supports reproducible manipulations and clearer separation from downstream psychological outcomes. A widely adopted operational framework is provided by \citet{wood_task_1986}, who decomposes task complexity into \emph{component} complexity (e.g., number of cues/acts), \emph{coordinative} complexity (e.g., interdependence and integration among cues/acts), and \emph{dynamic} complexity (e.g., changes over time). This decomposition has been used in HCI decision-making contexts to standardize tasks and enable controlled comparisons across studies \citep{salimzadeh_missing_2023}, or to define experimental complexity conditions  \citep{falschlunger_infovis_2016, salimzadeh_dealing_2024}.

In this study we operationalize task complexity primarily in objective terms following Wood’s component and coordinative dimensions ---dynamic complexity is held constant because tasks are static ---, by manipulating the number of information cues required and the acts upon those cues needed to reach a decision.  Therefore, subjective experience of difficulty/demand is treated as an outcome captured by MWL measures.

\subsection{Data Literacy}

Data literacy (DL) corresponds to the individual’s knowledge and skill set for understanding and using data effectively. Multiple definitions and frameworks emphasize that DL entails more than basic numeracy as it covers the ability to manage and interpret data, draw appropriate conclusions, and integrate data into decision-making \citep{gartner, pinto_strategic_2023} . DL is often positioned as part of (or adjacent to) information literacy, with overlaps with statistical/quantitative literacy but a stronger emphasis on data practices and data-driven reasoning in applied contexts (e.g., finding, assessing, interpreting, and using data for decisions) \citep{pinto_strategic_2023, portner_data_2024}. 

Beyond definitional work, information-systems research has examined DL as a capability that shapes how people experience and cope with data-intensive work. In particular, \cite{cezar_cognitive_2023} connect DL to overload-related constructs (e.g., perceived data overload/technostress and related strain) as well as to performance-relevant outcomes in organizational settings.

In Industry 4.0, decision-makers face continuous, heterogeneous operational data streams (e.g., IoT/sensor, production and maintenance data) and are increasingly expected to act on analytics outputs under time constraints \citep{dalenogare_expected_2018}. Reviews highlight the centrality of data-driven decision-making and near–real-time decision processes \citep{nouinou_decision-making_2023, bousdekis_review_2021}; precisely the kind of settings where limited data literacy can hinder effective use of analytics and decision-support outputs \citep{ghobakhloo_drivers_2022}.

The rise of GenAI-enabled interfaces can reduce some “access” barriers (e.g., allowing natural-language querying), but it can also shift demands toward formulating good questions, checking assumptions, and validating outputs—making DL (and adjacent AI/data-critique competencies) more, not less, consequential in practice \citep{koloski_data_2025}. 

In this study, DL is treated as a pre-existing covariate that may shape task performance and subjective outcomes independently of the interface manipulation. We measure DL before the experimental tasks using the Brief Data Literacy Instrument (BDLI) by \cite{kirby_developing_2024}, a short self-report scale developed as a streamlined version of the 25-item Global Data Literacy Benchmark (GDLB) and psychometrically evaluated using confirmatory factor analysis and item-response theory across student and professional samples. We use BDLI scores both (i) as a covariate in analyses and (ii) to support balanced assignment across interface conditions.

\subsection{Intended Reliance}
Reliance is often understood as a behavioural consequence of trust in the reliability of an automated decision-support system \citep{Muir1987,lee_trust_1994,lee_trust_2004, Madhavan2007}. In interactions with classifiers or detection systems, reliance is commonly operationalized as the rate at which users accept the model's suggestions. The widespread adoption of complex AI models in decision-making, and their recognized influence on users' reasoning processes, has therefore led the HCI research community to investigate appropriate reliance, under-reliance, and over-reliance \citep{lee_trust_2004,Schemmer2023}.

Since no unique acceptance rate is agreed upon as normative, these concepts are often defined in relation to the correctness of the AI's suggestion and, in more recent studies, the correctness of the user's own decision. For instance, reliance may be considered excessive when the acceptance rate: (i) exceeds the accuracy of the system; (ii) is higher for incorrect suggestions than for correct ones; or (iii) is higher for incorrect suggestions when the user would have been correct without support than for correct suggestions when the user would initially have been wrong.

Reliance is also influenced by several factors. On the system side, these include experienced reliability, the presence of explanations, confidence scores for suggestions, and the interaction pattern. On the user side, they include factors such as confidence in one's own judgement. Reliance is also shaped by contextual and external factors, including time pressure, training interventions, and speed of interaction. In particular, the difficulty of the overall AI-assisted decision process can affect reliance: stronger over-reliance has been associated with more challenging tasks \citep{vasconcelos_explanations_2023,zhang_you_2024,wohleber_impact_2016,salimzadeh_dealing_2024} and with more elaborate explanations \citep{vasconcelos_explanations_2023,wright_effect_2016}.

In this study, however, reliance is not measured as behavioural acceptance of discrete AI recommendations. The comparison concerns two decision-support interfaces, a dashboard-based GUI and an LLM-based CUI, only the latter of which relies on AI. For this reason, the study focuses on an antecedent-oriented measure of reliance, namely intended reliance, drawing on the role of intention formation in models of trust and reliance in automation \citep{lee_trust_2004}, and treats the GUI condition as the baseline. Intended reliance captures participants' stated willingness to rely on the system for subsequent decision tasks after interacting with it. It is therefore treated as a post-use evaluative outcome that precedes actual behavioural reliance, distinct from objective task performance, perceived workload, usability, and confidence, but potentially shaped by all of them.
\section{Research model and hypotheses}

\begin{figure*}[!t]
\centering
\includegraphics[width=1\textwidth]{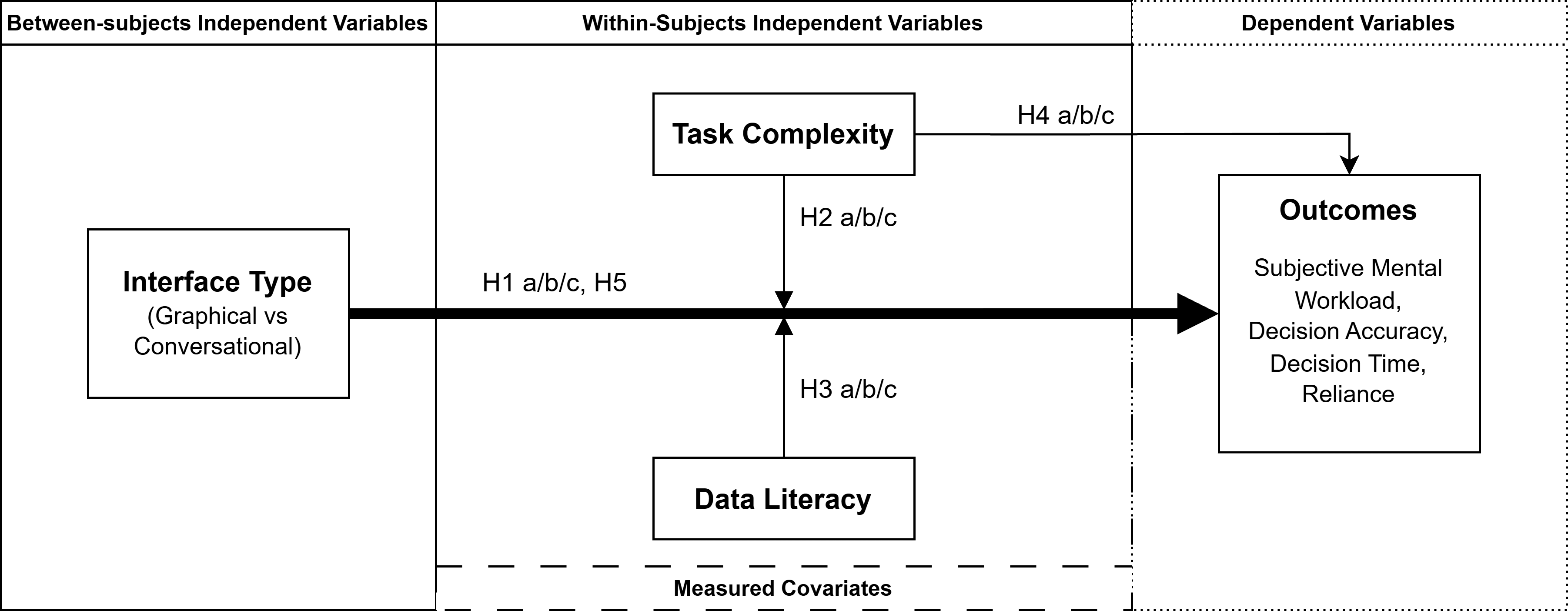}
\caption{Study Research Model}\label{fig:model}
\end{figure*}


%

\subsection{Primary hypotheses}

Our primary hypotheses concern (i) the overall effect of interface type and (ii) whether this effect varies as a function of task demands (complexity) and individual capabilities, specifically data literacy (DL).

\paragraph{Interface effects}
As argued above, an LLM-based CUI may reduce part of the cognitive effort involved in decision making by partially automating information access and supporting users through a stepwise interaction. Rather than requiring users to manually search across multiple views and integrate relevant cues on their own, the system can help narrow the information space and directly return task-relevant outputs \citep{flohr_chat_2021,kavaz_chatbot-based_2023, liu_conversational_2024}. In this sense, conversational interaction may reduce search, navigation, and intermediate integration costs.

At the same time, this benefit may come with countervailing costs. Because interaction is mediated through language, users have less direct manipulation of the underlying information space and may have to infer the system's current interpretation from sequential conversational outputs rather than from visual structures and explicit controls \citep{masson_directgpt_2024}. This may reduce inspectability, make cross-checking harder, and constrain the simultaneous consideration of multiple cues.

By contrast, GUIs --and dashboards as such-- offer persistent overview, simultaneous access to multiple information elements, and greater support for direct manipulation. These properties can be advantageous in decision-making contexts where users need to compare alternatives, inspect several indicators at once, or maintain awareness of the broader information space \citep{flohr_chat_2021,liu_conversational_2024}. However, these same properties can also increase cognitive demands when the display is dense, when many elements compete for attention, or when users must determine for themselves which information is relevant to the task at hand \citep{ke_effect_2023}.

Overall, the two interaction styles are therefore expected to redistribute cognitive costs differently, thus shifting workload in different ways. As suggested in a previous exploratory study, the conversational interface may reduce workload by compressing access to relevant information and guiding the interaction, whereas the dashboard may better support visual inspection, overview, and explicit comparison. These competing mechanisms suggest that differences in subjective MWL should be more pronounced than differences in decision accuracy. In particular, decision accuracy may differ between interfaces, but the direction of this difference is difficult to predict \textit{a priori} because each interface supports different aspects of the decision process. By contrast, completion time may favour the conversational interface overall, insofar as prior comparative work suggests that conversational systems can reduce upfront learning demands and limit the amount of information that must be processed at each step \citep{liu_conversational_2024}. We therefore hypothesize that:

\begin{hypothesis}
Overall, the conversational interface yields (a) lower subjective MWL, (b) different decision accuracy, and (c) shorter completion time than the graphical interface.
\end{hypothesis}

\paragraph{Interface $\times$ task complexity}
As objective task complexity increases, participants must process, retain, and coordinate a larger set of information cues. One possible expectation is that, if the LLM-based CUI functions as a cognitive assistant by partially automating information access and integration, its advantage should become more pronounced at higher complexity levels. However, prior work on Natural Language Interfaces (NLI) and CFT, together with findings from our first exploratory study, suggests that this expectation may be overly simplistic. In repeated use, initial advantages of conversational interaction may diminish relative to graphical interfaces \citep{liu_conversational_2024}; likewise, earlier work comparing textual and visual interfaces found that performance advantages can shift as task complexity increases \citep{speier_influence_2003}. This is also consistent with later CFT accounts, which suggest that representation effects are contingent not only on task--format fit, but also on task complexity and the resulting cognitive processing demands \citep{vessey_theory_2007,speier_influence_2006}.

Consistently, our first exploratory study suggested that interface differences were task-dependent rather than uniform across levels of demand. As tasks become more demanding, the benefits of NL interaction may be offset by the need to revisit prior outputs, compare multiple cues across turns, retain intermediate information in memory, and verify answers whose basis is not directly visible. In such situations, the persistent overview, simultaneous visibility of information, and greater inspectability provided by GUIs may become increasingly advantageous. Taken together, these mechanisms suggest that, even if the LLM-based CUI offers advantages at lower levels of task demand, its benefits may diminish as complexity rises. We therefore predict that:

\begin{hypothesis}
Task complexity moderates the effect of interface type on (a) MWL, (b) decision accuracy, and (c) completion time, such that the overall advantage of the conversational interface attenuates as task complexity rises.
\end{hypothesis}

\paragraph{Interface $\times$ data literacy} DL is expected to moderate interface effects because it shapes users' ability to interpret, evaluate, and use data in decision-making contexts \citep{pinto_strategic_2023,portner_data_2024}. Prior Information Systems research also links DL to how users experience data-intensive work, including perceived overload, and performance-relevant outcomes \citep{cezar_cognitive_2023}. Because the dashboard requires more self-directed data interpretation and integration than the chatbot, lower DL should be penalized more strongly in the GUI condition, increasing the interface gap at low DL. Conversely, higher DL should enable individuals to better leverage the dashboard’s visualizations and data manageability, attenuating the interface gap and potentially reversing it for decision accuracy. We therefore predict that:

\begin{hypothesis}
Data literacy moderates the effect of interface type on (a) MWL, (b) decision accuracy, and (c) completion time, such that lower data literacy disadvantages the graphical interface more than the conversational interface.
\end{hypothesis}

\subsection{Secondary hypothesis: objective task complexity}
We also test the direct effect of objective task complexity on subjective and objective indicators of information-processing demands. This secondary hypothesis functions as a manipulation-oriented check: it assesses whether the task complexity manipulation, designed according to Wood's model, captures meaningful differences in perceived and behavioural task demands. This is important for interpreting the moderation hypotheses, because any $interface \times complexity$ interaction is more informative if the complexity levels correspond to substantively different levels of task demand.

Objective task complexity theory suggests that more complex tasks require users to process more components and coordinate more relations among them \citep{wood_task_1986}. Consistently, overload accounts predict that increasing information-processing demands should increase subjective workload and impair performance when they exceed available cognitive resources \citep{kock_information_2011,chen_effects_2009}. We therefore expect higher task complexity to increase MWL, reduce decision accuracy, and extend completion time independently of the interface used:

\begin{hypothesis}
Higher task complexity leads to (a) higher MWL, (b) lower decision accuracy, and (c) longer completion time.
\end{hypothesis}

\subsection{Exploratory hypothesis: reliance}
Finally, we examine intended reliance as an exploratory post-use outcome. Reliance is commonly treated as a behavioural consequence of trust in automation, but it is shaped by several factors, including perceived system reliability, task difficulty, confidence, and contextual demands \citep{lee_trust_2004,Schemmer2023,vasconcelos_explanations_2023,zhang_you_2024}.

The direction of this effect is not specified \textit{a priori} because the two interfaces expose different reliance trade-offs. A previous exploratory study conducted by us suggested that a LLM-based CA can reduce interactional effort, but may also make the evidential basis of its answers less inspectable. Conversely, a dashboard may impose greater interactional effort while preserving a more visible path between data, interpretation, and decision. Since intended reliance may be shaped by both perceived support and perceived verifiability, the expected direction of the interface effect remains open.

\begin{hypothesis}
Interface type is associated with differences in reliance.
\end{hypothesis}

\subsection{Research model}
Figure~\ref{fig:model} summarizes the research model. The study examines how determinants of information processing demands---\textit{interface type} and \textit{objective task complexity}---affect decision outcomes and perceived cognitive demands in a manufacturing decision-making scenario. Interface type is treated as the representational and interactional condition through which these demands are managed. In line with the IO and MWL perspectives outlined above, we treat objective task complexity as an experimental manipulation of information load: higher-complexity tasks require processing and coordinating more information cues to reach a decision. DL is measured prior to the tasks and treated as a pre-existing individual difference. Outcomes are captured through (i) subjective MWL and (ii) decision performance, operationalized as decision accuracy and completion time. Additionally, intended reliance is examined as an exploratory post-use outcome.

\section{Method}



\subsection{Participants}
The number of participants required to test the hypotheses was determined by conducting power analysis tests on the data collected during the exploratory study.
Participants were recruited via Prolific using pre-screening criteria: (1) employment in manufacturing or closely related industrial sectors, (2) managerial role (junior, middle, or upper management), (3) decision-making responsibilities related to operations/production, supply chain/logistics, or business strategy, (4) fluency in English, and (5) residence in one of the targeted manufacturing economies\footnote{The countries of residence were determined among the top 10 manufacturing economies, with the European Union considered as a single entity.}.
In total,  $N=313$ individuals started the study.  Role seniority, industry, and decision responsibilities were re-assessed within the study to verify eligibility and to characterise the sample and participants whose in-study responses violated core eligibility criteria were excluded ($N= 79$),  leaving $N=234$ participants who entered the main study. 
Of these, $N=43$ returned their submission prior to completion and $N=6$ timed out. Among completed submissions, participants were excluded for non-compliance with study instructions ($N=20$), failing comprehension/attention checks ($N=24$), technical issues ($N=5$), or very low-effort responding evidenced by lack of verifiable task/interface engagement and repetitive questionnaire response patterns ($N=2$). The final analytic sample comprise $N=134$ participants. Sample characteristics are reported in Table~\ref{tab:sample}.
Participants received \pounds12 base compensation, with a performance-based bonus of up to \pounds1.50 (maximum total \pounds13.50). Participants excluded due to technical issues received partial compensation (\pounds6) to account for time spent, while those who were screened out received \pounds0.10. The median completion time was 26:22 minutes, corresponding to an average base reward rate of \pounds27.37/hr.

The study protocol was approved by the Bioethical Committee of the University of Pisa (approval no. 39/2025). Participants provided informed consent, could withdraw at any time without penalty, and data were stored securely in pseudonymized form on University of Pisa servers, in accordance with European General Data Protection Regulation (GDPR).

\begin{table}[!t]
\centering
\small 
\setlength{\tabcolsep}{4pt} 
\renewcommand{\arraystretch}{0.95} 
\begin{threeparttable}
\caption{Sample characteristics (final analytic sample, $N=134$).}
\label{tab:sample}
\begin{tabularx}{\columnwidth}{@{}l >{\raggedright\arraybackslash}X r r@{}}
\toprule
\textbf{Variable} & \textbf{Category / statistic} & \textbf{n} & \textbf{\%} \\
\midrule

Age & Mean (SD) & \multicolumn{2}{r}{39.44 (10.03)} \\
             & Median [IQR] & \multicolumn{2}{r}{37.5 [33, 45]} \\
\addlinespace

Sex & Male & 103 & 76.87  \\
       & Female & 31 & 23.13  \\
\addlinespace

Education Level & Secondary education  & 4 & 2.99 \\
                     & High School Diploma & 8 & 5.97 \\
                     & Technical / community college & 17 & 12.69 \\
                     & Undergraduate degree& 66 & 49.25 \\
                     & Graduate degree& 38 & 28.36 \\
                     & Doctorate degree& 1 & 0.75 \\
\addlinespace

Country/region\tnote{a} & United States & 49 & 36.57 \\
                        & United Kingdom & 41 & 30.60 \\
                        & EU (member states) & 27 & 20.15 \\
                        & Canada & 9 & 6.72 \\
                        & Other & 8 & 5.97 \\
\addlinespace

Managerial seniority & Junior management & 21 & 15.67 \\
                     & Middle management & 85 & 63.43 \\
                     & Upper management & 28 & 20.90 \\
\addlinespace

Responsibilities\tnote{b} & Operations/Production   & 113 & 84.33 \\
                    & Supply Chain/Logistics  & 73 & 54.48 \\
                    & Business Strategy  & 56 & 41.79\\
\addlinespace

Industry sector & Manufacturing & 88 & 65.67 \\
                         & Automotive & 6 & 4.48 \\
                         & Engineering & 6 & 4.48 \\
                         & Computer \& electronics manufacturing & 11 & 8.21 \\
                         & Oil \& gas / utilities / mining & 16 & 11.94 \\
                         & Pharmaceutical / bio-tech & 2 & 1.49 \\
                         & Other manufacturing-related & 5 & 3.73 \\
\addlinespace

Company size & Micro (1-9) & 5 & 3.73 \\
                            & Small (10-49) & 24 & 17.81 \\
                            & Medium (50-249) & 36 & 26.87 \\
                            & Large (250-999) & 21 & 15.67 \\
                            & Enterprise (1000+) & 48 & 35.82 \\

\bottomrule
\end{tabularx}

\begin{tablenotes}[flushleft]\footnotesize
\item \textit{Note.} Percentages may not sum to 100 due to rounding.
\item[a] Country/region is aggregated for readability. EU = member states; non-EU European countries are included in ``Other''.
\item[b] Multiple selections allowed; percentages refer to the share of participants endorsing each responsibility. Responses outside the study's focal domains are omitted for brevity; $n=114$ participants selected at least one such responsibility in addition to the reported ones.
\end{tablenotes}
\end{threeparttable}
\end{table}

\subsection{Experiment Design}
The study used a {2$\times$3} mixed experimental design with a measured moderator:

\textbf{Interface type} is a between-subjects factor with two levels: a visualization-heavy dashboard and a conversational interface. Drawing on CFT, the two interfaces offer different external representations and interaction styles for accessing and integrating information, which should translate into differences in perceived workload and decision performance even under comparable objective information demands.

\textbf{Task complexity} is manipulated within subjects by varying objective complexity, primarily through Wood's component and coordinative dimensions. Tasks are administered from lower to higher complexity for the methodological reasons detailed in \S~\ref{subsec:tasks}. 

\textbf{Data literacy} is measured prior to the experimental tasks and treated as a pre-existing individual difference variable. DL is included in the model for two reasons. First, DL is theoretically relevant: the ability to interpret, validate, and act on data is expected to shape both performance and perceived cognitive demands in data-intensive decision tasks. Second, DL was used to support balanced assignment across interface conditions, improving comparability between groups on a theoretically meaningful capability.

\subsubsection{Condition assignment} \label{par:assignment}
Participants were assigned at the individual level to one of two interface conditions with a target 1:1 allocation ratio. Because recruitment was sequential and fully online, we used a covariate-adaptive minimization procedure \citep{pocock_sequential_1975} to reduce the risk of chance imbalances on individual differences in DL and interface familiarity, that were expected to be prognostic for performance and perceived workload.

Assignment was performed \emph{after} the pre-task questionnaires. DL was operationalized as the participant's mean score on the BDLI and binned into three strata (low/mid/high) using two pre-specified cut-points decided after the pilot (5.0 and 6.0, respectively). Second, interface familiarity was derived from a brief technology-familiarity questionnaire administered separately for dashboard and chatbot. For each interface, familiarity was computed as the sum of (i) frequency of use and (ii) self-rated capability; these two interface-specific familiarity scores were then combined into a scale-free \textit{Relative Familiarity Index} (RFI):
\[
\mathrm{RFI} = \frac{f_{\mathrm{dashboard}} - f_{\mathrm{chatbot}}}{f_{\mathrm{dashboard}} + f_{\mathrm{chatbot}}},
\]
which ranges from $-1$ (leans chatbot) to $+1$ (leans dashboard), with $\mathrm{RFI}=0$ when both are equal. The combination of DL (3 levels) and RFI (3 levels) yielded 9 composite strata.

At assignment time, the server computed current arm counts within each composite stratum (eligible participants only) and compared the two candidate allocations for the incoming participant. The preferred arm was the one that minimized the absolute chatbot--dashboard imbalance across all strata and overall totals, while ties were broken at random. To retain allocation unpredictability, the preferred arm was selected with a biased coin ($p=0.80$), otherwise the alternative arm was assigned \citep{pocock_sequential_1975}. The assigned condition and covariate values were stored server-side and returned to the client.

\subsection{Materials}
The study materials consisted of a custom web application, two experimental interfaces, and a synthetic manufacturing dataset. 

\subsubsection{Study web application}
The study was delivered through a custom web-based platform implementing the full experimental flow. The back end was implemented in Python using FastAPI and a PostgreSQL database, while the front end was implemented in React. The platform served the two experimental interfaces, namely the chatbot and the dashboard, as dedicated components while keeping the remaining study elements identical across conditions (e.g., task instructions, scenario framing, navigation structure, and timing/transition logic). The application also implemented centralized logging to support data quality checks and post-hoc auditing (e.g., page visits, timestamps, and interaction events). Both interfaces provided read-only access to the same underlying manufacturing dataset and time window, ensuring that differences between conditions reflected the interaction style rather than differences in available information. Before deployment, the complete experimental platform, including both the conversational and graphical interfaces, was assessed during pilot testing to identify technical failures and interaction obstacles that could interfere with task completion independently of the intended comparison between interaction styles.

\subsubsection{Conversational interface}
The conversational condition consisted of an LLM-based chatbot embedded in the study web application corresponding to the refined version of the CA architecture described in a previous study~\citep{figlie_towards_2024} and based on OpenAI GPT-4o model. The chatbot was implemented as a separate FastAPI service receiving user prompts from the front end. Responses were generated by combining the LLM with a constrained set of domain-specific retrieval functions. These functions provided read-only access to the study dataset, including machine signals, alarms, and maintenance records, allowing the chatbot to ground its responses in the same information space available in the dashboard condition. Finally, all chatbot interactions were stored in the central database at the level of conversations and individual messages. This logging supported later inspection of participant interaction traces and data-quality checks.

\subsubsection{Graphical interface}
The graphical condition consisted of a custom analytics dashboard implemented in React and composed of reusable widgets for data exploration and decision support (e.g., filtering, drill-down, and time-series inspection). Although developed specifically for the experiment, its layout and interaction patterns were informed by selected task-relevant components of the industrial partner's analytics platform. In particular, the experimental dashboard adopted comparable conventions for organizing operational indicators, navigating among views, applying filters, and inspecting temporal trends. This design choice was made to improve ecological validity by grounding the graphical condition in interface patterns developed for industrial data exploration, while also establishing a credible graphical comparator rather than an \textit{ad hoc} baseline that could introduce unnecessary interface-specific issues. Experimental control was retained by implementing only the data, views, and analytical operations required for the study tasks. Dashboard interaction was monitored through the platform's event logger, which captured user actions for objective behavioural measures and for identifying non-engagement or technical failures.

\subsubsection{Simulated scenario data}
Real machine Key Performance Indicator (KPI) traces were used exclusively to characterize plausible baseline ranges, correlations among indicators, and temporal patterns associated with day and night shifts and maintenance events. Based on these characteristics, a one-year synthetic dataset was generated within the same feasible operating envelope. Controlled perturbations and noise were introduced to reproduce realistic variability and selected deviations relevant to the study tasks. The simulated decision scenario was anchored to a fixed reference point: 5 November 2025 at 11:00. For every participant, the simulated manufacturing data available through either interface extended only up to this timestamp, such that all participants examined the same operational state and the same information history. The synthetic dataset and scenario timestamp were held constant across interface conditions, providing a common ground truth for scoring task outcomes and ensuring that comparisons were not affected by differences in the underlying data or temporal context.

\subsection{Experimental Tasks}
\label{subsec:tasks}

\paragraph{Task design goals}
The task set was designed to emulate data-driven operational decision-making in a manufacturing setting while remaining solvable by proxy participants. Tasks were grounded in the study scenario and the same underlying dataset across conditions, so that any performance differences could be attributed to interface/individual characteristics rather than information availability. Consistent with prior work treating task properties as a key determinant of information-processing demands, we explicitly specified and quantified objective task complexity for each task (see \S\ref{sec:task_complexity}) \citep{wood_task_1986, salimzadeh_missing_2023}.

\paragraph{Task set and structure}
The tasks and success criteria were identical across interface conditions and participants were instructed to base answers exclusively on information available in the interface and scenario. Task prompts were iteratively refined during pilot testing to ensure clarity, solvability with the available information, and alignment with the intended decision-making construct.  Terminology and the plausibility of task constraints were further informed by stakeholder feedback from prior field work.

Participants completed a fixed sequence of three tasks in increasing objective complexity (\ref{app:task} reports full prompts, complexity computation, and reference solutions criteria). Task order was held constant because tasks were designed as successive stages of a single analytic workflow: later tasks presupposed familiarity with the scenario, terminology, and interface elements introduced in earlier tasks. In this setting, counterbalancing task order would systematically vary practice and carry-over across tasks, complicating interpretation of task-level differences. Following prior similar research that fixed lower complexity tasks first to establish a baseline and reduce order-driven inflation/deflation of completion times across complexity levels \citep{speier_influence_2003}, we used a low-to-high sequence.

To mitigate fatigue and motivational decline in this fixed-order protocol, the study was designed to remain short (median completion time $<30$ minutes) and performance-contingent bonuses increased across tasks. 

Each task was administered under a task-specific time cap to bound session duration and to standardize exposure to the interface, reducing unbounded exploration that could introduce large between-participant variability \citep{crescenzi_adaptation_2021}.  The countdown was not displayed during task execution to avoid introducing an additional exogenous time-pressure cue via a visible clock/countdown, which can itself alter performance and perceived pressure  \citep{hunt_endogenous_2017}. Instead, prior to each task we informed participants of the typical time required (based on pilot data) and applied a generous cap set above the pilot median completion time for that task (\ref{app:task:times} reports pilot times and caps). Participants received a brief warning shortly before expiry; if the limit was reached, a short grace period allowed final submission before the study advanced. Upon grace expiry, the study advanced automatically and the task was recorded as timed out.

Nevertheless, because complexity could be partially confounded with task position, any complexity-related trends should be interpreted with this limitation in mind.

\subsection{Procedure}
\label{subsec:procedure}
\begin{figure}[t]
\centering
\includegraphics[width=1\linewidth]{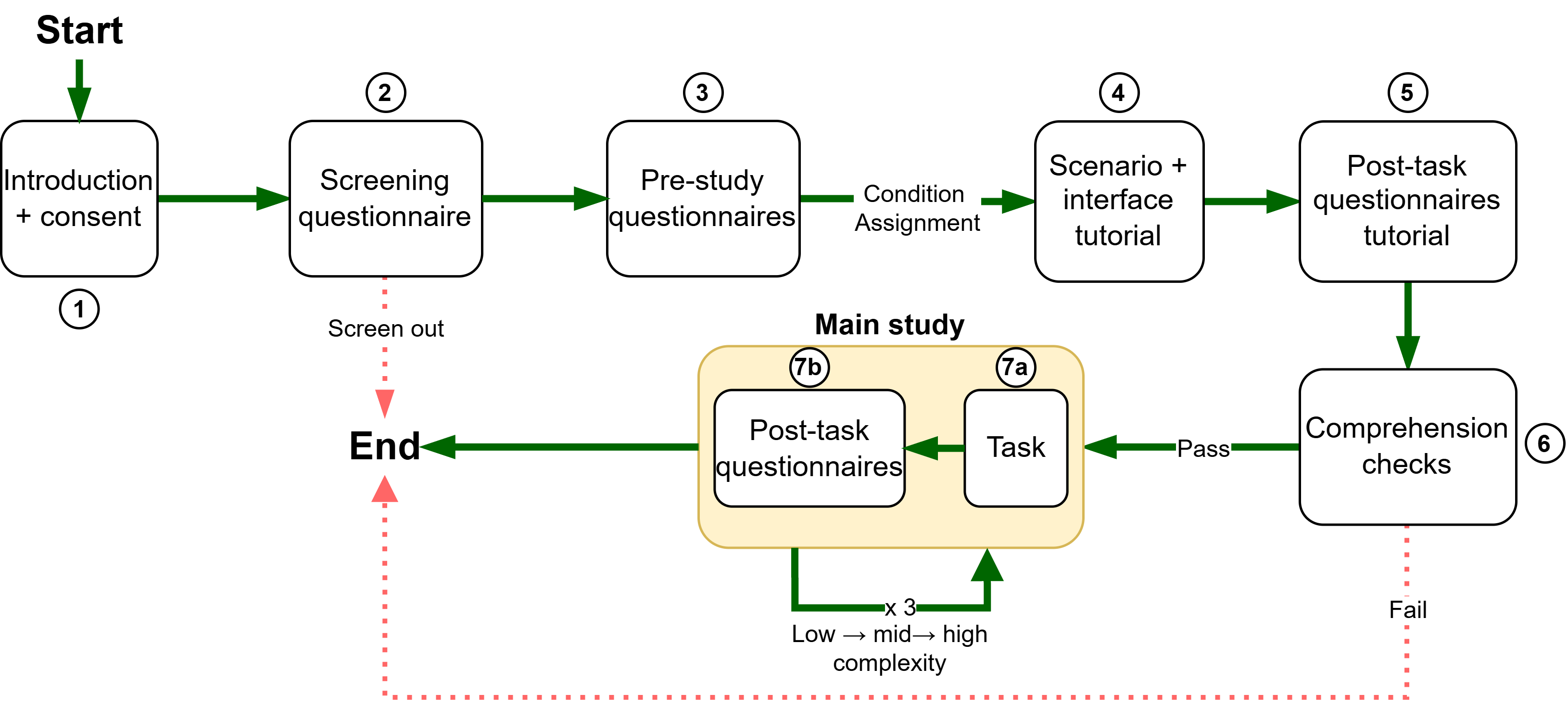}
\caption{Procedure followed by the participants during the study}\label{fig:procedure}
\end{figure}

Figure \ref{fig:procedure} summarizes the study flow. Data collection was conducted between 10 December 2025 and 2 February 2026. Participants first viewed an introduction page and provided informed consent (Step 1). They then completed a screening questionnaire (Step 2) that repeated the Prolific pre-screening items to verify eligibility\footnote{This procedure is suggested to ensure data quality by highlighting people who changed conditions (e.g., work, sector) since they took Prolific's questionnaires, or to identify possible liars.}; ineligible respondents were screened out and the study ended. Eligible participants completed BDLI and interface familiarity pre-study questionnaires (Step 3). The system then assigned participants to an interface condition using covariate-adaptive minimization (Step 3→4, see \S~\ref{par:assignment}). Next, participants read the study scenario and completed an interface tutorial explaining how tasks would be carried out  (Step 4). Participants then completed a brief instructional tutorial on the post-task questionnaires (NASA-TLX and intended reliance questionnaire, Step 5), followed by comprehension checks (Step 6); failing these checks ended the study. In addition, attention checks were embedded at multiple points during the study (e.g., within questionnaires and tutorials), and failing twice terminated the study . Participants who passed proceeded to the main study, completing three tasks presented in fixed order of increasing objective complexity (Step 7a). Each task was first shown in a reading phase, during which participants could inspect the task instructions but could not yet interact with the assigned interface. After reading the task, participants clicked the \emph{I read the task} button to indicate that they were ready to begin. At that point, the assigned interface became available and the task timer started. Participants then completed the task and submitted their answer through the task panel by clicking \emph{Submit \& Next}. When participants approached the task time limit, a grace period was applied, as described in \S~\ref{subsec:tasks}. Each task was followed immediately by the post-task questionnaires (Step 7b). After the third task and its corresponding post-task questionnaires, participants completed a final open-ended compliance check asking whether they had used any additional tools beyond those provided in the study interface. Participants were then redirected to a closing page thanking them for their participation (End). Where applicable, the page displayed any bonus payment awarded during the study before participants were returned to Prolific for completion confirmation.

\subsection{Measures}
\paragraph{Subjective Mental Workload}
Subjective mental workload was assessed using the NASA Task Load Index (NASA-TLX). Participants completed the NASA-TLX after each task. Following current recommendations, we used the "Raw TLX" scoring procedure (i.e., no pairwise weighting): subscale ratings were aggregated into a single overall workload score by averaging across the six subscales \citep{hart_nasa-task_2006, hertzum_reference_2021}.

\paragraph{Decision Accuracy}
Decision accuracy was computed by comparison with pre-specified ground-truth criteria (reported in \ref{app:task:decision}). Because tasks yielded raw scores on different ranges, accuracy was normalized using Percent of Maximum Possible (POMP) scoring \citep{cohen_problem_1999}. For each task $t$, a raw score $S_t$ was transformed into a normalized Decision Accuracy Score (DAS) as follows:

\[
DAS_t = \frac{S_t - S_{\min,t}}{S_{\max,t} - S_{\min,t}},
\]

yielding $DAS_t \in [0,1]$. This normalization places task-specific accuracy scores on a common scale, supporting comparison across tasks.

Skipped responses or non-answers were assigned the minimum possible score when participants had engaged with the assigned interface during the task window but did not provide a valid solution. In these cases, the normalized score was therefore $DAS_t=0$. By contrast, submissions showing no task-relevant interaction with the assigned interface were treated as non-compliance and excluded from the analysis (see \S~\ref{subsec:procedure}).

\paragraph{Task completion time}
This outcome was measured as time-on-task for each task, computed from client-side timestamps recorded by the study web application. Time-on-task was defined as the elapsed time between (a) the participant clicking the \emph{I read the task} button and (b) their final answer submission.

\paragraph{Data literacy}
Participants’ DL was measured using the BDLI \citep{kirby_developing_2024}, composed of 12 items covering core DL capabilities (e.g., analysis, acquisition/quality assessment, and wrangling). Responses were recorded on a 7-point Likert scale (1--7) and aggregated into a composite BDLI score by averaging across items (higher scores indicate higher DL).

\paragraph{Intended reliance}
Reliance on the assigned interface’s outputs was assessed after each task using three items adapted from \cite{ajzen_constructing_2006}. Items were rated on a 7-point scale from \emph{Unlikely} (1) to \emph{Likely} (7). The primary intended reliance measure for hypothesis testing was the \emph{sole reliance} item (``I would take an action or make a subsequent decision based solely on the information I received/gathered from the chat assistant/the dashboard''). Two additional items probed intended reliance conditional on external verification (seeking a colleague’s second opinion or verifying using other analytical tools).

\section{Statistical Analysis and Results}
\label{sec:results}

Statistical analyses and results are presented by outcome for convenience. A summary of the hypotheses and support decisions is shown in Table~\ref{tab:hypothesis-summary}. The primary analyses reported below use the model families specified for each outcome. Supplementary robustness analyses and industry-role confounding checks are reported in \ref{app:robustness} and \ref{app:sec.confounding}. These analyses did not alter the main interpretation of the confirmatory results.

\begin{table*}[t]
\centering
\small
\setlength{\tabcolsep}{3.5pt}
\renewcommand{\arraystretch}{1.05}
\caption{Summary of hypotheses, model families, and support decisions.}
\label{tab:hypothesis-summary}

\begin{tabularx}{\textwidth}{@{}
>{\bfseries}p{0.07\textwidth}
>{\raggedright\arraybackslash}p{0.40\textwidth}
>{\raggedright\arraybackslash}p{0.38\textwidth}
>{\centering\arraybackslash}p{0.08\textwidth}
@{}}
\toprule
\textbf{Hyp.} & \textbf{Outcome / effect tested} & \textbf{Model family} & \textbf{Supported} \\
\midrule

\multicolumn{4}{@{}l}{\textit{Interface main effects}} \\
H1a & MWL: CUI lower than GUI & CLMM & \cmark \\
H1b & Accuracy: CUI different from GUI & Fractional logit GEE & \xmark \\
H1c & Time: CUI shorter than GUI & Gamma GEE with log link & \xmark \\

\addlinespace[0.25em]
\multicolumn{4}{@{}l}{\textit{Moderation by task complexity}} \\
H2a & MWL: CUI advantage attenuates as complexity rises & CLMM with interface $\times$ task complexity interaction & \cmark \\
H2b & Accuracy: CUI advantage attenuates as complexity rises & Fractional logit GEE with interface $\times$ task complexity interaction & \cmark \\
H2c & Time: CUI advantage attenuates as complexity rises & Gamma GEE with interface $\times$ task complexity interaction & \cmark \\

\addlinespace[0.25em]
\multicolumn{4}{@{}l}{\textit{Moderation by data literacy}} \\
H3a & Lower data literacy disadvantages GUI more for MWL & CLMM with interface $\times$ data-literacy interaction & \xmark \\
H3b & Lower data literacy disadvantages GUI more for accuracy & Fractional logit GEE with interface $\times$ data-literacy interaction & \xmark \\
H3c & Lower data literacy disadvantages GUI more for time & Gamma GEE with interface $\times$ data-literacy interaction & \xmark \\

\addlinespace[0.25em]
\multicolumn{4}{@{}l}{\textit{Task complexity main effects}} \\
H4a & Higher complexity increases MWL & CLMM with task complexity effect & \cmark \\
H4b & Higher complexity reduces accuracy & Fractional logit GEE with task complexity effect & \cmark \\
H4c & Higher complexity increases time & Gamma GEE with task complexity effect & \cmark \\

\addlinespace[0.25em]
\multicolumn{4}{@{}l}{\textit{Intended reliance}} \\
H5 & Interface effect on intended reliance & Item-wise CLMMs adjusted for task complexity & \xmark \\

\bottomrule
\end{tabularx}
\end{table*}

\subsection{Effects on Mental Workload}
\label{subsec:results:mwl}
Table \ref{tab:results-mental-workload} presents the results of each hypothesis with \textit{MWL} as outcome. All MWL hypotheses were evaluated with probit cumulative link mixed models (CLMMs) fitted to \texttt{tlx\_mean} rounded to 5-point ordered categories, with participant random intercepts to account for repeated observations; adjusted comparisons were summarized with estimated marginal means and contrasts \citep{Wieditz2024,Hedeker2015,Lenth2016}. Although this ordinalization aligns the endpoint with the CLMM framework, categorizing a continuous TLX score may discard some information \citep{Naggara2011}.

\begin{table*}[!tb]
\centering
\scriptsize
\setlength{\tabcolsep}{3.5pt}
\renewcommand{\arraystretch}{1.08}
\caption{Primary results for mental workload hypotheses.}
\label{tab:results-mental-workload}

\begin{threeparttable}
\begin{tabularx}{\textwidth}{@{}
>{\bfseries}p{0.07\textwidth}
>{\raggedright\arraybackslash}p{0.25\textwidth}
>{\raggedright\arraybackslash}p{0.16\textwidth}
>{\raggedright\arraybackslash}X
@{}}
\toprule
\textbf{Hyp.} & \textbf{Test / contrast} & \textbf{$p$-value} & \textbf{Effect summary} \\
\midrule

\multicolumn{4}{@{}l}{\textit{Interface main effect}} \\
H1a &
GUI vs. CUI &
$p < .001^{***}$ &
Dashboard--chatbot latent contrast $= 1.096$, 95\% CI [0.708, 1.485], $z = 5.533$ \\

\addlinespace[0.35em]
\multicolumn{4}{@{}l}{\textit{Moderation by task complexity}} \\
H2a &
Interface $\times$ task complexity &
$p < .001^{***}$ &
Omnibus interaction: LR $= 16.460$, df $= 2$ \\

H2a &
GUI--CUI, T3 vs. T1 &
Holm $p < .001^{***}$ &
DiD $\beta = -0.962$, 95\% CI [$-1.463$, $-0.461$], $z = -3.764$; raw $p < .001$ \\

H2a &
GUI--CUI, T2 vs. T1 &
Holm $p = .002^{**}$ &
DiD $\beta = -0.824$, 95\% CI [$-1.322$, $-0.325$], $z = -3.238$; raw $p = .001$ \\

H2a &
GUI--CUI, T3 vs. T2 &
Holm $p = .577$ &
DiD $\beta = -0.138$, 95\% CI [$-0.625$, 0.348], $z = -0.557$; raw $p = .577$ \\

\addlinespace[0.35em]
\multicolumn{4}{@{}l}{\textit{Moderation by data literacy}} \\
H3a &
Interface $\times$ data literacy &
$p = .125$ &
Omnibus interaction: LR $= 4.152$, df $= 2$ \\

H3a &
GUI--CUI, high vs. low DL &
Holm $p = .361$ &
DiD $\beta = -0.576$, 95\% CI [$-1.812$, 0.660], $z = -0.914$; raw $p = .361$ \\

\addlinespace[0.35em]
\multicolumn{4}{@{}l}{\textit{Task complexity main effect}} \\
H4a &
Task complexity &
$p < .001^{***}$ &
Omnibus effect: LR $= 127.302$, df $= 2$ \\

H4a &
T3 vs. T1 &
$p < .001^{***}$ &
T3--T1 latent contrast $= 1.536$, 95\% CI [1.195, 1.876], $z = 10.797$ \\

\bottomrule
\end{tabularx}

\begin{tablenotes}[flushleft]
\footnotesize
\item \textit{Note.} CUI = chatbot condition; GUI = dashboard condition; DL = data literacy; T1 = first task (low complexity); T2 = second task (mid complexity); T3 = third task (high complexity); DiD = difference-in-differences. For prespecified follow-up contrasts, Holm--Bonferroni-adjusted $p$-values are reported. Asterisks denote $^{*}p < .05$, $^{**}p < .01$, and $^{***}p < .001$.
\end{tablenotes}
\end{threeparttable}
\end{table*}

\paragraph{H1a} The hypothesis was tested with the additive CLMM specified above. Interface condition was the focal predictor, with task complexity included as an additive adjustment covariate. Both factors were treatment-coded, so the estimated marginal contrast captured the adjusted difference between the dashboard and chatbot conditions \citep{Lenth2016}. The dashboard-minus-chatbot latent contrast was 1.096, 95\% CI [0.708, 1.485], $z = 5.533$, $p < .001$, indicating higher mental workload in the dashboard condition and lower mental workload in the chatbot condition. Because the interval excluded 0 and the effect had the predicted sign, H1a was supported.

\begin{figure}
    \centering
    \includegraphics[width=1\linewidth]{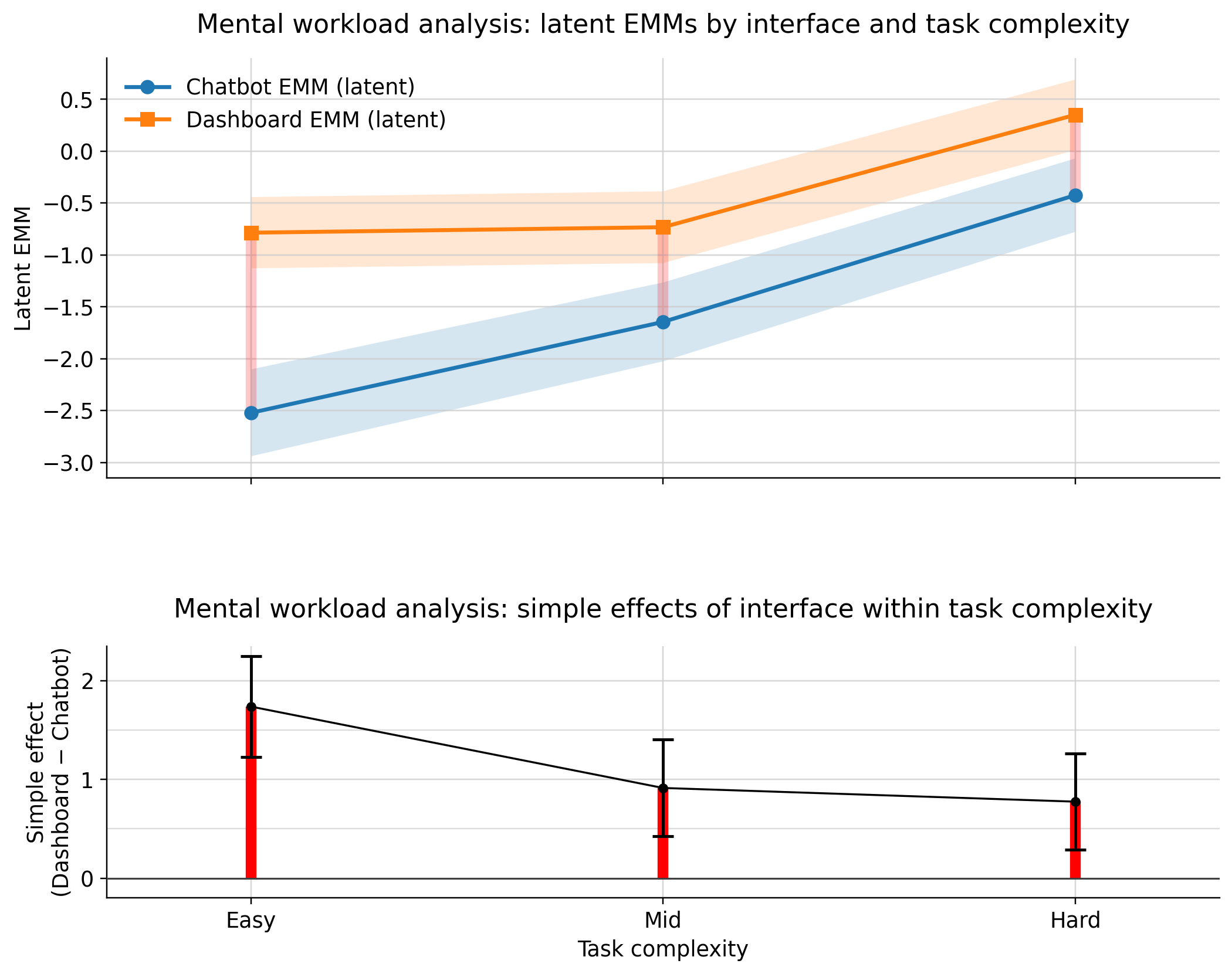}
    \caption[Model-based mental workload interaction plot]{
    Model-based mental workload by interface and task complexity. The top panel shows CLMM estimated marginal means on the latent probit scale, with shaded 95\% confidence intervals. The bottom panel shows simple interface effects expressed as \emph{Dashboard $-$ Chatbot}; positive values therefore indicate higher latent mental workload for the dashboard. Red vertical segments indicate the cell-wise gaps used to compute the simple effects.
    }
    \label{fig:h2a_interaction}
\end{figure}

\paragraph{H2a} The hypothesis was tested by comparing additive and interface-by-task complexity CLMMs with an omnibus Likelihood-Ratio Test (LRT) and then probing the interaction with model-based simple effects and exact DiD-like contrasts; follow-up tests were Holm-adjusted \citep{Chen2020LRT,Lenth2016,Mize2019,Aickin1996}. Accordingly, the primary quantities used for visualization and interpretation were latent-scale CLMM estimated marginal means and latent simple effects. This scale is appropriate because the hypothesis concerns moderation of the model-implied interface effect, rather than raw mean differences on an interval outcome. The figure therefore displays the latent Estimated Marginal Means (EMMs) and the simple effects expressed as \emph{Dashboard $-$ Chatbot}, so positive simple effects indicate higher latent mental workload for the dashboard than for the chatbot. The omnibus interaction was significant, LR = 16.460, df = 2, $p < .001$. On the \emph{Dashboard $-$ Chatbot} scale, the primary DiD for T3 versus T1 was $-0.962$, 95\% CI [$-1.463$, $-0.461$], $z = -3.764$, raw $p < .001$, Holm-adjusted $p < .001$. The additional follow-ups showed the same attenuation from T2 versus T1, DiD = $-0.824$, 95\% CI [$-1.322$, $-0.325$], Holm-adjusted $p = .002$, whereas the T3-versus-T2 contrast was not reliable, DiD = $-0.138$, 95\% CI [$-0.625$, 0.348], Holm-adjusted $p = .577$. The plotted simple effects were consistent with this pattern: T1 = 1.737, 95\% CI [1.226, 2.248]; T2 = 0.913, 95\% CI [0.425, 1.402]; T3 = 0.775, 95\% CI [0.291, 1.259]. The negative DiDs indicate that the positive \emph{Dashboard $-$ Chatbot} workload gap moved toward zero as complexity increased, meaning that the chatbot workload advantage attenuated with task complexity (Figure \ref{fig:h2a_interaction}); H2a was therefore supported.

\paragraph{H3a} The hypothesis was tested with an additive-versus-interaction probit CLMM for the interface-by-data literacy effect. DL was operationalized as the mean of the 12 BDLI items and grouped as low [1,5], medium (5,6], and high (6,7], respecting the pre-specified cut-points decided after the pilot (\S\ref{par:assignment}); although this facilitated interpretation, categorizing a continuous score may reduce information \citep{BeltranTarwater2024}. The omnibus interaction was not significant, LR = 4.152, df = 2, p = .125, and the prespecified high-versus-low DiD was -0.576, 95\% CI [-1.812, 0.660], z = -0.914, raw p = .361, Holm-adjusted p = .361. Thus, neither the omnibus interaction nor the primary follow-up showed reliable evidence that the interface slopes changed across data-literacy levels, so H3a was not supported.

\paragraph{H4a} The hypothesis was tested with a CLMM adding task complexity to a model that already controlled for interface condition; task effects were assessed with an omnibus LRT followed by estimated marginal contrasts \citep{Chen2020LRT,Lenth2016}. The omnibus task complexity effect was significant, LR = 127.302, df = 2, p < .001. Follow-up contrasts confirmed the expected monotonic pattern: T2 versus T1, beta = 0.415, 95\% CI [0.111, 0.720], p = .001; T3 versus T1, beta = 1.536, 95\% CI [1.195, 1.876], z = 10.797, p < .001; and T3 versus T2, beta = 1.121, 95\% CI [0.800, 1.441], p < .001. Together with the reported monotonic increase from T1 to T3, these results supported H4a.

\begin{figure}
    \centering
    \includegraphics[width=1\linewidth]{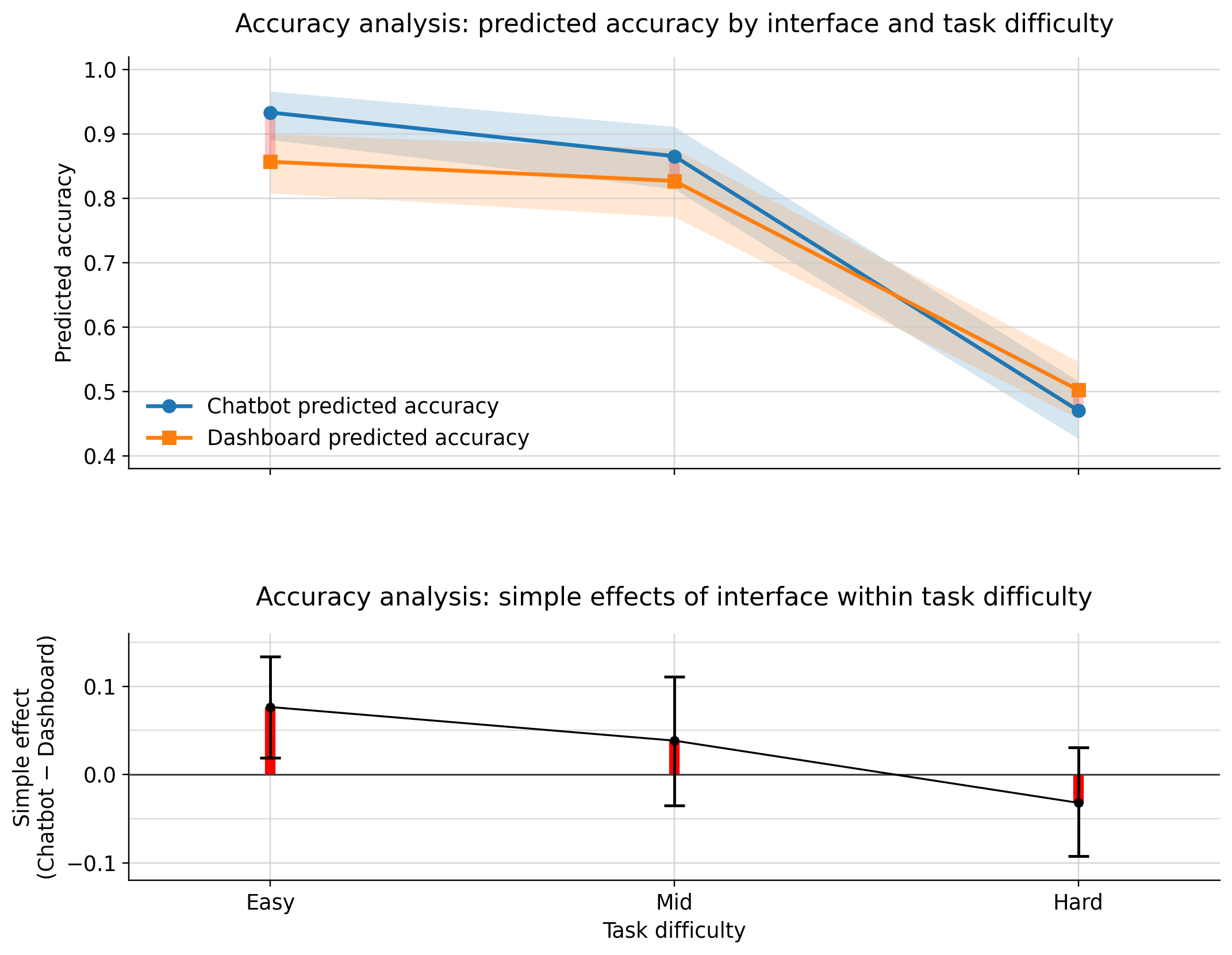}
    \caption[Predicted accuracy interaction plot]{
    Predicted decision accuracy by interface across task difficulty. The top panel shows predicted accuracy on the response scale, with shaded 95\% bootstrap confidence intervals. The bottom panel shows simple interface effects expressed as \emph{Chatbot $-$ Dashboard}; positive values therefore indicate higher predicted accuracy for the chatbot, whereas negative values indicate higher predicted accuracy for the dashboard. Red vertical segments indicate the cell-wise gaps used to compute the simple effects.
    }
    \label{fig:h2b_interaction}
\end{figure}

\begin{figure}
    \centering
    \includegraphics[width=1\linewidth]{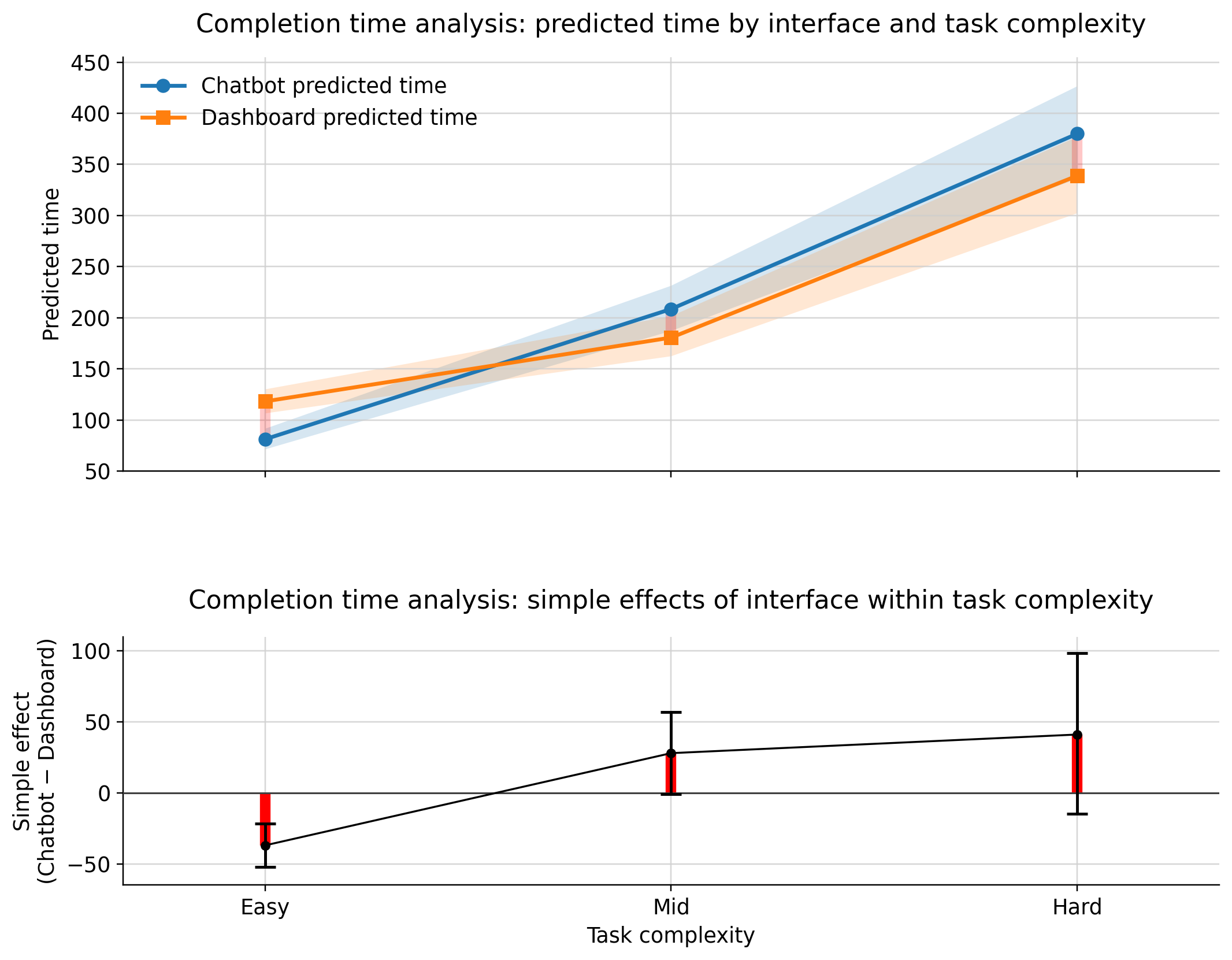}
    \caption[Predicted completion time interaction plot]{
    Predicted completion time by interface across task complexity levels. The top panel shows predicted completion time on the response scale, with shaded 95\% bootstrap confidence intervals. The bottom panel shows simple interface effects expressed as \emph{Chatbot $-$ Dashboard}; negative values therefore indicate faster chatbot completion, whereas positive values indicate slower chatbot completion. Red vertical segments indicate the cell-wise gaps used to compute the simple effects.
    }
    \label{fig:h2c_interaction}
\end{figure}

\subsection{Decision Accuracy}
\label{subsec:results:acc}

Table \ref{tab:results-decision-accuracy} presents the results of each hypothesis with \textit{decision accuracy} as outcome. All accuracy hypotheses were evaluated with population-averaged fractional-logit Generalized Estimating Equations (GEEs) because \texttt{accuracy\_score} is bounded in $[0,1]$; GEE accounted for repeated observations within participants, and response-scale adjusted means and confidence intervals were obtained with participant-level cluster bootstrapping when marginal effects or contrasts were summarized \cite{PapkeWooldridge1996,LiangZeger1986,Austin2024,Huang2018}.

\begin{table*}[t]
\centering
\scriptsize
\setlength{\tabcolsep}{3.5pt}
\renewcommand{\arraystretch}{1.08}
\caption{Primary results for decision accuracy hypotheses.}
\label{tab:results-decision-accuracy}

\begin{threeparttable}
\begin{tabularx}{\textwidth}{@{}
>{\bfseries}p{0.07\textwidth}
>{\raggedright\arraybackslash}p{0.21\textwidth}
>{\raggedright\arraybackslash}p{0.14\textwidth}
>{\raggedright\arraybackslash}X
@{}}
\toprule
\textbf{Hyp.} & \textbf{Test / contrast} & \textbf{$p$-value} & \textbf{Effect summary} \\
\midrule

\multicolumn{4}{@{}l}{\textit{Interface main effect}} \\
H1b &
CUI vs. GUI &
$p = .188$ &
Adjusted means: CUI $= 0.756$, GUI $= 0.730$; CUI--GUI $= 0.026$, 95\% CI [$-0.012$, 0.064], $z = 1.318$, OR $= 1.180$ \\

\addlinespace[0.35em]
\multicolumn{4}{@{}l}{\textit{Moderation by task complexity}} \\
H2b &
Interface $\times$ task complexity &
$p = .022^{*}$ &
Omnibus interaction: $\chi^2 = 7.635$, df $= 2$ \\

H2b &
CUI--GUI, T3 vs. T1 &
Holm $p = .029^{*}$ &
Predicted-accuracy DiD $= -0.109$, 95\% CI [$-0.188$, $-0.028$], logit $= -0.979$; raw $p = .010$ \\

H2b &
CUI--GUI, T2 vs. T1 &
Holm $p = .348$ &
Predicted-accuracy DiD $= -0.038$, 95\% CI [$-0.129$, 0.048], logit $= -0.554$; raw $p = .220$ \\

H2b &
CUI--GUI, T3 vs. T2 &
Holm $p = .348$ &
Predicted-accuracy DiD $= -0.070$, 95\% CI [$-0.164$, 0.033], logit $= -0.425$; raw $p = .174$ \\

\addlinespace[0.35em]
\multicolumn{4}{@{}l}{\textit{Moderation by data literacy}} \\
H3b &
Interface $\times$ data literacy &
$p = .045^{*}$ &
Omnibus interaction: $\chi^2 = 6.218$, df $= 2$ \\

H3b &
GUI--CUI, high vs. low DL &
Holm $p = .106$ &
Predicted-accuracy DiD $= 0.087$, 95\% CI [$-0.014$, 0.198], logit $= 0.564$; raw $p = .106$ \\

\addlinespace[0.35em]
\multicolumn{4}{@{}l}{\textit{Task complexity main effect}} \\
H4b &
Task complexity &
$p < .001^{***}$ &
Omnibus effect: $\chi^2 = 229.910$, df $= 2$ \\

H4b &
T3 vs. T1 &
Holm $p < .001^{***}$ &
Accuracy difference $= -0.406$, 95\% CI [$-0.447$, $-0.360$] \\

\bottomrule
\end{tabularx}

\begin{tablenotes}[flushleft]
\footnotesize
\item \textit{Note.} CUI = chatbot; GUI = dashboard; DL = data literacy; T1 = first task (low complexity); T2 = second task (mid complexity); T3 = task 3 (high complexity); DiD = difference-in-differences. For prespecified follow-up contrasts, Holm--Bonferroni-adjusted $p$-values are reported. Raw follow-up $p$-values are retained in the effect summaries. Asterisks denote $^{*}p < .05$, $^{**}p < .01$, and $^{***}p < .001$.
\end{tablenotes}
\end{threeparttable}
\end{table*}

\paragraph{H1b} The hypothesis was tested with the additive fractional-logit GEE specified above, controlling for task. The primary effect was expressed on the response scale as the adjusted difference in predicted accuracy, equally weighted across tasks, with uncertainty summarized by participant-level cluster bootstrap intervals \cite{Huang2018}. Adjusted accuracy was slightly higher in the chatbot condition (0.756) than in the dashboard condition (0.730), but the chatbot-minus-dashboard difference was 0.026, 95\% CI [-0.012, 0.064], $z = 1.318, p = .188$, corresponding to OR = 1.180. Because the confidence interval included 0, H1b was not supported.

\paragraph{H2b} The hypothesis was tested with a fractional-logit GEE  including an interaction between interface condition and task, to assess moderation \cite{Mize2019}. The omnibus interaction was assessed with a Wald test and followed by prespecified Holm-adjusted DiD contrasts on the predicted-accuracy scale \cite{Aickin1996,Huang2018}. Because accuracy is a bounded response-scale outcome, the figure reports predicted probabilities and probability differences, which directly represent the hypothesized chatbot advantage in interpretable accuracy units; logit-scale contrasts were retained for model-based inference. 

The omnibus interaction was significant, $\chi^2 = 7.635$, df = 2, $p = .022$. The primary DiD comparing T3 against T1 was $-0.109$ on the predicted-accuracy scale, 95\% CI [$-0.188$, $-0.028$], with a logit contrast of $-0.979$, raw $p = .010$, Holm-adjusted $p = .029$. The additional DiDs were not reliable for T2 versus T1, DiD = $-0.038$, 95\% CI [$-0.129$, 0.048], Holm-adjusted $p = .348$, or for T3 versus T2, DiD = $-0.070$, 95\% CI [$-0.164$, 0.033], Holm-adjusted $p = .348$. The plotted simple effects showed the same attenuation pattern: the chatbot advantage was 0.077, 95\% CI [0.019, 0.134], for T1; 0.038, 95\% CI [$-0.035$, 0.111], for T2; and $-0.032$, 95\% CI [$-0.092$, 0.031], for T3. Predicted accuracies were 0.933 versus 0.857 for T1, 0.866 versus 0.827 for T2, and 0.470 versus 0.502 for T3 for chatbot and dashboard, respectively. The negative T3-versus-T1 DiD therefore indicated convergence of the interface slopes across task difficulty, meaning that the chatbot accuracy advantage shrank and reversed descriptively in the hardest task (Figure \ref{fig:h2b_interaction}); H2b was supported.

\paragraph{H3b} The hypothesis was tested with a fractional-logit GEE including an interaction between interface condition and the three-level data-literacy grouping, while controlling for task. DL used the same grouping described for H3a. The omnibus interaction reached significance, $\chi^2 = 6.218$, df = 2, $p = .045$, but the prespecified high-versus-low DiD was 0.087, 95\% CI [-0.014, 0.198], with a logit contrast of 0.564, raw $p = .106$, Holm-adjusted $p = .106$ \cite{Mize2019,Aickin1996,Huang2018}. Additional simple effects suggested that the chatbot outperformed the dashboard at low DL by 0.124 points, 95\% CI [0.035, 0.216], Holm-adjusted $p = .009$, whereas no reliable interface difference emerged at medium or high data-literacy levels. Thus, although the omnibus interaction suggested some overall non-parallelism, the prespecified high-versus-low contrast did not support the hypothesized moderation pattern. H3b was therefore not supported.

\paragraph{H4b} The hypothesis was tested with a fractional-logit GEE including task as the focal predictor and interface as a control covariate. Task effects were assessed with an omnibus Wald test followed by Holm-adjusted pairwise contrasts on the predicted-accuracy scale, with uncertainty quantified by participant-level cluster bootstrap intervals \cite{Mize2019,Aickin1996,Huang2018}. The omnibus task effect was significant, $\chi^2 = 229.910$, df = 2, $p < .001$. Adjusted predicted accuracy declined monotonically from 0.894 for T1 to 0.846 for T2 and 0.488 for T3. The T2-versus-T1 contrast was $-0.048$, 95\% CI [$-0.094$, $-0.003$], Holm-adjusted $p = .037$; the T3-versus-T1 contrast was $-0.406$, 95\% CI [$-0.447$, $-0.360$], Holm-adjusted $p < .001$; and the T3-versus-T2 contrast was $-0.357$, 95\% CI [$-0.405$, $-0.308$], Holm-adjusted $p < .001$. This monotonic decline in accuracy from T1 to T3 supported H4b.

\subsection{Task Completion Time}
\label{subsec:results:time}

Table \ref{tab:results-completion-time} presents the results of each hypothesis with \textit{completion time} as outcome. All completion-time hypotheses were evaluated with Gamma GEEs with a log link because completion time was strictly positive and right-skewed; interface and task effects are therefore interpreted as time ratios, and response-scale summaries were accompanied by participant-level cluster-bootstrap intervals \cite{NgCribbie2017,Malehi2015,Austin2024,Huang2018}.

\begin{table*}[t]
\centering
\scriptsize
\setlength{\tabcolsep}{3.5pt}
\renewcommand{\arraystretch}{1.08}
\caption{Primary results for completion time hypotheses.}
\label{tab:results-completion-time}

\begin{threeparttable}
\begin{tabularx}{\textwidth}{@{}
>{\bfseries}p{0.07\textwidth}
>{\raggedright\arraybackslash}p{0.22\textwidth}
>{\raggedright\arraybackslash}p{0.14\textwidth}
>{\raggedright\arraybackslash}X
@{}}
\toprule
\textbf{Hyp.} & \textbf{Test / contrast} & \textbf{$p$-value} & \textbf{Effect summary} \\
\midrule

\multicolumn{4}{@{}l}{\textit{Interface main effect}} \\
H1c &
CUI vs. GUI &
$p = .534$ &
Adjusted means: CUI $= 213.606$, GUI $= 221.673$; time ratio $= 0.964$, 95\% CI [0.857, 1.083], $z = -0.622$ \\

\addlinespace[0.35em]
\multicolumn{4}{@{}l}{\textit{Moderation by task complexity}} \\
H2c &
Interface $\times$ task complexity &
$p < .001^{***}$ &
Omnibus interaction: $\chi^2 = 42.797$, df $= 2$ \\

H2c &
CUI--GUI, T3 vs. T1 &
Holm $p < .001^{***}$ &
Ratio-of-time-ratios $= 1.633$, 95\% CI [1.348, 1.977], log ratio $= 0.490$; raw $p < .001$ \\

H2c &
CUI--GUI, T2 vs. T1 &
Holm $p < .001^{***}$ &
Ratio-of-time-ratios $= 1.681$, 95\% CI [1.428, 1.979], log ratio $= 0.520$; raw $p < .001$ \\

H2c &
CUI--GUI, T3 vs. T2 &
Holm $p = .736$ &
Ratio-of-time-ratios $= 0.971$, 95\% CI [0.818, 1.153], log ratio $= -0.030$; raw $p = .736$ \\

\addlinespace[0.35em]
\multicolumn{4}{@{}l}{\textit{Moderation by data literacy}} \\
H3c &
Interface $\times$ data literacy &
$p = .899$ &
Omnibus interaction: $\chi^2 = 0.212$, df $= 2$ \\

H3c &
GUI--CUI, high vs. low DL &
Holm $p = 1.000$ &
Ratio-of-time-ratios $= 1.028$, 95\% CI [0.672, 1.557], log ratio $= 0.028$; raw $p = .890$ \\

\addlinespace[0.35em]
\multicolumn{4}{@{}l}{\textit{Task complexity main effect}} \\
H4c &
Task complexity &
$p < .001^{***}$ &
Omnibus effect: $\chi^2 = 562.997$, df $= 2$ \\

H4c &
T3 vs. T1 &
Holm $p < .001^{***}$ &
T3--T1 time ratio $= 3.583$, 95\% CI [3.219, 3.973], log ratio $= 1.276$; raw $p < .001$ \\

\bottomrule
\end{tabularx}

\begin{tablenotes}[flushleft]
\footnotesize
\item \textit{Note.} CUI = chatbot; GUI = dashboard; DL = data literacy; T1 = first task (low complexity); T2 = second task (mid complexity); T3 = third task (high complexity). For prespecified follow-up contrasts, Holm--Bonferroni-adjusted $p$-values are reported. Raw follow-up $p$-values are retained in the effect summaries. Asterisks denote $^{*}p < .05$, $^{**}p < .01$, and $^{***}p < .001$.
\end{tablenotes}
\end{threeparttable}
\end{table*}

\paragraph{H1c} The hypothesis was tested with the additive Gamma GEE specified above, controlling for task complexity. The primary interface effect was interpreted as a chatbot-to-dashboard Time Ratio (TR), so values below 1 indicate faster performance in the chatbot condition. Completion time was somewhat shorter in the chatbot condition (adjusted mean = 213.606 s) than in the dashboard condition (adjusted mean = 221.673 s), but the chatbot-to-dashboard TR was 0.964, 95\% CI [0.857, 1.083], $z = -0.622$, $p = .534$. Because the interval included 1, H1c was not supported.

\paragraph{H2c} The hypothesis was tested with a Gamma GEE including an interaction between interface condition and task complexity, to assess moderation \cite{Mize2019}. The omnibus interaction was assessed with a Wald test and followed by Holm-adjusted ratio-of-time-ratios contrasts \cite{Aickin1996,Huang2018}. Because completion time is positive and right-skewed, the Gamma log-link model makes multiplicative effects the primary inferential scale. For interpretability, however, the figure displays predicted completion time in seconds and the corresponding \emph{Chatbot $-$ Dashboard} time differences; negative values indicate that the chatbot is faster. The omnibus interaction was significant, $\chi^2 = 42.797$, df = 2, $p < .001$. The primary DiD comparing hard against easy tasks yielded a ratio-of-time-ratios of 1.633, 95\% CI [1.348, 1.977], with a log contrast of 0.490, raw $p < .001$, Holm-adjusted $p < .001$. The additional follow-ups showed the same attenuation from T2 versus T1, ratio-of-time-ratios = 1.681, 95\% CI [1.428, 1.979], raw $p < .001$, Holm-adjusted $p < .001$, whereas the T3-versus-T2 comparison was not reliable, ratio-of-time-ratios = 0.971, 95\% CI [0.818, 1.153], raw $p = .736$, Holm-adjusted $p = .736$. The plotted simple effects were consistent with this pattern: for T1, the chatbot was faster than the dashboard, difference = $-37.014$ seconds, 95\% CI [$-52.286$, $-21.521$], TR = 0.687, 95\% CI [0.585, 0.801]; for T2, the difference was 27.954 seconds, 95\% CI [$-0.945$, 57.011], TR = 1.155, 95\% CI [0.995, 1.343]; and for T3, the difference was 41.076 seconds, 95\% CI [$-14.733$, 98.251], TR = 1.121, 95\% CI [0.959, 1.311]. Predicted times were 81.166 versus 118.180 seconds for T1, 208.533 versus 180.579 seconds for T2, and 379.967 versus 338.890 seconds for T3 for chatbot and dashboard, respectively. Because the significant ratio-of-time-ratios contrasts exceeded 1, the interface slopes converged: the chatbot time advantage weakened as complexity increased (Figure \ref{fig:h2c_interaction}). H2c was therefore supported.

\paragraph{H3c} The hypothesis was tested with a Gamma GEE including an interaction between interface condition and categorized DL while controlling for task. DL was entered using the same low, medium, and high BDLI groups described above. The omnibus interaction was not significant, $\chi^2 = 0.212$, df = 2, $p = .899$, and the primary DiD comparing high versus low DL yielded a ratio-of-time-ratios of 1.028, 95\% CI [0.672, 1.557], with a log contrast of 0.028, raw $p = .890$, Holm-adjusted $p = 1.000$ \cite{Mize2019,Aickin1996,Huang2018}. Thus, neither the omnibus interaction nor the primary follow-up showed reliable convergence or divergence of the interface slopes across data-literacy levels, and H3c was not supported.

\paragraph{H4c} The hypothesis was tested with a Gamma GEE including task complexity as the focal predictor and interface as a control covariate; follow-up contrasts were interpreted as time ratios on the response scale. The omnibus complexity effect was significant, $\chi^2 = 562.997$, df = 2, $p < .001$. Adjusted mean completion time increased from 100.11 s for T1 to 194.09 s for T2 and 358.71 s for T3. The T2-versus-T1 contrast yielded TR = 1.939, 95\% CI [1.763, 2.120], Holm-adjusted $p < .001$; the T3-versus-T1 contrast yielded TR = 3.583, 95\% CI [3.219, 3.973], Holm-adjusted $p < .001$; and the T3-versus-T2 contrast yielded TR = 1.848, 95\% CI [1.687, 2.012], Holm-adjusted $p < .001$. Together with the reported T1-to-T2-to-T3 increase in completion time, these results supported H4c.

\subsection{Intended reliance}
\label{subsec:results:reliance}

Table \ref{tab:results-reliance-primary} presents the intended reliance results. Item r1 assessed willingness to rely on the interface output alone, whereas r2 and r3 assessed willingness to rely after obtaining a colleague's second opinion and after verification with another analytical tool, respectively. Thus, higher responses on r2 and r3 indicate greater willingness to rely under an additional verification condition, rather than greater unqualified reliance on the interface.

\begin{table*}[t]
\centering
\begin{minipage}{0.70\textwidth}
\centering
\scriptsize
\setlength{\tabcolsep}{4.5pt}
\renewcommand{\arraystretch}{1.08}

\caption{Item-wise H5 results for intended reliance.}
\label{tab:results-reliance-primary}

\begin{threeparttable}
\begin{tabularx}{\linewidth}{@{}
>{\bfseries}p{0.08\linewidth}
>{\raggedright\arraybackslash}p{0.23\linewidth}
>{\raggedright\arraybackslash}p{0.14\linewidth}
>{\raggedright\arraybackslash}X
@{}}
\toprule
\textbf{Item} & \textbf{Interface test} & \textbf{$p$-value} & \textbf{GUI--CUI latent contrast} \\
\midrule

r1 &
LR $= 0.601$, df $= 1$ &
$p = .438$ &
$\beta = 0.158$, 95\% CI [$-0.241$, 0.557], $z = 0.776$ \\

r2 &
LR $= 0.776$, df $= 1$ &
$p = .378$ &
$\beta = 0.144$, 95\% CI [$-0.177$, 0.465], $z = 0.881$ \\

r3 &
LR $= 6.847$, df $= 1$ &
$p = .009^{**}$ &
$\beta = -0.514$, 95\% CI [$-0.898$, $-0.131$], $z = -2.627$ \\

\bottomrule
\end{tabularx}

\begin{tablenotes}[flushleft]
\footnotesize
\item \textit{Note.} CUI = chatbot; GUI = dashboard. Contrasts are expressed as GUI--CUI; positive values indicate higher item ratings for the dashboard and negative values higher item ratings for the chatbot. Asterisks denote $^{*}p < .05$, $^{**}p < .01$, and $^{***}p < .001$.
\end{tablenotes}
\end{threeparttable}

\end{minipage}
\end{table*}

\paragraph{H5} The hypothesis was tested separately for each intended reliance item (r1--r3) using item-wise probit CLMM, because the responses were discrete ordered categories rather than continuous scores \cite{Gambarota2024,Wieditz2024}. For each item, the interface condition was entered as the focal fixed effect, with CUI as the reference level, and task complexity was included as an adjustment factor, with T1 as the reference level and T2 and T3 treatment-coded. Participant ID was modelled with a random intercept to account for repeated within-participant responses \cite{Magezi2015}. The primary test was the LRT of the interface main effect in the additive model. Dashboard--chatbot contrasts were interpreted on the latent ordinal-model scale, with positive estimates indicating higher item ratings in the dashboard condition and negative estimates indicating higher item ratings in the chatbot condition.

For r1, the interface effect was not significant, LR = 0.601, df = 1, $p = .438$, with a positive but imprecise dashboard--chatbot contrast, $\beta = 0.158$, 95\% CI [-0.241, 0.557], $z = 0.776$. For r2, the effect was also not significant, LR = 0.776, df = 1, $p = .378$, with a similarly positive but imprecise contrast, $\beta = 0.144$, 95\% CI [-0.177, 0.465], $z = 0.881$. For r3, the interface effect was significant, LR = 6.847, df = 1, $p = .009$, and the contrast was negative, $\beta = -0.514$, 95\% CI [-0.898, -0.131], $z = -2.627$.

Taken together, H5 was not supported on its primary intended-reliance item: interface type was not reliably associated with willingness to rely on the interface output alone. No reliable interface difference was observed either for reliance following a colleague's second opinion. However, participants reported greater willingness to rely on the chatbot after verification with another analytical tool. This item-specific finding may indicate a possible exploratory difference in intended reliance under verification conditions, but it does not support H5 as originally specified.

\section{Discussion}

Although LLM-based CUIs are often expected to improve decision support in data-rich settings, our findings do not support a simple superiority account relative to dashboards. Instead, the results reveal a differentiated pattern across subjective and objective outcomes. Subjectively, the CUI was associated with lower mental workload, as measured by the NASA-TLX, but this advantage attenuated and disappeared as task complexity increased. Objectively, completion time showed a similar pattern, whereas decision accuracy did not. In that case, the chatbot outperformed the dashboard only in the first task, and the overall evidence did not indicate a consistent accuracy advantage for either interface. Thus, subjective and objective outcomes showed only partial convergence: the conversational interface reduced perceived effort and, to some extent, task time, but these benefits did not generalize to decision accuracy. Overall, the findings point to a conditional rather than general advantage of the CUI, one that calls for theoretical explanation in relation to task demands and interface characteristics.

\subsection{The Relative Performance of CUI and GUI in Industrial Decision Support}
Our results suggest that LLM-based conversational systems represent a substantive advance over earlier conversational interfaces for decision support, although this advantage is conditional rather than uniform across outcomes and task demands. In our study, the conversational interface was not overall disadvantaged in decision accuracy, yielded lower reported MWL than the dashboard, and showed time advantages under lower complexity. 

\subsubsection{Effects on Subjective Workload}
Regarding subjective MWL, the conversational interface appeared to reduce the perceived mental cost of task execution relative to the dashboard, suggesting that interaction through the chatbot simplified information access and reduced the cognitive effort participants felt they had to invest (H1a supported). This interpretation is consistent with the qualitative evidence from our previous exploratory study, where participants described the chatbot as reducing the need to search across the interface and manually assemble the answer. It also differs from the findings of \citet{nguyen_user_2022}, where participants reported higher MWL in the CUI condition than in the GUI condition. One plausible explanation is that the conversational system evaluated in that study differed substantially from the LLM-based interface used here, since the study predates the widespread adoption of current genAI chatbots.

However, the advantage of the CUI was not uniform across task-demand levels. When task complexity was considered, the interaction model provided better explanatory power than the additive model alone. The overall direction remained unchanged, but the pattern across tasks showed that the CUI's MWL advantage attenuated as complexity increased (DiD $\beta = -0.962$), consistent with H2a. As shown in Figure \ref{fig:h2a_interaction}, MWL increased more steeply across tasks in the CUI condition than in the GUI condition. One plausible explanation is that the CUI imposed a lower initial onboarding cost, in accordance with the previous study and \citet{liu_conversational_2024}: users could begin with NL requests and progressively learn task criteria through the dialogue, whereas the GUI required earlier orientation to views, controls, and visual conventions. As task demands increased, however, the initial advantage became harder to sustain. More information had to be retained, integrated, and verified during the interaction, and these requirements may have been better supported by the persistent external representation provided by the dashboard. 

\subsubsection{Effects on Objective Performance: Decision Accuracy and Completion Time} 
The objective performance indicators showed a more limited advantage for the CUI than the subjective workload results. For decision accuracy, we did not find an overall difference between the two interface experimental conditions. Thus, the evidence does not indicate a general accuracy advantage for either the LLM-based CA or the dashboard. Likewise, the simple CUI-versus-GUI contrasts within each task were not significant, indicating that neither interface showed a reliable accuracy advantage at any single task level (see \ref{app:robustness}). This result should not be interpreted as evidence of equivalence, but it suggests that the LLM-based CUI did not reproduce the accuracy disadvantage reported in earlier work on non-LLM conversational or text-based interfaces \citep{speier_influence_2003, hettiachchi_hi_2020}.

At the same time, the significant interaction for H2b indicates that the null overall effect masked a complexity-dependent shift. The interaction model accounted for the data better than the additive model, suggesting that the relative accuracy of the two interfaces changed as task complexity increased. Specifically, the significant T3-versus-T1 DiD (DiD $=-0.109$) indicates that the CUI--GUI accuracy gap was 10.9 percentage points smaller in the highest-complexity task than in the lowest-complexity task. In other words, the descriptively favourable chatbot gap observed at low complexity did not hold at high complexity. By contrast, the T2-versus-T1 and T3-versus-T2 contrasts were not significant, so the results do not support a clearly stepwise attenuation across adjacent task levels. Rather, they support a broader low-to-high complexity attenuation of the chatbot's relative accuracy advantage.

A similar pattern emerged for task completion time. In the additive model, we did not find a significant overall difference between the two interfaces (H1c not supported). When task complexity was included as a moderator, however, the results indicated that the CUI's time advantage was conditional on task demands (H2c supported), similar to MWL and decision accuracy. The conversational interface was significantly faster than the dashboard in the easier task, but this advantage disappeared as complexity increased. Both the T2-versus-T1 and T3-versus-T1 ratio-of-time-ratios were significant, indicating attenuation of the chatbot's relative time advantage. However, the T3-versus-T2 contrast was not significant, suggesting that most of the attenuation occurred once the task moved beyond the lowest-complexity level, rather than progressively increasing from medium to high complexity.

In summary, the decision accuracy and completion time results suggest that the CUI's objective performance advantage was fragile. The chatbot appeared beneficial when the task could be handled through a relatively direct request, but this advantage weakened when participants had to coordinate more information. This is consistent with \citet{bacic_task-representation_2018}, who showed that perceived cognitive effort, time, and accuracy do not necessarily move together across task-representation conditions.

Descriptively, the pattern also suggests that further increases in task complexity might eventually favour the dashboard, although this interpretation remains tentative. One possible explanation is that repeated use becomes progressively more beneficial in the GUI condition, as users overcome the dashboard's higher initial orientation cost and become more efficient with its persistent visual structure, a pattern also reflected in the narrowing MWL gap as complexity increased. A similar pattern was observed by \citet{liu_conversational_2024} for completion time. However, more evidence is needed to establish whether the observed attenuation reflects a genuine crossover tendency at higher levels of task complexity.

\subsection{Information Overload and Cognitive Fit} 
One broader interpretation of these findings is that LLM-based CUIs may mitigate information burden more effectively than earlier conversational systems, while still retaining some scaling constraints of conversational interaction. A common expectation is that modern conversational systems can reduce IO by digesting information and offloading part of the user's information-processing burden. Our results are consistent with this expectation at lower complexity, where the chatbot was associated with lower MWL and faster completion times. However, as task complexity increased through a larger number of information cues and greater coordination demands, this relative advantage attenuated. 

This suggests that the limiting factor may not lie only in the underlying \textit{intelligence} of the system, but also in the representational nature of the interaction style. When the solution space becomes larger, visual interfaces may better support filtering, comparison, and verification because of their higher representational bandwidth and their ability to keep multiple cues simultaneously visible \citep{speier_influence_2003, larkin_why_1987}. Similarly, \citet{smerecnik_understanding_2010} found that visualizations can attract attention, support information extraction, and improve comprehension more effectively than text. By contrast, the sequential and transient nature of NL interaction makes these operations harder relative to simpler tasks. Accordingly, the present findings therefore suggest that LLM-based CAs can lower the cost of accessing information, but conversational interaction may still be less suited as increasing task complexity raises the information load that users must manage.

CFT \citep{vessey_cognitive_1991} helps explain why the same increase in informational demands may have affected the two interfaces differently. In general terms, cognitive fit depends on how well a representation supports the processing required by the task: better fit reduces the mental transformation needed to produce a usable internal representation. In the easier task (T1), participants had to identify the day with the lowest consumption by comparing values across a limited set. This type of task could in principle benefit from the dashboard's visual representation. Yet participants reported lower MWL and completed the task faster with the chatbot. One possible explanation is that, at this low level of complexity, any fit-related advantage of the dashboard's visual representation was too weak to dominate the broader interaction cost of using that interface. \citet{bacic_advancing_2022} argue that, in simple tasks, users may not consciously recognize or accurately report the cognitive-effort differences assumed by CFT, even when such differences exist. In our case, subjective MWL in T1 may therefore have been shaped more by the chatbot's lower interaction threshold and easier task initiation than by any subtle representational advantage of the dashboard.

Interestingly, the remaining tasks were analytical, and therefore predominantly symbolic according to CFT, yet this did not translate into a stable advantage for the CUI. This suggests that representation type alone was not sufficient to determine cognitive fit in our setting, and that task complexity moderated the effect of interface type. This interpretation is consistent with \citet{speier_influence_2006}, who found that task complexity can affect the relationship between information presentation format and decision performance. In line with Vessey's extension of CFT \citep{vessey_theory_2007}, increasing complexity within an analytical task may alter the relative advantage of different representations by increasing the amount of analytical evaluation required. As these demands grow, users may cope more effectively by relying on a more perceptual representation. Our study did not allow participants to freely choose their representation or strategy, since interaction style was experimentally constrained. Nevertheless, the observed attenuation of the CUI advantage, together with the descriptive crossover in objective performance from the medium-complexity task onward, may signal a difference in coping mechanisms between the two interaction styles, consistent with Vessey's prediction.

\subsection{Data Literacy, Intended Reliance, and the Conditions of Effective Use}
The DL analyses did not provide consistent evidence that participants benefited differently from the two interfaces. We expected lower DL to disadvantage participants more strongly in the dashboard condition, but this pattern was not reliably supported across outcomes (H3a--c). DL did not significantly moderate the interface effect on MWL or completion time. For accuracy, the omnibus $Interface \times Data\ Literacy$ interaction was significant, indicating some non-parallelism across data-literacy groups; however, the prespecified high-versus-low contrast was not significant, and therefore the hypothesized moderation pattern was not supported. The result should thus be interpreted as an exploratory indication that DL may have shaped accuracy in some way, although the specific form of this effect remains unclear.

However, this does not imply that user capabilities are irrelevant, but it suggests that the present operationalization of DL may have been too broad. The BDLI offered a pragmatic self-report measure of general DL confidence, while the dashboard condition may depend more directly on narrower and objectively measurable abilities, such as visualization literacy: the ability to read, interpret, and use graphical representations to answer data-related questions \citep{boy_principled_2014,lee_vlat_2017}. Future work should therefore use more task-specific and performance-based measures, distinguishing, for example, visualization literacy for dashboard use from linguistic and prompting abilities for CUI use.

A similar qualification applies to intended reliance. The primary item, which assessed whether participants would make a subsequent decision based solely on the information provided by the assigned interface, showed no reliable interface effect. Thus, the lower workload observed for the CUI did not translate into greater willingness to rely on it as a standalone basis for decision-making. The item concerning reliance after seeking a colleague's second opinion also showed no reliable interface effect. The only significant interface effect emerged for the third item, where reliance was explicitly conditional on verification with another analytical tool. This contrast is informative: participants in the CUI condition did not report higher sole intended reliance than participants in the dashboard condition reported for their respective interface (r1), but they did report an higher intended reliance when the chatbot output could be checked against another analytical source (r3). This suggests that external verification may be more important for conversational decision support than for dashboard-based interaction, and that in decision making contexts a CA may not be perceived as sufficient when used as a sole source of evidence.

This interpretation aligns with work showing that users' willingness to rely on algorithmic or AI-based support can depend on opportunities for control, adjustment, or independent checking \citep{dietvorst_overcoming_2018,lu_does_2024}, and on users' ability to evaluate the system's output \citep{zhang_you_2024}. It is also compatible with recent evidence that efficient AI assistance can improve task performance while raising concerns about accountability and over-reliance when users receive direct answers without sufficient safeguards \citep{spatola_efficiency-accountability_2024}. In the present study, however, reliance was measured as an antecedent-oriented intention, not as behavioural reliance or as evidence of appropriate reliance, as research literature usually do. The intended-reliance instrument was also newly adapted and has not yet been psychometrically validated. These results should therefore be treated as exploratory, while still offering useful indications for future research on reliance and related factors in LLM-based decision support.

\subsection{Objective task complexity effects on experimental outcomes}
The results also provide evidence that the task complexity manipulation behaved as intended. Following Wood's model \citep{wood_task_1986}, task complexity was operationalized as an objective property of the task rather than as a subjective feeling of difficulty. The underlying assumption is that increasing objective demands should translate into greater subjective and behavioural difficulty \citep{longo_human_2022}. The analysis therefore tested whether this objective manipulation predicted systematic changes in the main experimental outcomes after controlling for interface condition.

Across all three outcome families, the pattern was consistent. As expected, subjective MWL and completion time increased monotonically from T1 to T2 to T3, while decision accuracy showed the opposite pattern. Higher objective task complexity was therefore associated with higher MWL, lower accuracy, and longer completion time, supporting H4a--H4c. This pattern suggests that the task design captured behaviourally and perceptually meaningful differences in complexity. In this sense, Wood's framework was useful both as a design rationale for constructing the experimental tasks and as a predictive framework for anticipating how task demands would affect user performance.

At the same time, the effects were not perfectly linear across outcomes. The largest accuracy loss occurred between T2 and T3, while completion time increased substantially at each step. This suggests that increasing complexity may initially be absorbed through additional time investment, but beyond a certain point it begins to affect decision accuracy more sharply. This interpretation is consistent with the intended manipulation of information load: the progressively increasing levels of objective complexity were specifically designed to move participants closer to the IO threshold. The observed pattern therefore supports the role of task complexity as a mechanism through which information load becomes visible in measurable performance outcomes.

\subsection{Copy-Paste Use as a Task Strategy}
A possible behavior that could have simplified tasks for participants assigned to the CUI condition is \textit{copy-pasting}. Whereas dashboard users must reformulate the given problem through GUI interactions, chatbot users can copy-paste parts of the task request directly. In our data, this behavior was adopted (partly or fully in at least one task) by 29 CUI participants, while 35 did not use it. As reported in the \ref{app:copy-paste-checks}, \textit{copy-paste} use was not associated with significant differences in subjective workload (NASA-TLX) or accuracy. In contrast, it was associated with significantly faster completion times overall (about 59 seconds faster on average; moderate effect size, $d \approx -0.71$), with the strongest effect in Task 3 (Holm-corrected $p=0.015$). Overall, these results suggest that \textit{copy-paste} primarily acts as a time-efficiency strategy rather than a mechanism for reducing workload or improving accuracy. Because this comparison was observational, it should not be interpreted causally. Nevertheless, the association indicates that copy-paste use may explain part of the CUI time advantage, while it does not appear to account for the MWL or accuracy patterns.

\subsection{Implications}
\subsubsection{Theoretical Implications}

A first theoretical implication of this study concerns CFT and its relevance in the era of genAI. The study incrementally extends CFT research in two ways. First, it applies the theory to LLM-based CUIs, understood as symbolic, language-based interfaces for decision support. In doing so, it shows that the theory remains analytically useful when information is represented and mediated through NL. Second, it connects this application to objective task complexity, adding evidence on the relationship between task demands and cognitive fit.

A related implication concerns the value of treating task complexity as an objective property of the task, rather than only as a post-hoc subjective judgement of difficulty \citep{wood_task_1986, longo_human_2022}. Defining task complexity a priori can make comparisons between heterogeneous interfaces more theoretically grounded, because observed differences in performance may otherwise reflect uncontrolled differences in what the task required. This does not make tasks interface-independent in an absolute sense, since the interface can still reshape how users act on the task. However, it helps distinguish between the demands built into the task and the additional costs or benefits introduced by a given representation or interaction design.

Objective complexity may also support stronger cross-study comparison in HCI. Systematic reviews and meta-analyses often classify tasks qualitatively, using broad labels such as simple and complex. These labels are useful, but they are often too coarse to determine whether different studies are testing comparable task demands. As discussed by \citet{salimzadeh_missing_2023}, a quantitative or semi-quantitative task complexity specification could improve how studies are grouped, compared, and synthesized.

This implication has one caveat. Users may exercise discretion in how they complete a task, and high discretion can make the same nominal task produce different effective complexities because users may choose different resolution paths \citep{gill_task_2006}. This is especially relevant when tasks contain dynamic components because changing task states can alter the number and sequence of operations required. In this study, this component was intentionally fixed to zero and excluded from the complexity computation (see Appendix \ref{app:task:complexity}). Future HCI studies using objective complexity should therefore report the procedure used to estimate it and the degree of discretion allowed during task execution.

\subsubsection{Practical Implications}

Our findings suggest that LLM-based CUIs should not be treated as straightforward replacements for dashboards. Instead, interface choice should be aligned with task demands. The CUI was particularly effective when tasks required bounded information access. In such cases, NL interaction can reduce navigation costs and allow users to obtain relevant information without manually searching across multiple views. However, the weakening of the CUI's advantage as task complexity increased suggests that dashboards remain important for more demanding decision tasks. When users must compare several cues, the persistent visual representations offered by dashboards can still provide practical advantages. GUIs make information spatially available, support direct manipulation, and provide higher representational bandwidth than a pure CUI. Organizations should therefore be cautious about treating LLM-based assistants as general-purpose substitutes for graphical decision-support systems.

A more promising design direction is therefore a hybrid CUI--GUI interaction modality. In such systems, a conversational assistant could answer a user's initial question, identify relevant indicators, and direct the user to the corresponding dashboard view or visualization. Conversely, the dashboard could provide the visual context from which users launch more specific NL queries. This would allow each modality to support the part of the task for which it is better suited. However, further research is needed to understand when such hybrids outperform individual modalities, and when they may instead introduce additional coordination costs.

\subsection{Limitations}

This study involved several methodological trade-offs, partly due to the nature of the constructs under investigation, which are theoretically and operationally contested, and partly due to the constraints of an unmoderated online experimental design.

First, DL was measured using a brief self-report questionnaire (BDLI) rather than a comprehensive performance-based assessment. This choice was pragmatic, given the lack of a single agreed definition of data literacy and the need to keep the study short. Moreover, the population under investigation (i.e., industrial managers) may have presented very similar levels of DL, which could have levelled off BDLI scores. Another possibility is related to the exaggerated perception that participants may have had of their own capabilities \citep{zell_people_2014}. As a result, these factors may have contributed to the limited explanatory role of DL in the present results.

Second, MWL was measured through NASA-TLX. Although NASA-TLX remains widely used in HCI, mental workload is still conceptually contested \citep{longo_human_2022}, and recent work has questioned whether NASA-TLX fully captures the construct in contemporary interactive systems \citep{kosch_survey_2023,babaei_should_2025}. Because the study was administered online, physiological measures such as eye tracking or pupillometry could not be included. The workload results should therefore be interpreted as perceived or subjective MWL, not as a complete measure of the cognitive processes involved during task execution.

Third, decision accuracy is easier to define for well-defined tasks than for more ill-defined decision tasks. For example, in Task 1, the request to identify the highest number in a set of numbers logically implies and allows for a definite solution. Conversely, Tasks 2 and 3 depend on a variety of context-dependent constraints and situations. While the logic of identifying the highest value is straightforward and easily communicated to participants, the more complex tasks did not have an equally determinate solution. For this reason, to make scoring possible, we applied a trade-off: the simplest and most explicit task wording was chosen and the solution set was defined using the simplest defensible scoring rules. However, other reasonable solutions may still have existed. Therefore, decision accuracy results should be read as alignment with our scoring model rather than as absolute correctness.

Finally, intended reliance was measured with ad hoc items developed for this study. Although these items were informed by the Theory of Planned Behavior \citep{ajzen_constructing_2006}, they were not psychometrically validated. For this reason, we analysed them separately and did not compute a composite reliance score. The reliance findings should therefore be considered exploratory, as stated in the related hypothesis.

Overall, these limitations delimit the interpretation of the interface comparison, and the results should be understood as evidence about the two tested interfaces under the specific task, measurement, and scoring choices adopted in this study.

\subsection{Conclusions and Future Works}
This study examined whether an LLM-based CAs can provide an effective alternative to dashboards for industrial decision support. The findings do not support a general replacement account. In fact, although the conversational interface reduced perceived mental workload overall and offered efficiency advantages in less demanding tasks, these benefits attenuated as task complexity increased. Moreover, it did not produce a consistent advantage in decision accuracy, and its completion-time advantages diminished as task complexity increased. The expected moderating role of general self-reported data literacy was also not supported, and the conversational interface was not preferred as a sole basis for subsequent decisions.

These results suggest that LLM-based conversational interfaces can improve access to industrial data by reducing interactional and navigational demands, but their value depends on the demands of the task. Therefore, rather than treating conversational systems as substitutes for graphical decision-support tools, future industrial interfaces should investigate hybrid designs that combine conversational access with visual overview and verifiable evidence. Future research should also examine whether the patterns observed in this study generalize beyond industrial settings by extending comparisons between conversational and graphical interfaces to other domains, such as healthcare, and to broader populations beyond professionals.

\section*{CRediT authorship contribution statement}
\textbf{Roberto Figliè}: Writing -- original draft, Writing -- review \& editing, Visualization, Validation, Software, Project administration, Methodology, Investigation, Formal analysis, Data curation, Conceptualization. \textbf{Simone Caputo}: Writing -- original draft, Writing -- review \& editing, Methodology, Formal analysis, Data curation, Visualization, Conceptualization. \textbf{Alan Serrano}: Writing -- review \& editing, Conceptualization. \textbf{Daria Mikhaylova}: Writing -- review \& editing. \textbf{Tommaso Turchi}: Writing -- review \& editing, Supervision, Project administration. \textbf{Daniele Mazzei}: Supervision, Project administration, Funding acquisition.

\section*{Declaration of competing interest}
One author holds a senior leadership role at Zerynth.  All other authors declare no competing interests. 

\section*{Funding}
This work has been partially funded by Programme Erasmus+, Knowledge Alliances, application no. 621639-EPP-1-2020-1-IT-EPPKA2-KA, PLANET4: Practical Learning of Artificial iNtelligence on the Edge for indusTry 4.0. This work was also supported by the National PhD Programme in Artificial Intelligence funded under Italy’s PNRR, NextGenerationEU (CUP I51J22000590007).
Zerynth provided cloud compute resources enabling access to GPT-4o via Microsoft Azure. Selected layout and interaction patterns of the graphical interface were informed by Zerynth's industrial analytics platform. The sponsors had no role in the study design, collection, analysis, or interpretation of data, writing of the report, or the decision to submit the article for publication.








\bibliographystyle{elsarticle-harv} 
\bibliography{bibliography.bib}

@article{allen_data_2021,
	title = {Data visualization for {Industry} 4.0: {A} stepping-stone toward a digital future, bridging the gap between academia and industry},
	volume = {2},
	issn = {2666-3899},
	shorttitle = {Data visualization for {Industry} 4.0},
	url = {https://www.sciencedirect.com/science/article/pii/S2666389921000921},
	doi = {10.1016/j.patter.2021.100266},
	number = {5},
	urldate = {2026-03-13},
	journal = {Patterns},
	author = {Allen, Louis and Atkinson, Jack and Jayasundara, Dinusha and Cordiner, Joan and Moghadam, Peyman Z.},
	month = may,
	year = {2021},
	pages = {100266},
}

@article{lindner_behavioral_2025,
	title = {A behavioral perspective on visualization in manufacturing and operations management: a review, framework, and research agenda},
	volume = {18},
	issn = {1936-9743},
	shorttitle = {A behavioral perspective on visualization in manufacturing and operations management},
	url = {https://doi.org/10.1007/s12063-024-00534-9},
	doi = {10.1007/s12063-024-00534-9},
	language = {en},
	number = {1},
	urldate = {2026-03-12},
	journal = {Operations Management Research},
	author = {Lindner, Fabian and Reiner, Gerald and Keil, Sophia},
	month = mar,
	year = {2025},
	pages = {317--352},
}

@article{yigitbasioglu_review_2012,
	title = {A review of dashboards in performance management: {Implications} for design and research},
	volume = {13},
	issn = {1467-0895},
	shorttitle = {A review of dashboards in performance management},
	url = {https://www.sciencedirect.com/science/article/pii/S1467089511000443},
	doi = {10.1016/j.accinf.2011.08.002},
	number = {1},
	urldate = {2024-11-06},
	journal = {International Journal of Accounting Information Systems},
	author = {Yigitbasioglu, Ogan M. and Velcu, Oana},
	month = mar,
	year = {2012},
	pages = {41--59},
}

@article{hjelle_organizational_2024,
	title = {Organizational decision making and analytics: {An} experimental study on dashboard visualizations},
	volume = {61},
	issn = {0378-7206},
	shorttitle = {Organizational decision making and analytics},
	url = {https://www.sciencedirect.com/science/article/pii/S0378720624000934},
	doi = {10.1016/j.im.2024.104011},
	number = {6},
	urldate = {2025-05-09},
	journal = {Information \& Management},
	author = {Hjelle, Sara and Mikalef, Patrick and Altwaijry, Najwa and Parida, Vinit},
	month = sep,
	year = {2024},
	pages = {104011},
}

@article{handler_large_2024,
	title = {Large language models present new questions for decision support},
	volume = {79},
	issn = {0268-4012},
	url = {https://www.sciencedirect.com/science/article/pii/S0268401224000598},
	doi = {10.1016/j.ijinfomgt.2024.102811},
	urldate = {2026-03-12},
	journal = {International Journal of Information Management},
	author = {Handler, Abram and Larsen, Kai R. and Hackathorn, Richard},
	month = dec,
	year = {2024},
	pages = {102811},
}

@article{elbasheer_natural_2025,
	title = {Natural language-driven production planning: integrating large language models with automatic simulation model generation in manufacturing systems},
	issn = {1572-8145},
	shorttitle = {Natural language-driven production planning},
	url = {https://doi.org/10.1007/s10845-025-02732-z},
	doi = {10.1007/s10845-025-02732-z},
	language = {en},
	urldate = {2026-03-12},
	journal = {Journal of Intelligent Manufacturing},
	author = {Elbasheer, Mohaiad and Laili, Yuanjun and Longo, Francesco and Solina, Vittorio and Tao, Yiran and Veltri, Pierpaolo and Zhang, Yuteng and Zhang, Lin},
	month = nov,
	year = {2025},
}

@article{cimino_integrating_2025,
	title = {Integrating large language models with industrial simulation for multi-level decision support: an innovation management perspective in {Industry} 5.0},
	volume = {29},
	issn = {1460-1060},
	shorttitle = {Integrating large language models with industrial simulation for multi-level decision support},
	url = {https://doi.org/10.1108/EJIM-10-2024-1246},
	doi = {10.1108/EJIM-10-2024-1246},
	number = {11},
	urldate = {2026-03-12},
	journal = {European Journal of Innovation Management},
	author = {Cimino, Antonio and Longo, Francesco and Solina, Vittorio and Veltri, Pierpaolo},
	month = dec,
	year = {2025},
	pages = {27--53},
}

@article{schobel_charting_2024,
	title = {Charting the {Evolution} and {Future} of {Conversational} {Agents}: {A} {Research} {Agenda} {Along} {Five} {Waves} and {New} {Frontiers}},
	volume = {26},
	issn = {1572-9419},
	shorttitle = {Charting the {Evolution} and {Future} of {Conversational} {Agents}},
	url = {https://doi.org/10.1007/s10796-023-10375-9},
	doi = {10.1007/s10796-023-10375-9},
	language = {en},
	number = {2},
	urldate = {2026-03-13},
	journal = {Information Systems Frontiers},
	author = {Schöbel, Sofia and Schmitt, Anuschka and Benner, Dennis and Saqr, Mohammed and Janson, Andreas and Leimeister, Jan Marco},
	month = apr,
	year = {2024},
	pages = {729--754},
}

@article{shahrzadi_causes_2024,
	title = {Causes, consequences, and strategies to deal with information overload: {A} scoping review},
	volume = {4},
	issn = {2667-0968},
	shorttitle = {Causes, consequences, and strategies to deal with information overload},
	url = {https://www.sciencedirect.com/science/article/pii/S2667096824000508},
	doi = {10.1016/j.jjimei.2024.100261},
	number = {2},
	urldate = {2025-05-29},
	journal = {International Journal of Information Management Data Insights},
	author = {Shahrzadi, Leila and Mansouri, Ali and Alavi, Mousa and Shabani, Ahmad},
	month = nov,
	year = {2024},
	pages = {100261},
}

@article{roetzel_information_2019,
	title = {Information overload in the information age: a review of the literature from business administration, business psychology, and related disciplines with a bibliometric approach and framework development},
	volume = {12},
	issn = {2198-2627},
	shorttitle = {Information overload in the information age},
	url = {https://doi.org/10.1007/s40685-018-0069-z},
	doi = {10.1007/s40685-018-0069-z},
	language = {en},
	number = {2},
	urldate = {2023-09-08},
	journal = {Business Research},
	author = {Roetzel, Peter Gordon},
	month = dec,
	year = {2019},
	pages = {479--522},
}

@article{eppler_concept_2004,
	title = {The {Concept} of {Information} {Overload}: {A} {Review} of {Literature} from {Organization} {Science}, {Accounting}, {Marketing}, {MIS}, and {Related} {Disciplines}},
	volume = {20},
	issn = {0197-2243},
	shorttitle = {The {Concept} of {Information} {Overload}},
	url = {https://doi.org/10.1080/01972240490507974},
	doi = {10.1080/01972240490507974},
	number = {5},
	urldate = {2023-09-17},
	journal = {The Information Society},
	author = {Eppler, Martin J. and Mengis, Jeanne},
	month = nov,
	year = {2004},
	note = {Publisher: Routledge
\_eprint: https://doi.org/10.1080/01972240490507974},
	pages = {325--344},
}

@incollection{kock_information_2011,
	title = {The {Information}  {Overload} {Paradox}: {A} {Cross}-{Cultural} {Research} {Study}},
	url = {http://services.igi-global.com/resolvedoi/resolve.aspx?doi=10.4018/978-1-60960-605-3},
	doi = {10.4018/978-1-60960-605-3},
	language = {en},
	urldate = {2026-01-20},
	booktitle = {International {Enterprises} and {Global} {Information} {Technologies}: {Advancing} {Management} {Practices}},
	publisher = {IGI Global},
	author = {Kock, Ned and Del Aguila-Obra, Ana Rosa and Padilla-Meléndez, Antonio},
	year = {2011},
}

@article{speier_influence_1999,
	title = {The {Influence} of {Task} {Interruption} on {Individual} {Decision} {Making}: {An} {Information} {Overload} {Perspective}},
	volume = {30},
	issn = {0011-7315, 1540-5915},
	shorttitle = {The {Influence} of {Task} {Interruption} on {Individual} {Decision} {Making}},
	url = {https://onlinelibrary.wiley.com/doi/10.1111/j.1540-5915.1999.tb01613.x},
	doi = {10.1111/j.1540-5915.1999.tb01613.x},
	language = {en},
	number = {2},
	urldate = {2026-01-19},
	journal = {Decision Sciences},
	author = {Speier, Cheri and Valacich, Joseph S. and Vessey, Iris},
	month = mar,
	year = {1999},
	pages = {337--360},
}

@article{arnold_dealing_2023,
	title = {Dealing with information overload: a comprehensive review},
	volume = {14},
	shorttitle = {Dealing with information overload},
	url = {https://pmc.ncbi.nlm.nih.gov/articles/PMC10322198/},
	doi = {10.3389/fpsyg.2023.1122200},
	language = {en},
	urldate = {2024-11-11},
	journal = {Frontiers in Psychology},
	author = {Arnold, Miriam and Goldschmitt, Mascha and Rigotti, Thomas},
	month = jun,
	year = {2023},
	pmid = {37416535},
	pages = {1122200},
}

@article{falschlunger_infovis_2016,
	title = {{InfoVis}: {The} {Impact} of {Information} {Overload} on {Decision} {Making} {Outcome} in {High} {Complexity} {Settings}},
	language = {en},
	author = {Falschlunger, Lisa and Lehner, Othmar and Treiblmaier, Horst},
	year = {2016},
    journal = {SIGHCI 2016 Proceedings. 3.},
}

@article{phillips-wren_decision_2020,
	title = {Decision making under stress: the role of information overload, time pressure, complexity, and uncertainty},
	volume = {29},
	issn = {1246-0125},
	shorttitle = {Decision making under stress},
	url = {https://doi.org/10.1080/12460125.2020.1768680},
	doi = {10.1080/12460125.2020.1768680},
	number = {sup1},
	urldate = {2026-01-13},
	journal = {Journal of Decision Systems},
	author = {Phillips-Wren, Gloria and Adya, Monica},
	month = aug,
	year = {2020},
	note = {Publisher: Taylor \& Francis
\_eprint: https://doi.org/10.1080/12460125.2020.1768680},
	pages = {213--225},
}

@article{blair_information_2011,
	title = {Information {Overload}’s 2,300-{Year}-{Old} {History}},
	issn = {0017-8012},
	url = {https://hbr.org/2011/03/information-overloads-2300-yea},
	urldate = {2023-09-08},
	journal = {Harvard Business Review},
	author = {Blair, Ann},
	month = mar,
	year = {2011},
	note = {Section: Managing yourself},
}

@article{young_state_2015,
	title = {State of science: mental workload in ergonomics},
	volume = {58},
	issn = {0014-0139},
	shorttitle = {State of science},
	url = {https://doi.org/10.1080/00140139.2014.956151},
	doi = {10.1080/00140139.2014.956151},
	number = {1},
	urldate = {2024-11-11},
	journal = {Ergonomics},
	publisher = {Taylor \& Francis},
	author = {Young, Mark S. and Brookhuis, Karel A. and Wickens, Christopher D. and Hancock, Peter A.},
	month = jan,
	year = {2015},
	note = {\_eprint: https://doi.org/10.1080/00140139.2014.956151},
	pages = {1--17},
}

@article{kosch_survey_2023,
	title = {A {Survey} on {Measuring} {Cognitive} {Workload} in {Human}-{Computer} {Interaction}},
	volume = {55},
	issn = {0360-0300},
	url = {https://dl.acm.org/doi/10.1145/3582272},
	doi = {10.1145/3582272},
	number = {13s},
	urldate = {2024-11-11},
	journal = {ACM Comput. Surv.},
	author = {Kosch, Thomas and Karolus, Jakob and Zagermann, Johannes and Reiterer, Harald and Schmidt, Albrecht and Woźniak, Paweł W.},
	year = {2023},
	pages = {283:1--283:39},
}

@article{bacic_advancing_2022,
	title = {Advancing our understanding and assessment of cognitive effort in the cognitive fit theory and data visualization context: {Eye} tracking-based approach},
	volume = {163},
	issn = {0167-9236},
	shorttitle = {Advancing our understanding and assessment of cognitive effort in the cognitive fit theory and data visualization context},
	url = {https://www.sciencedirect.com/science/article/pii/S0167923622001336},
	doi = {10.1016/j.dss.2022.113862},
	urldate = {2024-11-11},
	journal = {Decision Support Systems},
	author = {Bačić, Dinko and Henry, Raymond},
	month = dec,
	year = {2022},
	pages = {113862},
}

@article{longo_human_2022,
	title = {Human {Mental} {Workload}: {A} {Survey} and a {Novel} {Inclusive} {Definition}},
	volume = {13},
	issn = {1664-1078},
	shorttitle = {Human {Mental} {Workload}},
	url = {https://www.frontiersin.org/journals/psychology/articles/10.3389/fpsyg.2022.883321/full},
	doi = {10.3389/fpsyg.2022.883321},
	language = {English},
	urldate = {2025-02-24},
	journal = {Frontiers in Psychology},
	publisher = {Frontiers},
	author = {Longo, Luca and Wickens, Christopher D. and Hancock, Gabriella and Hancock, P. A.},
	month = jun,
	year = {2022},
}

@techreport{casner_measuring_2010,
	title = {Measuring and {Evaluating} {Workload}: {A} {Primer}},
	shorttitle = {Measuring and {Evaluating} {Workload}},
	author = {Casner, Stephen and Gore, Brian},
	month = jul,
	year = {2010},
}

@article{babaei_should_2025,
	title = {Should we use the {NASA}-{TLX} in {HCI}? {A} review of theoretical and methodological issues around {Mental} {Workload} {Measurement}},
	volume = {201},
	issn = {1071-5819},
	shorttitle = {Should we use the {NASA}-{TLX} in {HCI}?},
	url = {https://www.sciencedirect.com/science/article/pii/S1071581925000722},
	doi = {10.1016/j.ijhcs.2025.103515},
	urldate = {2025-05-29},
	journal = {International Journal of Human-Computer Studies},
	author = {Babaei, Ebrahim and Dingler, Tilman and Tag, Benjamin and Velloso, Eduardo},
	month = jun,
	year = {2025},
	pages = {103515},
}

@article{mosaferchi_simple_2025,
	title = {A simple upgrade or a gradual retirement? {A} critical commentary on {NASA}-{TLX}},
	volume = {0},
	issn = {0014-0139},
	shorttitle = {A simple upgrade or a gradual retirement?},
	url = {https://doi.org/10.1080/00140139.2025.2596331},
	doi = {10.1080/00140139.2025.2596331},
	number = {0},
	urldate = {2025-12-04},
	journal = {Ergonomics},
	publisher = {Taylor \& Francis},
	author = {Mosaferchi, Saeedeh and Lowe, Robert and Mortezapour, Alireza},
	month = dec,
	year = {2025},
	note = {\_eprint: https://doi.org/10.1080/00140139.2025.2596331},
	pages = {1--7},
}

@article{gill_task_2006,
	title = {Task {Complexity} and {Informing} {Science}: {A} {Synthesis}},
	volume = {6},
	shorttitle = {Task {Complexity} and {Informing} {Science}},
	url = {https://www.informingscience.org/Publications/3045},
	language = {en},
	urldate = {2025-03-05},
	journal = {Informing Science + IT Education Conference},
	author = {Gill, Grandon and Hicks, Richard},
	year = {2006},
}

@article{wood_task_1986,
	title = {Task complexity: {Definition} of the construct},
	volume = {37},
	issn = {0749-5978},
	shorttitle = {Task complexity},
	url = {https://www.sciencedirect.com/science/article/pii/0749597886900440},
	doi = {10.1016/0749-5978(86)90044-0},
	number = {1},
	urldate = {2025-03-12},
	journal = {Organizational Behavior and Human Decision Processes},
	author = {Wood, Robert E},
	month = feb,
	year = {1986},
	pages = {60--82},
}

@article{liu_task_2012,
	title = {Task complexity: {A} review and conceptualization framework},
	volume = {42},
	issn = {0169-8141},
	shorttitle = {Task complexity},
	url = {https://www.sciencedirect.com/science/article/pii/S0169814112000868},
	doi = {10.1016/j.ergon.2012.09.001},
	number = {6},
	urldate = {2025-03-18},
	journal = {International Journal of Industrial Ergonomics},
	author = {Liu, Peng and Li, Zhizhong},
	month = nov,
	year = {2012},
	pages = {553--568},
}

@inproceedings{salimzadeh_missing_2023,
	address = {New York, NY, USA},
	series = {{UMAP} '23},
	title = {A {Missing} {Piece} in the {Puzzle}: {Considering} the {Role} of {Task} {Complexity} in {Human}-{AI} {Decision} {Making}},
	isbn = {978-1-4503-9932-6},
	shorttitle = {A {Missing} {Piece} in the {Puzzle}},
	url = {https://dl.acm.org/doi/10.1145/3565472.3592959},
	doi = {10.1145/3565472.3592959},
	urldate = {2025-06-17},
	booktitle = {Proceedings of the 31st {ACM} {Conference} on {User} {Modeling}, {Adaptation} and {Personalization}},
	publisher = {Association for Computing Machinery},
	author = {Salimzadeh, Sara and He, Gaole and Gadiraju, Ujwal},
	year = {2023},
	pages = {215--227},
}

@article{speier_influence_2003,
	title = {The {Influence} of {Query} {Interface} {Design} on {Decision}-{Making} {Performance}},
	volume = {27},
	issn = {0276-7783},
	url = {https://www.jstor.org/stable/30036539},
	doi = {10.2307/30036539},
	number = {3},
	urldate = {2024-11-14},
	journal = {MIS Quarterly},
	publisher = {Management Information Systems Research Center, University of Minnesota},
	author = {Speier, Cheri and Morris, Michael G.},
	year = {2003},
	pages = {397--423},
}

@article{hart_nasa-task_2006,
	title = {Nasa-{Task} {Load} {Index} ({NASA}-{TLX}); 20 {Years} {Later}},
	volume = {50},
	issn = {1071-1813},
	url = {https://doi.org/10.1177/154193120605000909},
	doi = {10.1177/154193120605000909},
	language = {EN},
	number = {9},
	urldate = {2026-01-21},
	journal = {Proceedings of the Human Factors and Ergonomics Society Annual Meeting},
	publisher = {SAGE Publications Inc},
	author = {Hart, Sandra G.},
	month = oct,
	year = {2006},
	pages = {904--908},
}

@article{nuamah_evaluating_2020,
	title = {Evaluating effectiveness of information visualizations using cognitive fit theory: {A} neuroergonomics approach},
	volume = {88},
	issn = {0003-6870},
	shorttitle = {Evaluating effectiveness of information visualizations using cognitive fit theory},
	url = {https://www.sciencedirect.com/science/article/pii/S0003687020301277},
	doi = {10.1016/j.apergo.2020.103173},
	urldate = {2024-11-13},
	journal = {Applied Ergonomics},
	author = {Nuamah, Joseph K. and Seong, Younho and Jiang, Steven and Park, Eui and Mountjoy, Daniel},
	month = oct,
	year = {2020},
	pages = {103173},
}

@article{padilla_decision_2018,
	title = {Decision making with visualizations: a cognitive framework across disciplines},
	volume = {3},
	issn = {2365-7464},
	shorttitle = {Decision making with visualizations},
	url = {https://doi.org/10.1186/s41235-018-0120-9},
	doi = {10.1186/s41235-018-0120-9},
	number = {1},
	urldate = {2024-10-24},
	journal = {Cognitive Research: Principles and Implications},
	author = {Padilla, Lace M. and Creem-Regehr, Sarah H. and Hegarty, Mary and Stefanucci, Jeanine K.},
	month = jul,
	year = {2018},
	pages = {29},
}

@incollection{vessey_theory_2007,
	title = {The {Theory} of {Cognitive} {Fit}: {One} {Aspect} of a {General} {Theory} of {Problem} {Solving}?},
	shorttitle = {The {Theory} of {Cognitive} {Fit}},
	booktitle = {Human-computer {Interaction} and {Management} {Information} {Systems}: {Foundations}},
	publisher = {Routledge},
	author = {Vessey, Iris},
	year = {2007},
	note = {Num Pages: 43},
}

@inproceedings{salimzadeh_dealing_2024,
	address = {Honolulu HI USA},
	title = {Dealing with {Uncertainty}: {Understanding} the {Impact} of {Prognostic} {Versus} {Diagnostic} {Tasks} on {Trust} and {Reliance} in {Human}-{AI} {Decision} {Making}},
	isbn = {979-8-4007-0330-0},
	shorttitle = {Dealing with {Uncertainty}},
	url = {https://dl.acm.org/doi/10.1145/3613904.3641905},
	doi = {10.1145/3613904.3641905},
	language = {en},
	urldate = {2025-08-12},
	booktitle = {Proceedings of the {CHI} {Conference} on {Human} {Factors} in {Computing} {Systems}},
	publisher = {ACM},
	author = {Salimzadeh, Sara and He, Gaole and Gadiraju, Ujwal},
	month = may,
	year = {2024},
	pages = {1--17},
}

@article{cezar_cognitive_2023,
	title = {Cognitive {Overload}, {Anxiety}, {Cognitive} {Fatigue}, {Avoidance} {Behavior} and {Data} {Literacy} in {Big} {Data} environments},
	volume = {60},
	issn = {0306-4573},
	url = {https://www.sciencedirect.com/science/article/pii/S0306457323002194},
	doi = {10.1016/j.ipm.2023.103482},
	number = {6},
	urldate = {2024-11-18},
	journal = {Information Processing \& Management},
	author = {Cezar, Bibiana Giudice da Silva and Maçada, Antônio Carlos Gastaud},
	month = nov,
	year = {2023},
	pages = {103482},
}

@article{portner_data_2024,
	series = {34th {CIRP} {Design} {Conference}},
	title = {Data {Literacy} {Assessment} - {Measuring} {Data} {Literacy} {Competencies} to {Leverage} {Data}-{Driven} {Organizations}},
	volume = {128},
	issn = {2212-8271},
	url = {https://www.sciencedirect.com/science/article/pii/S2212827124006620},
	doi = {10.1016/j.procir.2024.07.047},
	urldate = {2025-07-24},
	journal = {Procedia CIRP},
	author = {Pörtner, Lara and Riel, Andreas and Klaassen, Vivian and Sezgin, Dilara and Kievits, Ysaline},
	month = jan,
	year = {2024},
	pages = {78--83},
}

@article{kirby_developing_2024,
	title = {Developing a brief data literacy measure using {CFA} and {IRT}},
	issn = {0961-0006},
	url = {https://doi.org/10.1177/09610006241300738},
	doi = {10.1177/09610006241300738},
	language = {EN},
	urldate = {2025-07-24},
	journal = {Journal of Librarianship and Information Science},
	publisher = {SAGE Publications Ltd},
	author = {Kirby, Joseph and Lewis, Jerome},
	month = dec,
	year = {2024},
	pages = {09610006241300738},
}

@article{koloski_data_2025,
	title = {Data {Literacy} in {Industry}: {High} {Time} to {Focus} on {Operationalization} {Through} {Middle} {Managers}},
	volume = {7},
	issn = {2644-2353,},
	shorttitle = {Data {Literacy} in {Industry}},
	url = {https://hdsr.mitpress.mit.edu/pub/quig0v1l/release/2},
	doi = {10.1162/99608f92.6f5dfc6f},
	language = {en},
	number = {1},
	urldate = {2025-08-12},
	journal = {Harvard Data Science Review},
	publisher = {The MIT Press},
	author = {Koloski, Dan and Porter, Caitlin and Almand-Hunter, Berkeley and Gatchell, Stephen and Logan, Valerie},
	month = jan,
	year = {2025},
}

@misc{gartner,
	title = {Definition of {Data} {Literacy} - {Gartner} {Information} {Technology} {Glossary}},
	url = {https://www.gartner.com/en/information-technology/glossary/data-literacy},
	language = {en},
    author = {Gartner},
	urldate = {2026-01-19},
	journal = {Gartner},
}

@article{pinto_strategic_2023,
	title = {A strategic approach to information literacy: data literacy. {A} systematic review},
	issn = {16992407, 13866710},
	shorttitle = {A strategic approach to information literacy},
	url = {https://revista.profesionaldelainformacion.com/index.php/EPI/article/view/87425},
	doi = {10.3145/epi.2023.nov.09},
	language = {en},
	urldate = {2026-01-19},
	journal = {El Profesional de la información},
	author = {Pinto, María and Caballero-Mariscal, David and García-Marco, Francisco-Javier and Gómez-Camarero, Carmen},
	month = nov,
	year = {2023},
	pages = {e320609},
}

@article{nouinou_decision-making_2023,
	title = {Decision-making in the context of {Industry} 4.0: {Evidence} from the textile and clothing industry},
	volume = {391},
	issn = {0959-6526},
	shorttitle = {Decision-making in the context of {Industry} 4.0},
	url = {https://www.sciencedirect.com/science/article/pii/S0959652623003426},
	doi = {10.1016/j.jclepro.2023.136184},
	urldate = {2026-01-22},
	journal = {Journal of Cleaner Production},
	author = {Nouinou, Hajar and Asadollahi-Yazdi, Elnaz and Baret, Isaline and Nguyen, Nhan Quy and Terzi, Mourad and Ouazene, Yassine and Yalaoui, Farouk and Kelly, Russell},
	month = mar,
	year = {2023},
	pages = {136184},
}

@article{bousdekis_review_2021,
	title = {A {Review} of {Data}-{Driven} {Decision}-{Making} {Methods} for {Industry} 4.0 {Maintenance} {Applications}},
	volume = {10},
	copyright = {http://creativecommons.org/licenses/by/3.0/},
	issn = {2079-9292},
	url = {https://www.mdpi.com/2079-9292/10/7/828},
	doi = {10.3390/electronics10070828},
	language = {en},
	number = {7},
	urldate = {2026-01-22},
	journal = {Electronics},
	publisher = {Multidisciplinary Digital Publishing Institute},
	author = {Bousdekis, Alexandros and Lepenioti, Katerina and Apostolou, Dimitris and Mentzas, Gregoris},
	month = jan,
	year = {2021},
	pages = {828},
}

@article{dalenogare_expected_2018,
	title = {The expected contribution of {Industry} 4.0 technologies for industrial performance},
	volume = {204},
	issn = {0925-5273},
	url = {https://www.sciencedirect.com/science/article/pii/S0925527318303372},
	doi = {10.1016/j.ijpe.2018.08.019},
	journal = {International Journal of Production Economics},
	author = {Dalenogare, Lucas Santos and Benitez, Guilherme Brittes and Ayala, Néstor Fabián and Frank, Alejandro Germán},
	year = {2018},
	pages = {383--394},
}

@article{ghobakhloo_drivers_2022,
	title = {Drivers and barriers of {Industry} 4.0 technology adoption among manufacturing {SMEs}: a systematic review and transformation roadmap},
	volume = {33},
	issn = {1741-038X},
	shorttitle = {Drivers and barriers of {Industry} 4.0 technology adoption among manufacturing {SMEs}},
	url = {https://doi.org/10.1108/JMTM-12-2021-0505},
	doi = {10.1108/JMTM-12-2021-0505},
	number = {6},
	urldate = {2023-09-14},
	journal = {Journal of Manufacturing Technology Management},
	publisher = {Emerald Publishing Limited},
	author = {Ghobakhloo, Morteza and Iranmanesh, Mohammad and Vilkas, Mantas and Grybauskas, Andrius and Amran, Azlan},
	month = jan,
	year = {2022},
	pages = {1029--1058},
}

@article{pocock_sequential_1975,
	title = {Sequential {Treatment} {Assignment} with {Balancing} for {Prognostic} {Factors} in the {Controlled} {Clinical} {Trial}},
	volume = {31},
	issn = {0006-341X},
	url = {https://www.jstor.org/stable/2529712},
	doi = {10.2307/2529712},
	number = {1},
	urldate = {2026-01-30},
	journal = {Biometrics},
	publisher = {International Biometric Society},
	author = {Pocock, Stuart J. and Simon, Richard},
	year = {1975},
	pages = {103--115},
}

@article{cohen_problem_1999,
	title = {The {Problem} of {Units} and the {Circumstance} for {POMP}},
	volume = {34},
	issn = {0027-3171},
	url = {https://doi.org/10.1207/S15327906MBR3403_2},
	doi = {10.1207/S15327906MBR3403_2},
	number = {3},
	urldate = {2025-12-09},
	journal = {Multivariate Behavioral Research},
	publisher = {Routledge},
	author = {Cohen, Patricia and Cohen, Jacob and Aiken, Leona S. and West, Stephen G.},
	month = jul,
	year = {1999},
	note = {\_eprint: https://doi.org/10.1207/S15327906MBR3403\_2},
	pages = {315--346},
}

@misc{ajzen_constructing_2006,
	title = {Constructing a {Theory} of {Planned} {Behavior} {Questionnaire}},
	url = {https://www.researchgate.net/publication/235913732_Constructing_a_Theory_of_Planned_Behavior_Questionnaire},
	author = {Ajzen, Icek},
	month = jan,
	year = {2006},
	note = {Pages: 12},
}

@inproceedings{crescenzi_adaptation_2021,
	address = {Canberra ACT Australia},
	title = {Adaptation in {Information} {Search} and {Decision}-{Making} under {Time} {Constraints}},
	isbn = {978-1-4503-8055-3},
	url = {https://dl.acm.org/doi/10.1145/3406522.3446030},
	doi = {10.1145/3406522.3446030},
	language = {en},
	urldate = {2026-02-04},
	booktitle = {Proceedings of the 2021 {Conference} on {Human} {Information} {Interaction} and {Retrieval}},
	publisher = {ACM},
	author = {Crescenzi, Anita and Capra, Rob and Choi, Bogeum and Li, Yuan},
	month = mar,
	year = {2021},
	pages = {95--105},
}

@article{hunt_endogenous_2017,
	title = {Endogenous and exogenous time pressure: {Interactions} with mathematics anxiety in explaining arithmetic performance},
	volume = {82},
	issn = {0883-0355},
	shorttitle = {Endogenous and exogenous time pressure},
	url = {https://www.sciencedirect.com/science/article/pii/S0883035516310977},
	doi = {10.1016/j.ijer.2017.01.005},
	urldate = {2026-02-04},
	journal = {International Journal of Educational Research},
	author = {Hunt, Thomas E. and Sandhu, Kaljit K.},
	month = jan,
	year = {2017},
	pages = {91--98},
}

@article{liu_conversational_2024,
	title = {Conversational versus graphical user interfaces: the influence of rational decision style when individuals perform decision-making tasks repeatedly},
	issn = {1615-5297},
	shorttitle = {Conversational versus graphical user interfaces},
	url = {https://doi.org/10.1007/s10209-024-01122-1},
	doi = {10.1007/s10209-024-01122-1},
	language = {en},
	urldate = {2024-11-12},
	journal = {Universal Access in the Information Society},
	author = {Liu, Xuanhui and Rietz, Tim and Maedche, Alexander},
	month = jun,
	year = {2024},
}

@inproceedings{masson_directgpt_2024,
	address = {Honolulu HI USA},
	title = {{DirectGPT}: {A} {Direct} {Manipulation} {Interface} to {Interact} with {Large} {Language} {Models}},
	isbn = {979-8-4007-0330-0},
	shorttitle = {{DirectGPT}},
	url = {https://dl.acm.org/doi/10.1145/3613904.3642462},
	doi = {10.1145/3613904.3642462},
	language = {en},
	urldate = {2026-03-26},
	booktitle = {Proceedings of the {CHI} {Conference} on {Human} {Factors} in {Computing} {Systems}},
	publisher = {ACM},
	author = {Masson, Damien and Malacria, Sylvain and Casiez, Géry and Vogel, Daniel},
	month = may,
	year = {2024},
	pages = {1--16},
}

@article{nguyen_user_2022,
	title = {User interactions with chatbot interfaces vs. {Menu}-based interfaces: {An} empirical study},
	volume = {128},
	issn = {0747-5632},
	shorttitle = {User interactions with chatbot interfaces vs. {Menu}-based interfaces},
	url = {https://www.sciencedirect.com/science/article/pii/S0747563221004167},
	doi = {10.1016/j.chb.2021.107093},
	urldate = {2024-11-12},
	journal = {Computers in Human Behavior},
	author = {Nguyen, Quynh N. and Sidorova, Anna and Torres, Russell},
	month = mar,
	year = {2022},
	pages = {107093},
}

@inproceedings{flohr_chat_2021,
	address = {New York, NY, USA},
	series = {{MobileHCI} '21},
	title = {Chat or {Tap}? – {Comparing} {Chatbots} with ‘{Classic}’ {Graphical} {User} {Interfaces} for {Mobile} {Interaction} with {Autonomous} {Mobility}-on-{Demand} {Systems}},
	isbn = {978-1-4503-8328-8},
	shorttitle = {Chat or {Tap}?},
	url = {https://dl.acm.org/doi/10.1145/3447526.3472036},
	doi = {10.1145/3447526.3472036},
	urldate = {2025-03-11},
	booktitle = {Proceedings of the 23rd {International} {Conference} on {Mobile} {Human}-{Computer} {Interaction}},
	publisher = {Association for Computing Machinery},
	author = {Flohr, Lukas A. and Kalinke, Sofie and Krüger, Antonio and Wallach, Dieter P.},
	year = {2021},
	pages = {1--13},
}

@inproceedings{figlie_towards_2024,
	title = {Towards an llmbased intelligent assistant for industry 5.0},
	volume = {3701},
	copyright = {All rights reserved},
	url = {https://ceur-ws.org/Vol-3701/paper7.pdf},
	urldate = {2026-03-09},
	booktitle = {Proceedings of the 1st {International} {Workshop} on {Designing} and {Building} {Hybrid} {Human}–{AI} {Systems} ({SYNERGY} 2024)},
	publisher = {CEUR-WS},
	author = {Figliè, Roberto and Turchi, Tommaso and Baldi, Giacomo and Mazzei, Daniele},
	year = {2024},
}

@article{kavaz_chatbot-based_2023,
	title = {Chatbot-{Based} {Natural} {Language} {Interfaces} for {Data} {Visualisation}: {A} {Scoping} {Review}},
	volume = {13},
	copyright = {http://creativecommons.org/licenses/by/3.0/},
	issn = {2076-3417},
	shorttitle = {Chatbot-{Based} {Natural} {Language} {Interfaces} for {Data} {Visualisation}},
	url = {https://www.mdpi.com/2076-3417/13/12/7025},
	doi = {10.3390/app13127025},
	language = {en},
	number = {12},
	urldate = {2024-10-24},
	journal = {Applied Sciences},
	publisher = {Multidisciplinary Digital Publishing Institute},
	author = {Kavaz, Ecem and Puig, Anna and Rodríguez, Inmaculada},
	month = jan,
	year = {2023},
	note = {Number: 12},
	pages = {7025},
}

@article{ke_effect_2023,
	title = {Effect of information load and cognitive style on cognitive load of visualized dashboards for construction-related activities},
	volume = {154},
	issn = {0926-5805},
	url = {https://www.sciencedirect.com/science/article/pii/S0926580523002893},
	doi = {10.1016/j.autcon.2023.105029},
	urldate = {2024-11-14},
	journal = {Automation in Construction},
	author = {Ke, Jinjing and Liao, Pinchao and Li, Jie and Luo, Xiaowei},
	month = oct,
	year = {2023},
	pages = {105029},
}

@article{vessey_cognitive_1991,
	title = {Cognitive {Fit}: {A} {Theory}-{Based} {Analysis} of the {Graphs} {Versus} {Tables} {Literature}},
	volume = {22},
	issn = {1540-5915},
	shorttitle = {Cognitive {Fit}},
	url = {https://onlinelibrary.wiley.com/doi/abs/10.1111/j.1540-5915.1991.tb00344.x},
	doi = {10.1111/j.1540-5915.1991.tb00344.x},
	language = {en},
	number = {2},
	urldate = {2024-11-13},
	journal = {Decision Sciences},
	author = {Vessey, Iris},
	year = {1991},
	note = {\_eprint: https://onlinelibrary.wiley.com/doi/pdf/10.1111/j.1540-5915.1991.tb00344.x},
	pages = {219--240},
}

@article{bacic_task-representation_2018,
	title = {Task-{Representation} {Fit}’s {Impact} on {Cognitive} {Effort} in the {Context} of {Decision} {Timeliness} and {Accuracy}: {A} {Cognitive} {Fit} {Perspective}},
	volume = {10},
	issn = {1944-3900},
	shorttitle = {Task-{Representation} {Fit}’s {Impact} on {Cognitive} {Effort} in the {Context} of {Decision} {Timeliness} and {Accuracy}},
	url = {https://aisel.aisnet.org/thci/vol10/iss3/2},
	doi = {10.17705/1thci.00108},
	number = {3},
	journal = {AIS Transactions on Human-Computer Interaction},
	author = {Bacic, Dinko and Henry, Raymond},
	month = sep,
	year = {2018},
	pages = {164--187},
}

@article{smerecnik_understanding_2010,
	title = {Understanding the {Positive} {Effects} of {Graphical} {Risk} {Information} on {Comprehension}: {Measuring} {Attention} {Directed} to {Written}, {Tabular}, and {Graphical} {Risk} {Information}},
	volume = {30},
	copyright = {© 2010 Society for Risk Analysis},
	issn = {1539-6924},
	shorttitle = {Understanding the {Positive} {Effects} of {Graphical} {Risk} {Information} on {Comprehension}},
	url = {https://onlinelibrary.wiley.com/doi/abs/10.1111/j.1539-6924.2010.01435.x},
	doi = {10.1111/j.1539-6924.2010.01435.x},
	language = {en},
	number = {9},
	urldate = {2026-04-16},
	journal = {Risk Analysis},
	author = {Smerecnik, Chris M. R. and Mesters, Ilse and Kessels, Loes T. E. and Ruiter, Robert A. C. and De Vries, Nanne K. and De Vries, Hein},
	year = {2010},
	note = {\_eprint: https://onlinelibrary.wiley.com/doi/pdf/10.1111/j.1539-6924.2010.01435.x},
	pages = {1387--1398},
}

@article{zell_people_2014,
	title = {Do {People} {Have} {Insight} {Into} {Their} {Abilities}? {A} {Metasynthesis}},
	volume = {9},
	issn = {1745-6916},
	shorttitle = {Do {People} {Have} {Insight} {Into} {Their} {Abilities}?},
	url = {https://www.jstor.org/stable/44289970},
	number = {2},
	urldate = {2026-05-07},
	journal = {Perspectives on Psychological Science},
	publisher = {[Association for Psychological Science, Sage Publications, Inc.]},
	author = {Zell, Ethan and Krizan, Zlatan},
	year = {2014},
	pages = {111--125},
}

@article{babaei2025nasatlx,
  author  = {Babaei, Ebrahim and Dingler, Tilman and Tag, Benjamin and Velloso, Eduardo},
  title   = {Should we use the {NASA-TLX} in {HCI}? {A} review of theoretical and methodological issues around Mental Workload Measurement},
  journal = {International Journal of Human-Computer Studies},
  year    = {2025},
  volume  = {201},
  pages   = {103515},
  doi     = {10.1016/j.ijhcs.2025.103515}
}

@article{Magezi2015,
  author  = {Magezi, David A.},
  title   = {Linear mixed-effects models for within-participant psychology experiments: an introductory tutorial and free, graphical user interface ({LMMgui})},
  journal = {Frontiers in Psychology},
  year    = {2015},
  volume  = {6},
  pages   = {2},
  doi     = {10.3389/fpsyg.2015.00002}
}

@InProceedings{Schemmer2023,
  author       = {Schemmer, Max and Kuehl, Niklas and Benz, Carina and Bartos, Andrea and Satzger, Gerhard},
  booktitle    = {Proceedings of the 28th International Conference on Intelligent User Interfaces},
  title        = {Appropriate Reliance on AI Advice: Conceptualization and the Effect of Explanations},
  year         = {2023},
  month        = mar,
  pages        = {410--422},
  publisher    = {ACM},
  series       = {IUI ’23},
  collection   = {IUI ’23},
  creationdate = {2025-12-09T10:15:09},
  doi          = {10.1145/3581641.3584066},
}

@Article{Muir1987,
  author       = {Muir, Bonnie M.},
  journal      = {International Journal of Man-Machine Studies},
  title        = {Trust between humans and machines, and the design of decision aids},
  year         = {1987},
  issn         = {0020-7373},
  month        = nov,
  number       = {5},
  pages        = {527--539},
  volume       = {27},
  creationdate = {2024-08-09T20:01:36},
  doi          = {10.1016/S0020-7373(87)80013-5},
}

@Article{Madhavan2007,
  author       = {Poornima Madhavan and Douglas A. Wiegmann},
  journal      = {Theoretical Issues in Ergonomics Science},
  title        = {Similarities and differences between human–human and human–automation trust: an integrative review},
  year         = {2007},
  issn         = {1463-922X},
  month        = jul,
  number       = {4},
  pages        = {277--301},
  volume       = {8},
  creationdate = {2024-05-19T11:57:40},
  doi          = {10.1080/14639220500337708},
  publisher    = {Taylor \& Francis},
  shorttitle   = {Similarities and differences between human–human and human–automation trust},
  urldate      = {2024-05-19},
}

@misc{vasconcelos_explanations_2023,
	title = {Explanations {Can} {Reduce} {Overreliance} on {AI} {Systems} {During} {Decision}-{Making}},
	url = {http://arxiv.org/abs/2212.06823},
	doi = {10.48550/arXiv.2212.06823},
	language = {en},
	urldate = {2025-08-12},
	publisher = {arXiv},
	author = {Vasconcelos, Helena and Jörke, Matthew and Grunde-McLaughlin, Madeleine and Gerstenberg, Tobias and Bernstein, Michael and Krishna, Ranjay},
	month = jan,
	year = {2023},
	note = {arXiv:2212.06823 [cs]},
}

@inproceedings{zhang_you_2024,
	address = {Karlsruhe Germany},
	title = {You {Can} {Only} {Verify} {When} {You} {Know} the {Answer}: {Feature}-{Based} {Explanations} {Reduce} {Overreliance} on {AI} for {Easy} {Decisions}, but {Not} for {Hard} {Ones}},
	isbn = {979-8-4007-0998-2},
	shorttitle = {You {Can} {Only} {Verify} {When} {You} {Know} the {Answer}},
	url = {https://dl.acm.org/doi/10.1145/3670653.3670660},
	doi = {10.1145/3670653.3670660},
	language = {en},
	urldate = {2026-05-12},
	booktitle = {Proceedings of {Mensch} und {Computer} 2024},
	publisher = {ACM},
	author = {Zhang, Zelun Tony and Buchner, Felicitas and Liu, Yuanting and Butz, Andreas},
	month = sep,
	year = {2024},
	pages = {156--170},
}

@article{wohleber_impact_2016,
	title = {The {Impact} of {Automation} {Reliability} and {Operator} {Fatigue} on {Performance} and {Reliance}},
	volume = {60},
	issn = {1071-1813},
	url = {https://doi.org/10.1177/1541931213601047},
	doi = {10.1177/1541931213601047},
	language = {EN},
	number = {1},
	urldate = {2026-05-12},
	journal = {Proceedings of the Human Factors and Ergonomics Society Annual Meeting},
	publisher = {SAGE Publications Inc},
	author = {Wohleber, Ryan W. and Calhoun, Gloria L. and Funke, Gregory J. and Ruff, Heath and Chiu, C.-Y. Peter and Lin, Jinchao and Matthews, Gerald},
	month = sep,
	year = {2016},
	pages = {211--215},
}

@inproceedings{wright_effect_2016,
	address = {Cham},
	title = {The {Effect} of {Agent} {Reasoning} {Transparency} on {Automation} {Bias}: {An} {Analysis} of {Response} {Performance}},
	isbn = {978-3-319-39907-2},
	shorttitle = {The {Effect} of {Agent} {Reasoning} {Transparency} on {Automation} {Bias}},
	doi = {10.1007/978-3-319-39907-2_45},
	language = {en},
	booktitle = {Virtual, {Augmented} and {Mixed} {Reality}},
	publisher = {Springer International Publishing},
	author = {Wright, Julia L. and Chen, Jessie Y. C. and Barnes, Michael J. and Hancock, P. A.},
	editor = {Lackey, Stephanie and Shumaker, Randall},
	year = {2016},
	pages = {465--477},
}

@article{speier_influence_2006,
	title = {The influence of information presentation formats on complex task decision-making performance},
	volume = {64},
	issn = {1071-5819},
	url = {https://www.sciencedirect.com/science/article/pii/S1071581906001042},
	doi = {10.1016/j.ijhcs.2006.06.007},
	number = {11},
	urldate = {2024-11-14},
	journal = {International Journal of Human-Computer Studies},
	author = {Speier, Cheri},
	month = nov,
	year = {2006},
	pages = {1115--1131},
}

@article{chen_effects_2009,
	title = {The effects of information overload on consumers’ subjective state towards buying decision in the internet shopping environment},
	volume = {8},
	issn = {1567-4223},
	url = {https://www.sciencedirect.com/science/article/pii/S1567422308000367},
	doi = {10.1016/j.elerap.2008.09.001},
	number = {1},
	urldate = {2026-04-23},
	journal = {Electronic Commerce Research and Applications},
	author = {Chen, Yu-Chen and Shang, Rong-An and Kao, Chen-Yu},
	month = jan,
	year = {2009},
	pages = {48--58},
}

@article{lu_does_2024,
	title = {Does {More} {Advice} {Help}? {The} {Effects} of {Second} {Opinions} in {AI}-{Assisted} {Decision} {Making}},
	volume = {8},
	issn = {2573-0142},
	shorttitle = {Does {More} {Advice} {Help}?},
	url = {https://dl.acm.org/doi/10.1145/3653708},
	doi = {10.1145/3653708},
	language = {en},
	number = {CSCW1},
	urldate = {2026-05-14},
	journal = {Proceedings of the ACM on Human-Computer Interaction},
	author = {Lu, Zhuoran and Wang, Dakuo and Yin, Ming},
	month = apr,
	year = {2024},
	pages = {1--31},
}

@article{dietvorst_overcoming_2018,
	title = {Overcoming {Algorithm} {Aversion}: {People} {Will} {Use} {Imperfect} {Algorithms} {If} {They} {Can} ({Even} {Slightly}) {Modify} {Them}},
	volume = {64},
	issn = {0025-1909},
	shorttitle = {Overcoming {Algorithm} {Aversion}},
	url = {https://pubsonline.informs.org/doi/10.1287/mnsc.2016.2643},
	doi = {10.1287/mnsc.2016.2643},
	number = {3},
	urldate = {2026-05-14},
	journal = {Management Science},
	publisher = {INFORMS},
	author = {Dietvorst, Berkeley J. and Simmons, Joseph P. and Massey, Cade},
	month = mar,
	year = {2018},
	pages = {1155--1170},
}

@article{spatola_efficiency-accountability_2024,
	title = {The efficiency-accountability tradeoff in {AI} integration: {Effects} on human performance and over-reliance},
	volume = {2},
	issn = {2949-8821},
	shorttitle = {The efficiency-accountability tradeoff in {AI} integration},
	url = {https://www.sciencedirect.com/science/article/pii/S2949882124000598},
	doi = {10.1016/j.chbah.2024.100099},
	number = {2},
	urldate = {2026-04-17},
	journal = {Computers in Human Behavior: Artificial Humans},
	author = {Spatola, Nicolas},
	month = aug,
	year = {2024},
	pages = {100099},
}

@article{boy_principled_2014,
	title = {A {Principled} {Way} of {Assessing} {Visualization} {Literacy}},
	volume = {20},
	copyright = {https://ieeexplore.ieee.org/Xplorehelp/downloads/license-information/IEEE.html},
	issn = {1077-2626},
	url = {http://ieeexplore.ieee.org/document/6875906/},
	doi = {10.1109/tvcg.2014.2346984},
	language = {en},
	number = {12},
	urldate = {2025-07-24},
	journal = {IEEE Transactions on Visualization and Computer Graphics},
	publisher = {Institute of Electrical and Electronics Engineers (IEEE)},
	author = {Boy, Jeremy and Rensink, Ronald A. and Bertini, Enrico and Fekete, Jean-Daniel},
	month = dec,
	year = {2014},
	pages = {1963--1972},
}

@article{lee_vlat_2017,
	title = {{VLAT}: {Development} of a {Visualization} {Literacy} {Assessment} {Test}},
	volume = {23},
	issn = {1941-0506},
	shorttitle = {{VLAT}},
	url = {https://ieeexplore.ieee.org/document/7539634},
	doi = {10.1109/TVCG.2016.2598920},
	number = {1},
	urldate = {2026-05-14},
	journal = {IEEE Transactions on Visualization and Computer Graphics},
	author = {Lee, Sukwon and Kim, Sung-Hee and Kwon, Bum Chul},
	month = jan,
	year = {2017},
	pages = {551--560},
}

@inproceedings{hettiachchi_hi_2020,
	address = {Honolulu HI USA},
	title = {"{Hi}! {I} am the {Crowd} {Tasker}" {Crowdsourcing} through {Digital} {Voice} {Assistants}},
	isbn = {978-1-4503-6708-0},
	url = {https://dl.acm.org/doi/10.1145/3313831.3376320},
	doi = {10.1145/3313831.3376320},
	language = {en},
	urldate = {2026-05-14},
	booktitle = {Proceedings of the 2020 {CHI} {Conference} on {Human} {Factors} in {Computing} {Systems}},
	publisher = {ACM},
	author = {Hettiachchi, Danula and Sarsenbayeva, Zhanna and Allison, Fraser and Van Berkel, Niels and Dingler, Tilman and Marini, Gabriele and Kostakos, Vassilis and Goncalves, Jorge},
	month = apr,
	year = {2020},
	pages = {1--14},
}

@article{Wieditz2024,
  author  = {Wieditz, Johannes and Miller, Clemens and Scholand, Jan and Nemeth, Marcus},
  title   = {A Brief Introduction on Latent Variable Based Ordinal Regression Models With an Application to Survey Data},
  journal = {Statistics in Medicine},
  year    = {2024},
  volume  = {43},
  number  = {29},
  pages   = {5618--5634},
  doi     = {10.1002/sim.10208}
}

@article{Hedeker2015,
  author  = {Hedeker, Donald},
  title   = {Methods for Multilevel Ordinal Data in Prevention Research},
  journal = {Prevention Science},
  year    = {2015},
  volume  = {16},
  number  = {7},
  pages   = {997--1006},
  doi     = {10.1007/s11121-014-0495-x}
}

@article{Lenth2016,
  author  = {Lenth, Russell V.},
  title   = {Least-Squares Means: The {R} Package lsmeans},
  journal = {Journal of Statistical Software},
  year    = {2016},
  volume  = {69},
  number  = {1},
  pages   = {1--33},
  doi     = {10.18637/jss.v069.i01}
}

@article{Naggara2011,
  author  = {Naggara, Olivier and Raymond, Jean and Guilbert, Fran{\c{c}}ois and Roy, Daniel and Weill, Alain and Altman, Douglas G.},
  title   = {Analysis by Categorizing or Dichotomizing Continuous Variables Is Inadvisable: An Example from the Natural History of Unruptured Aneurysms},
  journal = {AJNR. American Journal of Neuroradiology},
  year    = {2011},
  volume  = {32},
  number  = {3},
  pages   = {437--440},
  doi     = {10.3174/ajnr.A2425}
}

@article{Chen2020LRT,
  author  = {Chen, Yunxiao and Moustaki, Irini and Zhang, Haoran},
  title   = {A Note on Likelihood Ratio Tests for Models with Latent Variables},
  journal = {Psychometrika},
  year    = {2020},
  volume  = {85},
  pages   = {996--1012},
  doi     = {10.1007/s11336-020-09735-0}
}

@article{Mize2019,
  author  = {Mize, Trenton D.},
  title   = {Best Practices for Estimating, Interpreting, and Presenting Nonlinear Interaction Effects},
  journal = {Sociological Science},
  year    = {2019},
  volume  = {6},
  pages   = {81--117},
  doi     = {10.15195/v6.a4}
}

@article{Aickin1996,
  author  = {Aickin, Mikel and Gensler, Helen},
  title   = {Adjusting for multiple testing when reporting research results: the Bonferroni vs Holm methods},
  journal = {American Journal of Public Health},
  year    = {1996},
  volume  = {86},
  number  = {5},
  pages   = {726--728},
  doi     = {10.2105/AJPH.86.5.726}
}

@article{BeltranTarwater2024,
  author  = {Beltran, Roxanne S. and Tarwater, Corey E.},
  title   = {Overcoming the pitfalls of categorizing continuous variables in ecology, evolution and behaviour},
  journal = {Proceedings of the Royal Society B: Biological Sciences},
  year    = {2024},
  volume  = {291},
  number  = {2032},
  pages   = {20241640},
  doi     = {10.1098/rspb.2024.1640}
}

@article{PapkeWooldridge1996,
  author  = {Papke, Leslie E. and Wooldridge, Jeffrey M.},
  title   = {Econometric Methods for Fractional Response Variables with an Application to 401(k) Plan Participation Rates},
  journal = {Journal of Applied Econometrics},
  year    = {1996},
  volume  = {11},
  number  = {6},
  pages   = {619--632},
  doi     = {10.1002/(SICI)1099-1255(199611)11:6<619::AID-JAE418>3.0.CO;2-1}
}

@article{LiangZeger1986,
  author  = {Liang, Kung-Yee and Zeger, Scott L.},
  title   = {Longitudinal Data Analysis Using Generalized Linear Models},
  journal = {Biometrika},
  year    = {1986},
  volume  = {73},
  number  = {1},
  pages   = {13--22},
  doi     = {10.1093/biomet/73.1.13}
}

@article{Austin2024,
  author  = {Austin, Peter C. and Kapral, Moira K. and Vyas, Manav V. and Fang, Jim and Yu, Amy Y. X.},
  title   = {Using Multilevel Models and Generalized Estimating Equation Models to Account for Clustering in Neurology Clinical Research},
  journal = {Neurology},
  year    = {2024},
  volume  = {103},
  number  = {9},
  pages   = {e209947},
  doi     = {10.1212/WNL.0000000000209947}
}

@article{Huang2018,
  author  = {Huang, Francis L.},
  title   = {Using Cluster Bootstrapping to Analyze Nested Data With a Few Clusters},
  journal = {Educational and Psychological Measurement},
  year    = {2018},
  volume  = {78},
  number  = {2},
  pages   = {297--318},
  doi     = {10.1177/0013164416678980}
}

@article{NgCribbie2017,
  author  = {Ng, Victoria K. Y. and Cribbie, Robert A.},
  title   = {Using the Gamma Generalized Linear Model for Modeling Continuous, Skewed and Heteroscedastic Outcomes in Psychology},
  journal = {Current Psychology},
  year    = {2017},
  volume  = {36},
  number  = {2},
  pages   = {225--235},
  doi     = {10.1007/s12144-015-9404-0}
}

@article{Malehi2015,
  author  = {Malehi, Amal Saki and Pourmotahari, Fatemeh and Angali, Kambiz Ahmadi},
  title   = {Statistical models for the analysis of skewed healthcare cost data: a simulation study},
  journal = {Health Economics Review},
  year    = {2015},
  volume  = {5},
  pages   = {11},
  doi     = {10.1186/s13561-015-0045-7}
}

@article{Gambarota2024,
  author  = {Gambarota, Filippo and Alto{\`e}, Gianmarco},
  title   = {Ordinal regression models made easy: A tutorial on parameter interpretation, data simulation and power analysis},
  journal = {International Journal of Psychology},
  year    = {2024},
  volume  = {59},
  number  = {6},
  pages   = {1263--1292},
  doi     = {10.1002/ijop.13243}
}

@article{larkin_why_1987,
	title = {Why a {Diagram} is ({Sometimes}) {Worth} {Ten} {Thousand} {Words}},
	volume = {11},
	issn = {1551-6709},
	url = {https://onlinelibrary.wiley.com/doi/abs/10.1111/j.1551-6708.1987.tb00863.x},
	doi = {10.1111/j.1551-6708.1987.tb00863.x},
	language = {en},
	number = {1},
	urldate = {2026-05-20},
	journal = {Cognitive Science},
	author = {Larkin, Jill H. and Simon, Herbert A.},
	year = {1987},
	note = {\_eprint: https://onlinelibrary.wiley.com/doi/pdf/10.1111/j.1551-6708.1987.tb00863.x},
	pages = {65--100},
}

@article{hertzum_reference_2021,
	title = {Reference values and subscale patterns for the task load index ({TLX}): a meta-analytic review},
	volume = {64},
	issn = {0014-0139, 1366-5847},
	shorttitle = {Reference values and subscale patterns for the task load index ({TLX})},
	url = {https://www.tandfonline.com/doi/full/10.1080/00140139.2021.1876927},
	doi = {10.1080/00140139.2021.1876927},
	language = {en},
	number = {7},
	urldate = {2026-03-17},
	journal = {Ergonomics},
	author = {Hertzum, Morten},
	month = jul,
	year = {2021},
	pages = {869--878},
}

@article{lee_trust_1994,
	title = {Trust, self-confidence, and operators' adaptation to automation},
	volume = {40},
	issn = {1071-5819},
	url = {https://www.sciencedirect.com/science/article/pii/S107158198471007X},
	doi = {10.1006/ijhc.1994.1007},
	number = {1},
	urldate = {2026-05-12},
	journal = {International Journal of Human-Computer Studies},
	author = {Lee, John D. and Moray, Neville},
	month = jan,
	year = {1994},
	pages = {153--184},
}

@article{lee_trust_2004,
	title = {Trust in {Automation}: {Designing} for {Appropriate} {Reliance}},
	volume = {46},
	issn = {0018-7208},
	shorttitle = {Trust in {Automation}},
	url = {https://journals.sagepub.com/action/showAbstract},
	doi = {10.1518/hfes.46.1.50_30392},
	language = {EN},
	number = {1},
	urldate = {2026-04-17},
	journal = {Human Factors},
	publisher = {SAGE Publications Inc},
	author = {Lee, John D. and See, Katrina A.},
	month = mar,
	year = {2004},
	pages = {50--80},
}

\appendix
\counterwithin{table}{section}
\counterwithin{figure}{section}
\section{Task prompts, complexity and reference solutions}
\label{app:task}

\subsection{Task Prompts} \label{app:task:prompts}
\paragraph{Task 1 (low complexity)}
\begin{quote}\small
On which day in May did the factory consume the most?
\end{quote}

\paragraph{Task 2 (medium complexity)}
\begin{quote}\small
From September to October, your factory has noticed excessive energy waste from machines sitting idle unnecessarily for too long. Management is considering installing an automatic power-down feature on one cutting machine to reduce energy costs.
During a recent team meeting, it was decided that we should compare how long different machines sit idle and how much energy they draw while idle.
Based on these observations, on which cutting machine would this feature have the greatest impact? To give the process crew some room for manoeuvre when planning operations, please also indicate your second-best choice.
\end{quote}

\paragraph{Task 3 (high complexity)}

\begin{quote}\small
Over the next 3 working days, we can deploy three specialist crews for one half-day each. Using data since the start of October, consider the period as a whole and pick three different machines, one for each crew to work on:
\begin{itemize}
\item Maintenance crew: choose the machine where reducing stops and downtime would most improve overall output and stability. Check: Availability, Alarm time and number of Alarms, Idle time, Offline time, Utilization rate.
\item Process optimization crew: choose the machine where improving efficiency and flow would produce the biggest gain. Check: Performance, Average cycle time, Good cycles, Utilization rate, Consumption while working and Cost.
\item Quality crew: choose the machine where improving first-pass yield would save the most good parts. Check: Quality, Bad cycles, Average cycle cost, Good cycles.
\end{itemize}
Assume safety is currently under control and comparable across machines.
\textbf{Constraint:} to avoid disrupting operations and over-concentrating work on a single line, you must not select more than one machine from the same family (e.g., you can’t pick two Medium Cutting machines, or two Assembly machines).
\end{quote}

\subsection{Objective complexity} \label{app:task:complexity}
We operationalized objective task complexity using Wood’s task analytic model, which decomposes complexity into \emph{component} complexity (acts and information cues), \emph{coordinative} complexity (dependencies among acts), and \emph{dynamic} complexity (changes in cue--product relationships over time) \citep{wood_task_1986}. 

Component complexity ($TC_1$) reflects (a) the number of distinct acts required and (b) the number of \emph{non-redundant} information cues, with repeated use of the same cue (or cues implied by others) counted once.
Therefore for $TC_1$, we summed the number of distinct cues required within each act across the task. 

Coordinative complexity ($TC_2$) was operationalized as the number of required cross-act dependencies (e.g., integrating cues, applying conditional constraints, or comparing alternatives before selecting an answer) and was computed by counting directed dependencies between acts (e.g., one link per required dependency between acts $i\rightarrow j$).

Dynamic complexity was designed to be negligible by freezing the scenario time window and keeping the dataset static during each task. Accordingly, we set $TC_3=0$ for all tasks.

Total complexity was computed as
\[
TC=\alpha TC_1+\beta TC_2.
\]
with $\alpha=1$ and $\beta=2$ as suggested by \citet{wood_task_1986}, assigning greater weight to coordinative than component complexity.

Table~\ref{tab:task_complexity} reports the resulting complexity values for each task, computed from the finalized task scripts and interface instrumentation logs.

\begin{table}[!t]
\centering
\caption{Objective task complexity computed.}
\label{tab:task_complexity}
\small
\renewcommand{\arraystretch}{1.05}
\begin{tabular*}{\columnwidth}{@{\extracolsep{\fill}} l r r r r r @{}}
\toprule
Task & Cues & Acts & $TC_1$ & $TC_2$ & $TC$ \\
\midrule
Task 1 & 3  & 1 & 3  & 0 & 3  \\
Task 2 & 4  & 2 & 7  & 1 & 9  \\
Task 3 & 16 & 5 & 19 & 6 & 31 \\
\bottomrule
\end{tabular*}
\end{table}

\subsection{Tasks times and caps}\label{app:task:times}

Table~\ref{tab:task_timecaps} reports the per-task timing parameters implemented in the study web application well over the measured \emph{pilot median} times. Therefore, for each task the system enforced a nominal \emph{time-cap}. A single warning message was shown when the remaining time reached the \emph{pre-warning} threshold (without showing a countdown). If the budget elapsed before submission, the task entered a \emph{grace} period during which participants could still submit. Upon grace expiry, the study advanced automatically and the task was recorded as timed out.

\begin{table}[t]
\centering
\small
\setlength{\tabcolsep}{3.5pt}
\renewcommand{\arraystretch}{1.05}
\caption{Pilot completion times and time-cap parameters by task.}
\label{tab:task_timecaps}

\begin{threeparttable}
\begin{tabularx}{\columnwidth}{@{}l *{3}{>{\raggedleft\arraybackslash}X}@{}}
\toprule
Task & Pilot median (s) & Time-cap (s) & Pre-warning (s)\tnote{a} \\
\midrule
Task 1 & 91.36  & 180 & 60 \\
Task 2 & 183.95 & 360 & 90  \\
Task 3 & 357.35 & 600 & 120  \\
\bottomrule
\end{tabularx}

\begin{tablenotes}[flushleft]\footnotesize
\item[a] Values indicate the seconds \emph{before} the time-cap at which the warning is shown.
\end{tablenotes}
\end{threeparttable}
\end{table}

\subsection{Task decision reference solutions} \label{app:task:decision}

\subparagraph{Task 1.}
For each day $d$ in May, we computed daily consumption $E(d)=\sum_{t\in d} c_t$.
The task’s \emph{reference solutions} are the full ordering of days obtained by sorting
$E(d)$ in descending order where the top-ranked day is $d^*=\arg\max_d E(d)$.

\subparagraph{Task 2.}
Let $\mathcal{M}$ be the set of eligible cutting machines in the September--October window. For each machine $m\in\mathcal{M}$ we computed
\begin{align*}
T_m=\text{total idle time (hours)},  \\
\qquad E_m=\text{total idle energy (kWh)}.
\end{align*}
We then min--max normalized idle time and idle energy across machines:
\begin{align*}
\tilde{T}_m&=\frac{T_m-\min_{k\in\mathcal{M}}T_k}{\max_{k\in\mathcal{M}}T_k-\min_{k\in\mathcal{M}}T_k},\\\\
\tilde{E}_m&=\frac{E_m-\min_{k\in\mathcal{M}}E_k}{\max_{k\in\mathcal{M}}E_k-\min_{k\in\mathcal{M}}E_k}.
\end{align*}
and defined the single-machine impact score as
\[
I_m=\frac{\tilde{T}_m+\tilde{E}_m}{2}.
\]
The task’s \emph{reference solutions} correspond to the ordering of \emph{ordered pairs} $(m_1,m_2)$ (first and second choice, with $m_1\neq m_2$) by
\[
S(m_1,m_2)=0.6\cdot\frac{I_{m_1}}{\max_{k\in\mathcal{M}} I_k}\;+\;0.4\cdot\frac{I_{m_2}}{\max_{k\in\mathcal{M}} I_k}.
\]
Pairs were ranked by descending $S(m_1,m_2)$; the top-ranked pair is the reference best+second-best response.

\subparagraph{Task 3.}
Let $\mathcal{M}$ be the set of machines in the October--November analysis window, and let $f(m)$ denote machine $m$'s family. We computed three crew-specific machine scores from window aggregates, using family-wise POMP scaling to map each indicator to $[0,1]$ (higher = stronger candidate within family):
\[
\mathrm{POMP}(x_m)=\frac{x_m-\min_{k:f(k)=f(m)}x_k}{\max_{k:f(k)=f(m)}x_k-\min_{k:f(k)=f(m)}x_k}.
\]

\emph{Maintenance.} Using availability loss $1-\mathrm{Avail}_m$,
downtime share $\mathrm{Down}_m$, alarm rate $\mathrm{AlarmRate}_m$,
and utilisation rate $\mathrm{UR}_m$:
\begin{equation*}
\begin{aligned}
\mathrm{Maint}(m)
={}& \Big(0.40\,\operatorname{POMP}(1-\mathrm{Avail}_m) \\
 &\quad + 0.40\,\operatorname{POMP}(\mathrm{Down}_m) \\
 &\quad + 0.15\,\operatorname{POMP}(\mathrm{AlarmRate}_m) \\
 &\quad + 0.05\,\operatorname{POMP}(\mathrm{UR}_m)\Big)^2 .
\end{aligned}
\end{equation*}

\emph{Process.} Using performance loss $1-\mathrm{Perf}_m$,
energy per good part $\mathrm{EnergyGood}_m$, cost per good part
$\mathrm{CostGood}_m$, and utilisation rate $\mathrm{UR}_m$:
\begin{equation*}
\begin{aligned}
\mathrm{Proc}(m)
={}& \Big(0.35\,\operatorname{POMP}(1-\mathrm{Perf}_m) \\
 &\quad + 0.35\,\operatorname{POMP}(\mathrm{EnergyGood}_m) \\
 &\quad + 0.20\,\operatorname{POMP}(\mathrm{CostGood}_m) \\
 &\quad + 0.10\,\operatorname{POMP}(\mathrm{UR}_m)\Big)^2 .
\end{aligned}
\end{equation*}

\emph{Quality.} Using quality loss $1-\mathrm{Qual}_m$, scrap rate
$\mathrm{Scrap}_m$, cost per cycle $\mathrm{CostCycle}_m$, and volume
leverage $\mathrm{Good}_m$:
\begin{equation*}
\begin{aligned}
\mathrm{Qual}(m)
={}& \Big(0.30\,\operatorname{POMP}(1-\mathrm{Qual}_m) \\
 &\quad + 0.40\,\operatorname{POMP}(\mathrm{Scrap}_m) \\
 &\quad + 0.20\,\operatorname{POMP}(\mathrm{CostCycle}_m) \\
 &\quad + 0.10\,\operatorname{POMP}(\mathrm{Good}_m)\Big)^2 .
\end{aligned}
\end{equation*}

Machines were ranked globally for each crew by descending score (ranks $r^{M}_m$, $r^{P}_m$, $r^{Q}_m$), then mapped to utilities in $[0,1]$:
\[
u(r)=1-\frac{r-1}{N-1}, \qquad N=|\mathcal{M}|.
\]

The task’s \emph{reference solutions} correspond to the ordering of all admissible triples $(m,p,q)$ (maintenance, process, quality), subject to $m,p,q$ being distinct machines and all from distinct families, by:
\begin{multline*}
\mathrm{Score}(m,p,q)=\lambda\cdot\frac{u(r^M_m)+u(r^P_p)+u(r^Q_q)}{3}\\
+(1-\lambda)\cdot\min\{u(r^M_m),u(r^P_p),u(r^Q_q)\},
\end{multline*}
with $\lambda=0.75$. Triples were ranked by descending $\mathrm{Score}(m,p,q)$ and the top-ranked triple is the reference maintenance+process+quality assignment.

\section{Robustness Analyses}
\label{app:robustness}

This appendix reports robustness analyses for the tested hypotheses. The aim was to assess whether the conclusions reported in the Results section depended on the primary model families used there. Overall, the robustness checks led to the same substantive interpretation as the main analyses: the MWL, task complexity moderation, and task complexity main-effect hypotheses were supported, whereas the overall accuracy and completion time interface effects, the DL moderation hypotheses, and H5 were not robustly supported.

\subsection{Mental Workload}
\label{app:subsec.robustness-mwl}

The MWL robustness checks used participant-level non-parametric tests and ordinal-model alternatives to assess sensitivity to the primary CLMM analysis. The results were consistent with the main conclusions. Participant-level NASA-TLX ordinal-class sums were lower in the chatbot condition than in the dashboard condition, supporting H1a. The chatbot workload advantage was larger at T1 than at T3, and T3-minus-T1 workload increases were larger in the chatbot condition, supporting the attenuation pattern predicted in H2a. The robustness checks for H3a did not provide reliable support for the prespecified data-literacy moderation contrast. Finally, Friedman and paired Wilcoxon tests confirmed the expected monotonic increase in workload from easy to medium to hard tasks, supporting H4a. Table~\ref{tab:robustness-mwl-a} reports the detailed results.

\begin{table*}[!t]
\centering
\scriptsize
\setlength{\tabcolsep}{4pt}
\renewcommand{\arraystretch}{1.06}
\caption{Robustness checks for Mental Workload hypotheses.}
\label{tab:robustness-mwl-a}

\begin{threeparttable}
\begin{tabularx}{\linewidth}{@{}
>{\bfseries}p{0.055\linewidth}
>{\raggedright\arraybackslash}p{0.18\linewidth}
>{\raggedright\arraybackslash}p{0.24\linewidth}
>{\centering\arraybackslash}p{0.105\linewidth}
>{\raggedright\arraybackslash}X
@{}}
\toprule
\textbf{Hyp.} &
\textbf{Contrast} &
\textbf{Robustness check} &
\textbf{$p$-value} &
\textbf{Effect summary} \\
\midrule

\multicolumn{5}{@{}l}{\textit{Interface main effect}} \\
H1a &
CUI vs. GUI &
Mann--Whitney U and Brunner--Munzel tests on participant-level TLX ordinal-class sums &
$p < .001^{***}$ &
Means: CUI $= 18.27$, GUI $= 26.79$; $U = 1128.00$; BM $= 5.86$; Cliff's $\delta = -0.496$, 95\% CI [$-0.654$, $-0.333$]; HL shift $= -9.00$, 95\% CI [$-12.00$, $-6.00$] \\

\addlinespace[0.35em]
\multicolumn{5}{@{}l}{\textit{Moderation by task complexity}} \\
H2a &
CUI vs. GUI at T1 &
Brunner--Munzel simple interface effect &
$p < .001^{***}$ &
Median: CUI $= 3.00$, GUI $= 7.00$; mean: CUI $= 3.73$, GUI $= 7.73$; BM $= 7.17$; Cliff's $\delta = -0.556$, 95\% CI [$-0.699$, $-0.404$] \\

H2a &
CUI vs. GUI at T3 &
Brunner--Munzel simple interface effect &
$p = .006^{**}$ &
Median: CUI $= 9.00$, GUI $= 11.50$; mean: CUI $= 8.98$, GUI $= 11.20$; BM $= 2.80$; Cliff's $\delta = -0.274$, 95\% CI [$-0.464$, $-0.080$] \\

H2a &
T3--T1 change: CUI vs. GUI &
Participant-level ordinal change and permutation check &
$p = .032^{*}$ &
Mean change: CUI $= 5.25$, GUI $= 3.47$; difference in change $= 1.78$, 95\% CI [0.27, 3.30]; BM $= -2.17$; permutation $p = .028$; Cliff's $\delta = 0.211$, 95\% CI [0.016, 0.396] \\

\addlinespace[0.35em]
\multicolumn{5}{@{}l}{\textit{Moderation by data literacy}} \\
H3a &
Interface $\times$ data literacy &
Ordered-logit robustness model with task as covariate &
$p = .036^{*}$ &
Omnibus interaction: LR $= 6.67$, df $= 2$; the model indicates some non-parallelism across data-literacy groups \\

H3a &
GUI vs. CUI, high vs. low DL &
Prespecified DiD  &
$p = .371$ &
DID on expected ordinal class $= -1.31$, 95\% CI [$-5.61$, 1.64]; ratio-of-ORs $= 0.575$, 95\% CI [0.065, 2.520] \\

\addlinespace[0.35em]
\multicolumn{5}{@{}l}{\textit{Task complexity main effect}} \\
H4a &
Task complexity &
Friedman omnibus test &
$p < .001^{***}$ &
$\chi^2 = 100.13$, df $= 2$; Kendall's $W = 0.374$; means increased monotonically: T1 $= 24.19$, T2 $= 28.73$, T3 $= 45.84$ \\

H4a &
T2 vs. T1 &
One-sided paired Wilcoxon signed-rank test &
Holm $p < .001^{***}$ &
Mean difference $= 4.55$, 95\% CI [1.35, 7.61]; rank-biserial $r = 0.382$, 95\% CI [0.184, 0.563] \\

H4a &
T3 vs. T1 &
One-sided paired Wilcoxon signed-rank test &
Holm $p < .001^{***}$ &
Mean difference $= 21.65$, 95\% CI [17.56, 25.59]; rank-biserial $r = 0.813$, 95\% CI [0.705, 0.900] \\

H4a &
T3 vs. T2 &
One-sided paired Wilcoxon signed-rank test &
Holm $p < .001^{***}$ &
Mean difference $= 17.11$, 95\% CI [14.11, 20.26]; rank-biserial $r = 0.872$, 95\% CI [0.780, 0.945] \\

\bottomrule
\end{tabularx}

\begin{tablenotes}[flushleft]
\scriptsize
\item \textit{Note.} CUI = chatbot; GUI = dashboard; DL = data literacy; T1 = first task (low complexity); T2 = medium; T3 = third task (high complexity); BM = Brunner--Munzel statistic; HL = Hodges--Lehmann shift. H1a and H2a use TLX ordinal-class summaries. H3a uses an ordered-logit model with task included as a covariate. H4a uses complete within-participant task triples ($N = 134$). For Cliff's $\delta$, negative values indicate lower workload in the chatbot condition. For the H2a change-score contrast, positive values indicate a larger T3-minus-T1 increase in the chatbot condition. Asterisks denote $^{*}p < .05$, $^{**}p < .01$, and $^{***}p < .001$.
\end{tablenotes}
\end{threeparttable}
\end{table*}

\subsection{Decision Accuracy}
\label{app:subsec.robustness-accuracy}

The decision accuracy robustness checks used participant-level and non-parametric alternatives to the primary fractional-logit analyses. The results were again consistent with the main conclusions. Participant-level mean accuracy was descriptively higher in the chatbot condition, but the uncertainty intervals included zero; H1b therefore remained unsupported. For H2b, the chatbot showed a clearer accuracy advantage at T1 than at T3, and the T1-to-T3 accuracy drop was larger in the chatbot condition, consistent with attenuation of the chatbot advantage as task complexity increased. The robustness checks did not support the prespecified data-literacy moderation contrast for H3b. For H4b, the Friedman test confirmed an overall task complexity effect, with accuracy declining most clearly in the hard task. Table~\ref{tab:robustness-accuracy-b} reports the detailed results.

\begin{table*}[!t]
\centering
\scriptsize
\setlength{\tabcolsep}{4pt}
\renewcommand{\arraystretch}{1.06}
\caption{Robustness checks for Decision Accuracy hypotheses.}
\label{tab:robustness-accuracy-b}

\begin{threeparttable}
\begin{tabularx}{\linewidth}{@{}
>{\bfseries}p{0.055\linewidth}
>{\raggedright\arraybackslash}p{0.18\linewidth}
>{\raggedright\arraybackslash}p{0.24\linewidth}
>{\centering\arraybackslash}p{0.105\linewidth}
>{\raggedright\arraybackslash}X
@{}}
\toprule
\textbf{Hyp.} &
\textbf{Contrast} &
\textbf{Robustness check} &
\textbf{$p$-value} &
\textbf{Effect summary} \\
\midrule

\multicolumn{5}{@{}l}{\textit{Interface main effect}} \\
H1b &
CUI vs. GUI &
Welch test and permutation test on participant-level mean accuracy &
$p = .144$ &
Means: CUI $= 0.759$, GUI $= 0.729$; mean difference $= 0.030$, 95\% Welch CI [$-0.010$, 0.070]; permutation $p = .148$; bootstrap mean difference $= 0.030$, 95\% CI [$-0.009$, 0.071]; Hedges' $g = 0.252$, 95\% CI [$-0.081$, 0.593] \\

\addlinespace[0.35em]
\multicolumn{5}{@{}l}{\textit{Moderation by task complexity}} \\
H2b &
CUI vs. GUI at T1 &
Simple interface effect with permutation test &
$p = .021^{*}$ &
Mean accuracy: CUI $= 0.931$, GUI $= 0.857$; difference $= 0.074$, 95\% CI [0.013, 0.135]; Cohen's $h = 0.245$, 95\% CI [0.039, 0.450]; Hedges' $g = 0.408$, 95\% CI [0.070, 0.780] \\

H2b &
CUI vs. GUI at T3 &
Simple interface effect with permutation test &
$p = .360$ &
Mean accuracy: CUI $= 0.471$, GUI $= 0.502$; difference $= -0.031$, 95\% CI [$-0.094$, 0.035]; Cohen's $h = -0.062$, 95\% CI [$-0.184$, 0.063]; Hedges' $g = -0.164$, 95\% CI [$-0.517$, 0.179] \\

H2b &
T1--T3 accuracy drop: CUI vs. GUI &
Participant-level drop and permutation check &
$p = .018^{*}$ &
Mean drop: CUI $= 0.460$, GUI $= 0.355$; drop difference $= 0.105$, 95\% CI [0.019, 0.186]; equivalent DID $= -0.105$, 95\% CI [$-0.189$, $-0.022$]; Hedges' $g = 0.422$, 95\% CI [0.094, 0.768] \\

\addlinespace[0.35em]
\multicolumn{5}{@{}l}{\textit{Moderation by data literacy}} \\
H3b &
Interface $\times$ data literacy &
Fractional-logit GLM with participant cluster-robust standard errors &
$p = .049^{*}$ &
Omnibus interaction: Wald $\chi^2 = 6.039$, df $= 2$; the model suggests some overall non-parallelism across data-literacy groups \\

H3b &
GUI vs. CUI, high vs. low DL &
Prespecified DiD  &
$p = .109$ &
DID on predicted accuracy $= 0.087$, 95\% CI [$-0.016$, 0.196]; ratio-of-ORs $= 1.769$, 95\% CI [0.880, 3.555] \\

\addlinespace[0.35em]
\multicolumn{5}{@{}l}{\textit{Task complexity main effect}} \\
H4b &
Task complexity &
Friedman omnibus test &
$p < .001^{***}$ &
$\chi^2 = 148.190$, df $= 2$; Kendall's $W = 0.566$; means declined across complexity: T1 $= 0.892$, T2 $= 0.851$, T3 $= 0.487$ \\

H4b &
T2 vs. T1 &
One-sided paired Wilcoxon signed-rank test &
Holm $p = .097$ &
Mean difference $= -0.041$, 95\% CI [$-0.086$, 0.003]; rank-biserial $r = -0.141$, 95\% CI [$-0.345$, 0.078]; Cohen's $h = -0.123$, 95\% CI [$-0.258$, 0.010] \\

H4b &
T3 vs. T1 &
One-sided paired Wilcoxon signed-rank test &
Holm $p < .001^{***}$ &
Mean difference $= -0.405$, 95\% CI [$-0.446$, $-0.361$]; rank-biserial $r = -0.948$, 95\% CI [$-0.993$, $-0.875$]; Cohen's $h = -0.928$, 95\% CI [$-1.041$, $-0.811$] \\

H4b &
T3 vs. T2 &
One-sided paired Wilcoxon signed-rank test &
Holm $p < .001^{***}$ &
Mean difference $= -0.364$, 95\% CI [$-0.412$, $-0.315$]; rank-biserial $r = -0.888$, 95\% CI [$-0.966$, $-0.782$]; Cohen's $h = -0.804$, 95\% CI [$-0.927$, $-0.684$] \\

\bottomrule
\end{tabularx}

\begin{tablenotes}[flushleft]
\scriptsize
\item \textit{Note.} CUI = chatbot; GUI = dashboard; DL = data literacy; T1 = first task (low complexity); T2 = medium; T3 = third task (high complexity). H1b uses participant-level mean accuracy across the three tasks. H2b uses complete participant triples and compares the interface gap at low and high complexity. H3b uses a fractional-logit robustness model with participant cluster-robust standard errors. H4b uses complete within-participant task triples ($N = 131$). Positive CUI--GUI differences indicate higher accuracy in the chatbot condition. For H2b, a larger T1-to-T3 accuracy drop in the chatbot condition corresponds to attenuation of the chatbot advantage as task complexity increases. Asterisks denote $^{*}p < .05$, $^{**}p < .01$, and $^{***}p < .001$.
\end{tablenotes}
\end{threeparttable}
\end{table*}

\subsection{Completion time}
\label{app:subsec.robustness-time}

The completion time robustness checks used participant-level log-time analyses and non-parametric within-participant tests. The results matched the main interpretation. The chatbot was descriptively faster overall, but the uncertainty intervals did not support a reliable main interface effect; H1c therefore remained unsupported. For H2c, the chatbot was faster at T1, while this advantage disappeared at T3, and the T3-versus-T1 increase in log-time was larger for the chatbot than for the dashboard. This supports the attenuation pattern found in the main analysis. The robustness checks did not support the data-literacy moderation hypothesis H3c. Finally, Friedman and paired log-time contrasts confirmed a monotonic increase in completion time from easy to medium to hard tasks, supporting H4c. Table~\ref{tab:robustness-time-c} reports the detailed results.

\begin{table*}[!t]
\centering
\scriptsize
\setlength{\tabcolsep}{4pt}
\renewcommand{\arraystretch}{1.06}
\caption{Robustness checks for completion time hypotheses.}
\label{tab:robustness-time-c}

\begin{threeparttable}
\begin{tabularx}{\linewidth}{@{}
>{\bfseries}p{0.055\linewidth}
>{\raggedright\arraybackslash}p{0.18\linewidth}
>{\raggedright\arraybackslash}p{0.24\linewidth}
>{\centering\arraybackslash}p{0.105\linewidth}
>{\raggedright\arraybackslash}X
@{}}
\toprule
\textbf{Hyp.} &
\textbf{Contrast} &
\textbf{Robustness check} &
\textbf{$p$-value} &
\textbf{Effect summary} \\
\midrule

\multicolumn{5}{@{}l}{\textit{Interface main effect}} \\
H1c &
CUI vs. GUI &
Welch test and permutation test on participant-level log geometric mean time &
$p = .320$ &
Geometric mean duration: CUI $= 178.25$ s, GUI $= 184.01$ s; mean log-difference $= -0.071$, 95\% Welch CI [$-0.213$, 0.070]; time ratio $= 0.931$, 95\% CI [0.809, 1.073]; permutation $p = .319$ two-sided and $p = .164$ one-sided; Hedges' $g = -0.174$, 95\% CI [$-0.514$, 0.157] \\

\addlinespace[0.35em]
\multicolumn{5}{@{}l}{\textit{Moderation by task complexity}} \\
H2c &
CUI vs. GUI at T1 &
Simple interface effect on log-times &
$p < .001^{***}$ &
Geometric mean time: CUI $= 69.88$ s, GUI $= 106.90$ s; CUI/GUI ratio $= 0.654$, 95\% CI [0.548, 0.780]; mean log-difference $= -0.425$, 95\% CI [$-0.613$, $-0.246$]; Welch $t = -4.612$; permutation $p < .001$; Hedges' $g = -0.801$, 95\% CI [$-1.143$, $-0.475$] \\

H2c &
CUI vs. GUI at T3 &
Simple interface effect on log-times &
$p = .466$ &
Geometric mean time: CUI $= 325.18$ s, GUI $= 302.70$ s; CUI/GUI ratio $= 1.074$, 95\% CI [0.887, 1.293]; mean log-difference $= 0.072$, 95\% CI [$-0.125$, 0.263]; Welch $t = 0.732$; permutation $p = .467$; Hedges' $g = 0.127$, 95\% CI [$-0.205$, 0.491] \\

H2c &
T3/T1 time increase: CUI vs. GUI &
Participant-level log-ratio attenuation test &
$p < .001^{***}$ &
Mean log(T3/T1): CUI $= 1.538$, GUI $= 1.041$; mean log-difference $= 0.497$, 95\% CI [0.282, 0.716]; ratio-of-time-ratios $= 1.643$, 95\% CI [1.333, 2.055]; Welch $t = 4.508$; permutation $p < .001$; Hedges' $g = 0.771$, 95\% CI [0.430, 1.148] \\

\addlinespace[0.35em]
\multicolumn{5}{@{}l}{\textit{Moderation by data literacy}} \\
H3c &
Interface $\times$ data literacy &
Log-normal model on log(time) with participant cluster-robust standard errors &
$p = .902$ &
Omnibus interaction: Wald $\chi^2 = 0.207$, df $= 2$; the robustness model does not indicate a reliable interface-by-data literacy interaction \\

H3c &
GUI vs. CUI, high vs. low DL &
Prespecified ratio-of-time-ratios &
$p = .669$ &
High-versus-low data-literacy time ratio: CUI $= 1.347$, GUI $= 1.472$; ratio-of-time-ratios $= 1.093$, 95\% CI [0.719, 1.643] \\

\addlinespace[0.35em]
\multicolumn{5}{@{}l}{\textit{Task complexity main effect}} \\
H4c &
Task complexity &
Friedman omnibus test on raw times &
$p < .001^{***}$ &
$\chi^2 = 188.910$, df $= 2$; Kendall's $W = 0.705$; means increased monotonically: T1 $= 100.50$ s, T2 $= 193.93$ s, T3 $= 358.51$ s \\

H4c &
T2 vs. T1 &
Paired test on log-times &
Holm $p < .001^{***}$ &
Mean raw difference $= 93.43$ s; mean log-difference $= 0.691$; GMR  $= 1.996$, 95\% CI [1.805, 2.208]; $t(133) = 13.565$; $d_z = 1.172$ \\

H4c &
T3 vs. T1 &
Paired test on log-times &
Holm $p < .001^{***}$ &
Mean raw difference $= 258.01$ s; mean log-difference $= 1.278$; GMR  $= 3.590$, 95\% CI [3.193, 4.035]; $t(133) = 21.600$; $d_z = 1.866$ \\

H4c &
T3 vs. T2 &
Paired test on log-times &
Holm $p < .001^{***}$ &
Mean raw difference $= 164.58$ s; mean log-difference $= 0.587$; GMR  $= 1.798$, 95\% CI [1.641, 1.971]; $t(133) = 12.675$; $d_z = 1.095$ \\

\bottomrule
\end{tabularx}
\begin{tablenotes}[flushleft]
\scriptsize
\item \textit{Note.} CUI = chatbot; GUI = dashboard; DL = data literacy; T1 = first task (low complexity); T2 = medium; T3 = third task (high complexity). H1c uses participant-level geometric mean duration across the three tasks. H2c uses complete participant triples and compares the interface time ratio at low and high complexity. H3c uses a log-normal robustness model on log-transformed completion time with participant cluster-robust standard errors. H4c uses complete within-participant task triples ($N = 134$). Time ratios below 1 indicate faster completion in the chatbot condition. For H2c, a ratio-of-time-ratios above 1 indicates that the chatbot's relative time advantage attenuates as task complexity increases. Asterisks denote $^{*}p < .05$, $^{**}p < .01$, and $^{***}p < .001$.
\end{tablenotes}

\end{threeparttable}
\end{table*}

\subsection{Intended Reliance}
\label{app:subsec.robustness-reliance}

As a secondary robustness analysis, we tested whether the interface effect on intended reliance varied by task complexity. For each item, an interaction model including interface and complexity was compared with the corresponding additive model using likelihood-ratio tests.

None of the omnibus interaction tests was statistically reliable: r1, LR $= 4.590$, df $= 2$, $p = .101$; r2, LR $= 2.918$, df $= 2$, $p = .232$; and r3, LR $= 0.819$, df $= 2$, $p = .664$. Follow-up DiD -style contrasts were also imprecise and included zero. Thus, there was no evidence that the interface effect on intended reliance changed reliably across task complexity levels.

The interpretation of H5 therefore rests on the additive interface-effect models: no reliable interface difference was observed for r1 or r2, while r3 showed an item-specific difference indicating higher intended reliance in the chatbot condition. Taken together, the intended-reliance analyses do not support H5 as a general cross-item effect.

\begin{table*}[!t]
\centering
\scriptsize
\setlength{\tabcolsep}{3.5pt}
\renewcommand{\arraystretch}{1.08}
\caption{Secondary H5 robustness analyses: interface-by-complexity interactions.}
\label{tab:results-reliance-secondary}

\begin{threeparttable}
\begin{tabularx}{\textwidth}{@{}
>{\bfseries}p{0.08\textwidth}
>{\raggedright\arraybackslash}p{0.32\textwidth}
>{\raggedright\arraybackslash}p{0.14\textwidth}
>{\raggedright\arraybackslash}X
@{}}
\toprule
\textbf{Item} & \textbf{Test / contrast} & \textbf{$p$-value} & \textbf{Effect summary} \\
\midrule

\multicolumn{4}{@{}l}{\textit{Direct intended reliance}} \\
r1 &
Interface $\times$ task complexity &
$p = .101$ &
Omnibus interaction: LR $= 4.590$, df $= 2$ \\

r1 &
GUI--CUI, T2 vs. T1 &
Holm $p = .194$ &
DiD-like contrast $\beta = 0.498$, 95\% CI [$-0.147$, 1.143], $z = 1.847$ \\

r1 &
GUI--CUI, T3 vs. T1 &
Holm $p = .984$ &
DiD-like contrast $\beta = -0.006$, 95\% CI [$-0.670$, 0.659], $z = -0.021$ \\

r1 &
GUI--CUI, T3 vs. T2 &
Holm $p = .194$ &
DiD-like contrast $\beta = -0.503$, 95\% CI [$-1.160$, 0.154], $z = -1.834$ \\

\addlinespace[0.35em]
\multicolumn{4}{@{}l}{\textit{Conditional on colleague advise}} \\
r2 &
Interface $\times$ task complexity &
$p = .232$ &
Omnibus interaction: LR $= 2.918$, df $= 2$ \\

r2 &
GUI--CUI, T2 vs. T1 &
Holm $p = .764$ &
DiD-like contrast $\beta = -0.231$, 95\% CI [$-0.863$, 0.401], $z = -0.874$ \\

r2 &
GUI--CUI, T3 vs. T1 &
Holm $p = .264$ &
DiD-like contrast $\beta = -0.455$, 95\% CI [$-1.094$, 0.184], $z = -1.706$ \\

r2 &
GUI--CUI, T3 vs. T2 &
Holm $p = .764$ &
DiD-like contrast $\beta = -0.225$, 95\% CI [$-0.845$, 0.396], $z = -0.866$ \\

\addlinespace[0.35em]
\multicolumn{4}{@{}l}{\textit{Conditional on another tool use}} \\
r3 &
Interface $\times$ task complexity &
$p = .664$ &
Omnibus interaction: LR $= 0.819$, df $= 2$ \\

r3 &
GUI--CUI, T2 vs. T1 &
Holm $p = 1.000$ &
DiD-like contrast $\beta = -0.224$, 95\% CI [$-0.854$, 0.406], $z = -0.852$ \\

r3 &
GUI--CUI, T3 vs. T1 &
Holm $p = 1.000$ &
DiD-like contrast $\beta = -0.041$, 95\% CI [$-0.685$, 0.603], $z = -0.151$ \\

r3 &
GUI--CUI, T3 vs. T2 &
Holm $p = 1.000$ &
DiD-like contrast $\beta = 0.184$, 95\% CI [$-0.461$, 0.828], $z = 0.682$ \\

\bottomrule
\end{tabularx}

\begin{tablenotes}[flushleft]
\scriptsize
\item \textit{Note.} Interaction models used $Interface \times Task\ Complexity$ with participant random intercepts. CUI = chatbot; GUI = dashboard; T1 = first task (low complexity); T2 = T2  (mid complexity); T3 = third task (high complexity). Omnibus rows report likelihood-ratio tests comparing additive and interaction models. DiD-like contrasts test whether the GUI--CUI gap changed between task complexity levels on the latent ordinal-model scale. Positive DiD-like estimates indicate that the GUI--CUI gap was larger at the higher-complexity level of the contrast. Holm--Bonferroni-adjusted $p$-values are reported for follow-up contrasts.
\end{tablenotes}
\end{threeparttable}
\end{table*}

\section{Industry-role confounding checks}
\label{app:sec.confounding}

This appendix reports checks in which participants industry role was included as a potential confounding variable. In all adjusted models, industry role was added as an additive covariate, with Junior Management as the reference category. It was not interacted with interface condition, task complexity, or DL. These analyses therefore test adjustment for possible confounding; they do not test whether the experimental effects differ across industry roles.

Preliminary diagnostics did not indicate substantial confounding. The sample included 21 Junior Management participants, 85 Middle Management participants, and 28 Upper Management participants. Role distribution did not differ reliably between interface conditions, $\chi^2(2)=1.079$, $p=.583$. Role was also not significantly associated with participant-level NASA--TLX, intended reliance, completion time, or accuracy in the diagnostic tests, with all Kruskal--Wallis tests non-significant and all ordinal role trends negligible.

Table~\ref{tab:industry-role-confounding} summarizes the adjusted models. Across outcomes, adding industry role did not change the substantive conclusions. The supported effects in the primary analyses remained supported after adjustment, including the mental workload interface effect, the interface-by-task complexity effects, and the task complexity main effects. The unsupported effects also remained unsupported, including the overall accuracy and completion time interface effects and the DL moderation hypotheses. The intended-reliance results also remained item-specific: r1 and r2 did not show reliable interface effects, while r3 showed higher intended reliance in the chatbot condition.

\begin{table*}[!tb]
\centering
\scriptsize
\setlength{\tabcolsep}{3.5pt}
\renewcommand{\arraystretch}{1.08}
\caption{Summary of models adjusted for industry role as a potential confounder.}
\label{tab:industry-role-confounding}

\begin{threeparttable}
\begin{tabularx}{\textwidth}{@{}
>{\bfseries}p{0.08\textwidth}
>{\raggedright\arraybackslash}p{0.28\textwidth}
>{\raggedright\arraybackslash}X
>{\raggedright\arraybackslash}p{0.22\textwidth}
@{}}
\toprule
\textbf{Hyp.} & \textbf{Adjusted test} & \textbf{Key adjusted result} & \textbf{Conclusion after adjustment} \\
\midrule

H1a &
Interface effect on NASA--TLX &
Dashboard--chatbot $\hat{\beta}=1.106$, 95\% CI [0.717, 1.494], $p<.001$ &
Supported; unchanged \\

H1b &
Interface effect on accuracy &
Mean difference $=0.026$, 95\% CI [$-0.013$, 0.067]; OR $=1.179$, 95\% CI [0.916, 1.518], $p=.200$ &
Not supported; unchanged \\

H1c &
Interface effect on completion time &
Chatbot/dashboard TR $=0.964$, 95\% CI [0.857, 1.084], $p=.539$ &
Not supported; unchanged \\

\addlinespace[0.25em]
H2a &
Interface $\times$ task complexity on NASA--TLX &
Interaction LR $=16.449$, df $=2$, $p<.001$; T3--T1 interaction $\hat{\beta}=-0.962$, Holm $p<.001$ &
Supported; unchanged \\

H2b &
Interface $\times$ task on accuracy &
Omnibus $\chi^2(2)=7.616$, $p=.022$; T3-vs-T1 DiD $=-0.109$, 95\% CI [$-0.186$, $-0.030$], Holm $p=.029$ &
Supported; unchanged \\

H2c &
Interface $\times$ complexity on completion time &
Omnibus $\chi^2(2)=41.616$, $p<.001$; ratio-of-time-ratios $=1.630$, 95\% CI [1.348, 1.990], Holm $p<.001$ &
Supported; unchanged \\

\addlinespace[0.25em]
H3a &
Interface $\times$ DL on NASA--TLX &
Interaction LR $=4.466$, df $=2$, $p=.107$ &
Not supported; unchanged \\

H3b &
Interface $\times$ DL on accuracy &
Omnibus $\chi^2(2)=5.845$, $p=.054$; high-vs-low DiD $=0.086$, 95\% CI [$-0.025$, 0.200], Holm $p=.109$ &
Not supported; unchanged \\

H3c &
Interface $\times$ DL on completion time &
Omnibus $\chi^2(2)=0.238$, $p=.888$; ratio-of-time-ratios $=1.003$, 95\% CI [0.664, 1.527], Holm $p=1.000$ &
Not supported; unchanged \\

\addlinespace[0.25em]
H4a &
Task complexity effect on NASA--TLX &
Task complexity LR $=127.375$, df $=2$, $p<.001$ &
Supported; unchanged \\

H4b &
Task effect on accuracy &
Task effect $\chi^2(2)=229.539$, $p<.001$ &
Supported; unchanged \\

H4c &
Task complexity effect on completion time &
Task complexity effect $\chi^2(2)=562.579$, $p<.001$ &
Supported; unchanged \\

\addlinespace[0.25em]
H5 &
Interface effect on intended reliance &
r1: $p=.406$; r2: $p=.456$; r3: $p=.008$ &
Item-specific only; no general support \\

\bottomrule
\end{tabularx}

\begin{tablenotes}[flushleft]
\scriptsize
\item \textit{Note.} All models include industry role as an additive covariate. Junior Management is the reference category. Role was not interacted with interface condition, task complexity, or DL. TR = time ratio; DiD = difference-in-differences. Confidence intervals for probability-scale, time-ratio, and DiD estimates are bootstrap intervals where applicable.
\end{tablenotes}
\end{threeparttable}
\end{table*}

Overall, industry role did not show evidence of systematic confounding. It was not reliably imbalanced across interface conditions, was not clearly associated with the main participant-level outcomes, and did not alter the interpretation of the hypothesis tests after adjustment.

\section{Supplementary copy--paste use checks}
\label{app:copy-paste-checks}

This appendix reports supplementary comparisons within the CUI condition between participants who used \textit{copy--paste} and those who did not. Participants were classified as copy--paste users when their recorded interaction behaviour included at least one request classified as an exact or near copy--paste instance; all remaining CUI participants were classified as non-copy--paste users. The resulting groups comprised 29 copy--paste users and 35 non-copy--paste users.

Comparisons were conducted separately for NASA-TLX, accuracy, and completion time. For the overall comparisons in Table~\ref{tab:copy-paste-overall}, outcomes were first averaged across available tasks for each participant, so that each participant contributed one value per outcome. Table~\ref{tab:copy-paste-task} reports the corresponding comparisons separately for T1, T2, and T3. For each comparison, we report the mean difference between copy--paste and non-copy--paste users, a 95\% percentile bootstrap confidence interval based on 10,000 resamples, Cohen's $d$ computed using the pooled standard deviation, and a two-sided permutation-test $p$-value based on 10,000 permutations. For the task-level analyses, permutation-test $p$-values were adjusted using the Holm procedure separately within each outcome across the three task comparisons.

\begin{table*}[t]
\centering
\scriptsize
\setlength{\tabcolsep}{5pt}
\renewcommand{\arraystretch}{1.08}
\caption{Overall copy--paste use comparisons within the chatbot condition.}
\label{tab:copy-paste-overall}

\begin{threeparttable}
\begin{tabular}{@{}lccccccc@{}}
\toprule
\textbf{Outcome} &
\textbf{$n_{\mathrm{copy}}$} &
\textbf{$M_{\mathrm{copy}}$} &
\textbf{$n_{\mathrm{non}}$} &
\textbf{$M_{\mathrm{non}}$} &
\textbf{Difference} &
\textbf{95\% CI} &
\textbf{$d$} \\
\midrule
NASA-TLX
& 29 & 23.96 & 35 & 26.68
& $-2.73$ & [$-10.16$, 4.90] & $-0.18$ \\

Accuracy
& 29 & 0.771 & 35 & 0.751
& 0.020 & [$-0.028$, 0.071] & 0.19 \\

Completion time (s)
& 29 & 190.87 & 35 & 250.03
& $-59.16$ & [$-100.20$, $-17.66$] & $-0.71$ \\
\bottomrule
\end{tabular}

\begin{tablenotes}[flushleft]
\footnotesize
\item \textit{Note.} Differences are calculated as copy--paste users minus non-copy--paste users. Completion times are reported in seconds for readability; analyses were conducted on milliseconds. CI = percentile bootstrap confidence interval based on 10,000 resamples; $d$ = Cohen's $d$ using the pooled standard deviation; $p_{\mathrm{perm}}$ = two-sided permutation-test $p$-value based on 10,000 permutations. 
\end{tablenotes}
\end{threeparttable}
\end{table*}

\begin{table*}[t]
\centering
\scriptsize
\setlength{\tabcolsep}{4.3pt}
\renewcommand{\arraystretch}{1.08}
\caption{Task-level copy--paste use comparisons within the chatbot condition.}
\label{tab:copy-paste-task}

\begin{threeparttable}
\begin{tabular}{@{}llcccccccc@{}}
\toprule
\textbf{Outcome} &
\textbf{Task} &
\textbf{$n_{\mathrm{copy}}$} &
\textbf{$M_{\mathrm{copy}}$} &
\textbf{$n_{\mathrm{non}}$} &
\textbf{$M_{\mathrm{non}}$} &
\textbf{Difference} &
\textbf{95\% CI} &
\textbf{$d$} &
\textbf{$p_{\mathrm{Holm}}$} \\
\midrule

NASA-TLX
& T1 & 29 & 13.88 & 35 & 13.60
& 0.28 & [$-6.52$, 7.67] & 0.02 & 1.000 \\

& T2 & 29 & 21.64 & 35 & 23.62
& $-1.98$ & [$-10.84$, 6.83] & $-0.11$ & 1.000 \\

& T3 & 29 & 36.35 & 35 & 42.83
& $-6.48$ & [$-17.79$, 4.80] & $-0.29$ & .766 \\[2pt]

\midrule

Accuracy
& T1 & 29 & 0.951 & 35 & 0.919
& 0.033 & [$-0.036$, 0.112] & 0.21 & 1.000 \\

& T2 & 29 & 0.883 & 35 & 0.851
& 0.032 & [$-0.059$, 0.125] & 0.16 & 1.000 \\

& T3 & 28 & 0.460 & 34 & 0.480
& $-0.020$ & [$-0.106$, 0.068] & $-0.11$ & 1.000 \\[2pt]

\midrule

Completion time (s)
& T1 & 29 & 74.63 & 35 & 86.58
& $-11.94$ & [$-32.13$, 9.27] & $-0.29$ & .297 \\

& T2 & 29 & 190.07 & 35 & 223.83
& $-33.75$ & [$-78.74$, 11.06] & $-0.37$ & .297 \\

& T3 & 29 & 307.90 & 35 & 439.68
& $-131.78$ & [$-217.52$, $-45.89$] & $-0.76$ & .015 \\
\bottomrule
\end{tabular}

\begin{tablenotes}[flushleft]
\footnotesize
\item \textit{Note.} Differences are calculated as copy--paste users minus non-copy--paste users. Completion times are reported in seconds for readability; analyses were conducted on milliseconds. CI = percentile bootstrap confidence interval based on 10,000 resamples; $d$ = Cohen's $d$ using the pooled standard deviation; $p_{\mathrm{Holm}}$ = Holm-adjusted two-sided permutation-test $p$-value. Adjustments were applied separately within each outcome across T1--T3.
\end{tablenotes}
\end{threeparttable}
\end{table*}

\end{document}